\documentclass[11pt,a4paper]{article}
\pdfoutput=1
\usepackage{jheppub}
\usepackage{amsfonts}
\usepackage{amssymb}
\usepackage{bbm}
\usepackage{graphicx}
\usepackage{caption}
\usepackage{subcaption}

\usepackage[normalem]{ulem}

\def\be{\begin{equation}}
\def\ee{\end{equation}}
\def\bea{\begin{eqnarray}}
\def\eea{\end{eqnarray}}

\newcommand{\nn}{\nonumber}

\newcommand\diff{\mathrm{d}}

\numberwithin{equation}{section}       

\def\rme{{\rm e}}
\def\elpot{\varphi}
\def\rBPS{r_{*}}
\def\OmBPS{\Omega^{*}}
\def\PhiBPS{\Phi^{*}}
\def\EBPS{E^{*}}
\def\JBPS{J^{*}}
\def\QBPS{Q^{*}}
\def\SBPS{S^{*}}
\def\IBPS{I^{*}}
\def\omBPS{\omega^{*}}
\def\elpotBPS{\varphi^{*}}

\def\coeff{\mu}

\renewcommand{\=}{\,= \,}

\renewcommand{\d}{\delta}
\renewcommand{\a}{\alpha}
\renewcommand{\b}{\beta}
\newcommand{\g}{\gamma}
\newcommand\G{\Gamma}
\renewcommand{\nn}{\nonumber}
\newcommand{\wh}{\widehat}
\newcommand{\wt}{\widetilde}

\newcommand{\ve}{\varepsilon}
\newcommand{\CN}{\mathcal{N}}

\newcommand{\CH}{\mathcal{H}}

\newcommand{\p}{\partial}

\newcommand{\CV}{\mathcal V}
\newcommand{\CI}{\mathcal I}
\newcommand{\CF}{\mathcal F}

\renewcommand{\v}{\varphi}

\newcommand\st{\psi_{\ve} }
\newcommand\sh{\psi_{\wh{\ve}} }
\newcommand\stb{\overline{\psi}_{\ve} }
\newcommand\shb{\overline{\psi}_{\wh{\ve}} }

\newcommand\bareR{
 \, {\ve^R}^* \gamma_1}
\newcommand\bareL{
\, {\ve^L}^* \gamma_1}
\newcommand\TwistV{\{\phi,\st,\sh,F_{\phi}\}}

\newcommand \Xel{\phi, \overline{\phi},\st,\stb}
\newcommand \XelB{\phi, \overline{\phi}}
\newcommand\XelF{\st,\stb}
\newcommand \QXelB{\sqrt{2} \,i  \sh,\sqrt{2} \,i \shb}
\newcommand \QXelF{\sqrt{2} \, F_\phi, \sqrt{2} \, \overline{F}_{\overline{\phi}}}

\newcommand\evpar{\frac{2 \pi n_0 }{\beta} + i \,\frac{ n_1}{\beta}\, \omega_1  + i\, \frac{n_2}{\beta}\, \omega_2 }

\newcommand\regL{ \qquad n_1 \,,  n_2 \geq 0}
\newcommand\regR{\qquad n_1 \,, n_2 \leq 0}
\newcommand\Gth[3]{\Gamma_3(#1\, | \, 1,#2,#3)}

\newcommand\pr[2]{\prod_{n_t=#1}^{#2}\prod_{n_1, \, n_2\in \mathbb{N}}}

\newcommand\Gell[3]{\prod_{j,k=0}^{\infty}\frac{1-\rme^{-2 \pi i #1}\rme^{
2 \pi i #2 (j+1)}\rme^{
2 \pi i #3(k+1)}}{1-\rme^{2 \pi i #1} \, \rme^{
2 \pi i #2 j} \, \rme^{2 \pi i #3 k}}}

\newcommand\uVartwJ{  \bigl( (r_I-1)^3-(r_I -1) \bigr) \frac{1}{6 } }
\newcommand\prADJ{\prod_{\rho \in \text{Adj(G)}}}

\newcommand\pexp[3]{\exp \left( \sum\limits_{n=1}^{\infty }\frac{1}{n} \frac{(#3)^n}{(1-#1^n)(1-#2^n)}\right)}

\usepackage{array}

\newcommand{\m}[4]{\biggl( \begin{matrix} 
#1 &#2 \\
#3 & #4 
\end{matrix} \biggr)}

\newcommand{\Lv}[2]{\bigl( \begin{matrix} 
#1 &#2 
\end{matrix} \bigr)}

\newcommand{\Rv}[2]{\biggl( \begin{matrix} 
#1 \\
#2 \end{matrix} \biggr)}

\newcommand{\dmat}{\m{D_{00}}{D_{01}}{D_{10}}{D_{11}}}

\newcommand\Gththr[4]{\Gamma_3(#1\, | \,#2,#3,#4)}
\newcommand\oarrow[1]{\overset{\rightarrow}{#1}}
\newcommand\zetathr[3]{\zeta_3(0,#1 \, | \,1,#2,#3)}
\newcommand\prID[2]{\prod^{(\infty,\infty)}_{\overset{\rightarrow}{m}=(0,0)
}\frac{(1-\rme^{2 \pi i ((#1)+\oarrow{m}\cdot \oarrow{\a})})}{(1-\rme^{2 \pi i ((#2)+\oarrow{m}\cdot \oarrow{\a})})}}
\newcommand\Rca{\sum_{I}  \text{dim(R$_I$)} \left(3 (r_I-1)^3-(r_I-1)\right)}
\newcommand\Rcc{\sum_{I}  \text{dim(R$_I$)} \left(9 (r_I-1)^3-5 (r_I-1)\right)}

\renewcommand\pr[2]{\prod_{n_0,\,n_1,\,n_2\,=\,#1}^{#2}
}

\newcommand{\newt}{x}
\def\ccc{c}

\title{Microscopic origin of the Bekenstein-Hawking entropy of supersymmetric AdS$_5$ black holes}

\author[a]{Alejandro Cabo-Bizet,}
\emailAdd{alejandro.cabo\_bizet@kcl.ac.uk}
\author[b]{Davide Cassani,}
\emailAdd{davide.cassani@pd.infn.it}
\author[a,c]{Dario Martelli,\footnotetext[3]{On leave at the Galileo Galilei Institute, Largo Enrico Fermi, 2, 50125 Firenze, Italy.}}
\emailAdd{dario.martelli@kcl.ac.uk}
\author[a]{Sameer Murthy}
\emailAdd{sameer.murthy@kcl.ac.uk}
\affiliation[a]{Department of Mathematics, King's College London,\\
The Strand, London WC2R 2LS, U.K.}
\affiliation[b]{\it INFN, Sezione di Padova, \\
Via Marzolo 8, 35131 Padova, Italy}

\abstract{We present a holographic derivation of the entropy of supersymmetric asymptotically AdS$_5$ black holes. 
We define a BPS limit of black hole thermodynamics by first focussing on a supersymmetric family of complexified solutions 
and then reaching extremality. We show that in this limit the black hole entropy is the Legendre transform of the on-shell 
gravitational action with respect to three chemical potentials subject to a constraint. This constraint follows from supersymmetry 
and regularity in the Euclidean bulk geometry. 
Further, we calculate, using localization, the exact partition function of the dual~$\CN=1$ SCFT on a twisted~$S^1\times S^3$ 
with complexified chemical potentials obeying this constraint. This defines a generalization of the supersymmetric Casimir energy, 
whose Legendre transform at large $N$ exactly reproduces the Bekenstein-Hawking entropy of the black hole. 
}

\begin{document}

\maketitle


\section{Introduction}

One of the striking successes of string theory is the microscopic description of certain classes of black 
holes~\cite{Sen:1995in,Strominger:1996sh}. Key to these advances is the realization of supersymmetric 
black holes in terms of configurations of microscopic constituents of string theory, like strings and branes. 
The degenerate excitations of these microscopic constituents give rise to a statistical Boltzmann entropy, 
which reproduces the thermodynamic Bekenstein-Hawking entropy of the black hole in a class of supersymmetric 
situations. These developments led to the formulation of the far-reaching AdS/CFT correspondence, which 
gives a very general approach to understanding the microscopic nature of black holes, including thermal black 
holes which do not preserve supersymmetry. Our current understanding of black holes, quite generally, is to 
view them as an ensemble of microscopic constituents in a dual quantum field theory. 

During the past twenty years several black hole solutions in Anti de Sitter (AdS) space-time in various dimensions 
have been constructed, including rotating and charged black holes. The broad picture of these black holes as being 
made up of microscopic excitations of the boundary theory has been checked in a variety of cases, and typically the 
accuracy of these checks is only limited by 
 the technology available to us
for computing strongly coupled quantities in the dual field theories. 
The leading behavior of the entropy at large~$N$, for example, bears out quite beautifully in most examples. In particular, in~\cite{Benini:2015eyy} the 
Bekenstein-Hawking entropy of a class of supersymmetric  AdS$_4$ black holes was successfully reproduced  from an exact 
localization computation in the dual ABJM theory~\cite{Benini:2015noa}.
One situation, however, sticks out---the dual field theoretic understanding of supersymmetric black holes in the original, 
emblematic,~AdS$_5/$CFT$_4$ correspondence has remained an enigma. 
Despite the fact that supersymmetric black holes in AdS$_5$ were constructed almost fifteen years 
ago~\cite{Gutowski:2004ez} and many generalizations have been found in the gravitational theory since 
then~\cite{Gutowski:2004yv,Chong:2005hr,Chong:2005da,Kunduri:2006ek}, they have so far eluded all attempts to 
obtain a statistical explanation of their entropy.  In this paper we will present a  derivation of the statistical entropy of a 
general class of supersymmetric AdS$_5$ black holes constructed in~\cite{Chong:2005hr}, which contains, as a 
special case, the black holes of~\cite{Gutowski:2004ez}. Since these black holes are solutions to five-dimensional 
minimal gauged supergravity, the holographically dual state should exist in any four-dimensional $\mathcal{N}=1$ 
SCFT with a weakly coupled supergravity dual. As a specific case, we will also analyze $\mathcal{N}=4$ SYM and in this way reproduce the entropy of the black holes of~\cite{Gutowski:2004yv,Chong:2005da,Kunduri:2006ek}, which have an embedding in type IIB supergravity on $S^5$.

Recall that the Bekenstein-Hawking entropy of five-dimensional black holes, expressed in terms of dual field theory variables, 
scales like $N^2$ in the large $N$ limit.  One could hope that the degeneracies of BPS states contributing to the 
entropy of the black hole can be estimated by evaluating the large $N$ limit of certain supersymmetric indices, as first 
investigated in~\cite{Kinney:2005ej}. However, the index defined in~\cite{Kinney:2005ej,Romelsberger:2005eg} and some of its
generalizations \cite{Janik:2007pm,Grant:2008sk,Chang:2013fba}
were found to behave as~$\mathcal{O}(1)$ in the large $N$ limit, thus failing 
to account for the correct~$\mathcal{O}(N^2)$ behaviour of the black hole entropy. One possible explanation that has been offered 
for this discrepancy is that the index counts bosonic and fermionic states with a sign, and there could be large cancellations
which reduce the index drastically. On the other hand, it has proven to be very difficult to directly rule out the contribution of the 
supersymmetric black hole from the gravitational ensemble, leaving this problem in a somewhat inconclusive state for a long time.

From the point of view of holography, it is most natural to relate the physics of the bulk to that of the boundary field theory 
through the ``master formula'' of~AdS/CFT, that instructs us to identify the gravitational on-shell action of a given solution 
with the corresponding functional integral of the dual field theory at large~$N$. This will lead us to perform a localization calculation~\cite{Witten:1988ze,Nekrasov:2002qd,Pestun:2007rz} on the field theory side and derive a generalization of the supersymmetric Casimir energy~\cite{Assel:2014paa}. We find that in the large $N$ limit not only this generalization scales as $\mathcal{O}(N^2)$, but it also precisely reproduces the entropy of the black holes. The supersymmetric Casimir energy has been shown to be independent of the regularization scheme and therefore a physical observable \cite{Assel:2015nca}.

In order to implement this idea there are some technical issues  that we need to address. 
As a starting point we consider the thermal ensemble, for which it is clear that the finite temperature Euclidean black hole 
is a saddle point of the gravitational functional integral~\cite{Witten:1998qj}. The issue becomes more subtle for the supersymmetric black 
hole at zero temperature because the geometry near the horizon of a supersymmetric black hole looks locally like an infinite throat,
and it is not obvious how to regulate this problem while keeping the asymptotic AdS$_5$ 
geometry.\footnote{The method of~\cite{Sen:2008yk} regulates 
this problem for generic supersymmetric black holes by taking a~$T\to 0$ limit of the thermal black hole and keeping the 
horizon always at finite distance. While this gives a nice definition of BPS black hole entropy, it does so by zooming in on the 
near-horizon~AdS$_2$ region, thus throwing away the asymptotics of the black hole solution. It is therefore not useful if 
our goal is to explain the black hole entropy as excitations in the~CFT$_4$ dual to 
the asymptotic~AdS$_5$ space. The corresponding problem from the holographic dual point of view is that the definition 
of~\cite{Sen:2008yk} casts the supersymmetric black hole entropy as the degeneracy of the~CFT$_1$ dual to the near-horizon~AdS$_2$, 
but one does not have good control over this~CFT$_1$.}
Ideally we would like to define a gravitational ensemble on asymptotic AdS$_5$ space which includes the 
supersymmetric black hole as a saddle point. We would expect that the on-shell action of that saddle point would then be 
related to the black hole entropy. Here we face a problem because the so-called quantum statistical relation that relates the black hole entropy to the Euclidean on-shell action is a priori valid for black holes at non-zero temperature with non-degenerate Killing horizons.

In this paper we will define a ``supersymmetric quantum statistical relation'' of the BPS black holes by taking an extremal limit 
of a family of supersymmetric, complexified solutions, for which a finite on-shell action and a formal ``temperature'' 
can be defined. For AdS$_5$ black holes the notions of extremality and supersymmetry do not coincide, therefore there are 
different ways to approach the BPS solution.\footnote{In this paper we take care distinguishing between supersymmetry 
and extremality, and will use the term ``BPS'' to denote a quantity evaluated after both supersymmetry and extremality have been imposed.} 
Indeed, an analogous limiting procedure, which served as an inspiration for us, was proposed  in~\cite{Silva:2006xv},
but as we will explain this procedure does not reach the BPS locus following a supersymmetric trajectory. 
Our limit, on the other hand, is taken along a supersymmetric trajectory by construction, 
at the expense of having to work with \emph{complex} charges and chemical potentials. 
With this limiting procedure, we find that we have to deal with a cigar-like geometry that is capped off in the 
infra-red. 
As we shall see, this  is a crucial feature which imposes a constraint that can be expressed as
\be \label{crucialconstraint}
 \omega_1 + \omega_2 -2 \elpot \=  2\pi  i n \, ,
\ee
 where $\omega_1,\omega_2$ and  $\elpot$  denote the chemical potentials and $n=\pm1$.
Relatedly, the fermions which are charged under the symmetries conjugate to the chemical potentials have 
anti-periodic boundary conditions on the~$S^1$.
We will show that in our BPS limit the on-shell action $I$ takes the particularly simple form
\be  \label{Iresult}
I = \frac{2\pi}{27 G g^3}\frac{\elpot^3}{\omega_1\omega_2}\ ,
\ee
 and that its (constrained) Legendre transform equals the Bekenstein-Hawking entropy $S$ of the black hole, i.e.,
\be \label{mainconstraint}
S(J_1,J_2,Q) \= -I(\omega_1,\omega_2,\elpot) - \omega_1  J_1 - \omega_2  J_2 - \elpot \, Q \ ,
\ee
where $J_1, J_2$ are the angular momenta and $Q$ is (dual to) the R-charge. 
Here the chemical potentials are saddle points of the constrained extremization.

Our regulated geometry in the bulk identifies a boundary geometry that corresponds to the background in which the dual SCFT is defined. 
We perform an exact localization computation and show that the partition function takes the form $Z = \rme^{-\mathcal{F}}\mathcal{I}$, 
where both factors depend on complex chemical potentials $\omega_1,\omega_2,\varphi$ satisfying \eqref{crucialconstraint}. As we will show,  $\mathcal{I}$ is essentially 
the superconformal index, while at large $N$ the prefactor reads $\mathcal{F}=-\frac{16}{27}\frac{\varphi^3}{\omega_1\omega_2}{\bf c}$, where ${\bf c}$ 
is the central charge that appears in the Weyl anomaly, which scales as $\mathcal{O}(N^2)$. This prefactor matches {\it minus} the 
supergravity on-shell action. 
Therefore $-\mathcal{F}$ also yields the black hole entropy upon Legendre transform. From a field theory perspective, one may think that the right hand side of \eqref{crucialconstraint} can be reabsorbed by a large gauge transformation (or by a large diffeomorphism) redefining the chemical potentials, so that the value of $n$ does not affect the physics. However both the result for $\mathcal{F}$ and its Legendre transform do depend on $n$. For $n=\pm 1$ we are led to the black hole entropy, while for $n=0$ $\mathcal{F}$ is proportional to the supersymmetric Casimir energy \cite{Assel:2014paa,Assel:2015nca} and yields a vanishing entropy.

The fact that the BPS~AdS$_5$ black hole entropy can be rewritten as an extremization of the 
quantity~\eqref{Iresult} with respect to the three chemical potentials constrained by~\eqref{crucialconstraint},
was first observed in the interesting paper~\cite{Hosseini:2017mds}. 
This crucial observation begs the question of a physical interpretation, and it was an important starting point for our investigations. 
Using the ideas sketched above, in this paper we provide an independent derivation of this extremization principle. 
It is also important to note the analogies with the problem in one dimension lower~\cite{Benini:2015eyy,Benini:2016rke}. Here the Bekenstein-Hawking entropy of a class of supersymmetric AdS$_4$ black holes is obtained by computing the Legendre transform of a topologically twisted index of the 
three-dimensional theory~\cite{Benini:2015noa}, evaluated in the large-$N$ limit. This Legendre transformation was related to the attractor mechanism for the scalar fields in the bulk.
Based on these considerations, the authors of~\cite{Hosseini:2017mds} related the extremization principle for AdS$_5$ black holes to the attractor mechanism in one dimension lower. The relation between the entropy of supersymmetric AdS$_4$ black holes and their on-shell supergravity action was discussed in~\cite{Azzurli:2017kxo,Halmagyi:2017hmw,Cabo-Bizet:2017xdr}.

The rest of this paper is organized as follows. In Section~\ref{sec:CCLPsolQSR} we review the asymptotically AdS$_5$ solutions 
of~\cite{Chong:2005hr}, emphasizing the features that will be important for deriving our  results. In particular, we will consider a 
one-parameter family extension of the supersymmetric black holes that will be instrumental for relating the entropy to the  Euclidean 
action of the  physical solution, defined through a particular BPS limit. In Section~\ref{sec:BPSlimit} we discuss this limit and obtain 
a quantum statistical relation for the supersymmetric black holes. We will also explain how this limiting procedure defines the rigid 
background of the field theory at the boundary, through the interplay of supersymmetry and regularity conditions in the interior, 
thus informing the dual field theory computation. This is performed in Section~\ref{field_theory_section}, which can be read 
independently of the rest of the paper. Here we compute a variant of the localized partition function on twisted $S^1\times S^3$ manifolds, 
that gives rise to a generalization of the supersymmetric Casimir energy. The results of this section are valid for 
arbitrary ${\cal N}=1$ field theories with a Lagrangian description and an $R$-symmetry. In Section~\ref{sec:discussion} we summarise our findings and discuss several possible extensions of our work. Four appendices contain further details on calculations both in gravity and in field theory.

\vskip 5mm 

\paragraph{Note added in v3:} After the first version of this  paper was submitted to  the arXiv, a number  of  interesting 
papers have discussed the problem of reproducing the  entropy of supersymmetric AdS$_5$ black holes from a dual field theory calculation. Notably, \cite{Choi:2018hmj, Honda:2019cio, ArabiArdehali:2019tdm} showed that for  $\mathcal{N}=4$ SYM the entropy function that leads to the entropy of the dual 
AdS$_5$ black holes can be extracted from a Cardy-like limit of the supersymmetric index, in suitable ranges of the fugacities.  
See also~\cite{Choi:2018vbz} for related work. 
In addition, \cite{Benini:2018ywd} showed that the same entropy function can arise in the large $N$ limit of the index. 
In this version of our paper we will make some comments on the relation of our results to these papers, 
leaving a more detailed investigation for future research.


\section{Review of AdS$_5$ black hole solutions \label{sec:CCLPsolQSR}}

In this section we review the asymptotically AdS$_5$ black hole solutions to minimal five-dimensional gauged supergravity found in \cite{Chong:2005hr}, emphasizing the features that will be important in the following.
 
In the conventions of \cite{Chong:2005hr}, the Lagrangian for the
bosonic sector of minimal five-dimensional gauged supergravity is\footnote{The gauge field appearing here is related to the one in \cite{Chong:2005hr} as $A^{\rm CCLP} = \frac{2}{\sqrt 3\,g}A$. Although this gives a non-canonically normalized kinetic term, the boundary value of our $A$ couples to the R-current of the dual $\mathcal{N}=1$ SCFT canonically. Our electric charge $Q$ is therefore related to the one in \cite{Chong:2005hr} by $Q^{\rm CCLP} = \frac{\sqrt 3 \,g}{2}Q$ and is dimensionless.\label{foot:rescaleA}}
\be
{\cal L} = (R+ 12g^2)\, {*1} - \tfrac{2}{3g^2} F\wedge{*F} + \tfrac{8}{27g^3}
      F\wedge F\wedge A\ ,
\ee
where $A$ is the graviphoton with field strength $F=\diff A$, while $g>0$ controls the cosmological constant, normalized so that the AdS solution has radius $1/g$.   
Any solution to this theory can be uplifted locally to type IIB supergravity on $S^5$ or, more generally, on a Sasaki-Einstein five-fold \cite{Buchel:2006gb,Gauntlett:2007ma}.

It was found in \cite{Chong:2005hr} that an asymptotically AdS solution to the equations of motion is given by:
\begin{eqnarray}
\diff s^2 &=& -\frac{\Delta_\theta\, [(1+g^2 r^2)\rho^2 \diff t + 2q \nu]
\, \diff t}{\Xi_a\, \Xi_b \, \rho^2} + \frac{2q\, \nu\omega}{\rho^2}
+ \frac{f}{\rho^4}\Big(\frac{\Delta_\theta \, \diff t}{\Xi_a\Xi_b} -
\omega\Big)^2 + \frac{\rho^2 \diff r^2}{\Delta_r} +
\frac{\rho^2 \diff \theta^2}{\Delta_\theta}\nn\\
&& + \frac{r^2+a^2}{\Xi_a}\sin^2\theta\, \diff \phi^2 +
      \frac{r^2+b^2}{\Xi_b} \cos^2\theta\, \diff \psi^2\ ,\label{5met}\\
A &=& \frac{3g q}{2\rho^2}\,
         \Big(\frac{\Delta_\theta\, \diff t}{\Xi_a\, \Xi_b}
       - \omega\Big) + \alpha\, \diff t\ ,\label{gaugepot}
\end{eqnarray}
where
\begin{align}\label{CCLPfunctions}
\nu &= b\sin^2\theta\, \diff \phi + a\cos^2\theta\, \diff \psi\,,\qquad\quad
\omega = a\sin^2\theta\, \frac{\diff \phi}{\Xi_a} +
              b\cos^2\theta\, \frac{\diff \psi}{\Xi_b}\ ,\nn\\
\Delta_r &= \frac{(r^2+a^2)(r^2+b^2)(1+g^2 r^2) + q^2 +2ab q}{r^2} - 2m
\ ,\nn\\
\Delta_\theta &= 1 - a^2 g^2 \cos^2\theta -
b^2 g^2 \sin^2\theta\,,\qquad
\rho^2 = r^2 + a^2 \cos^2\theta + b^2 \sin^2\theta\,,\nn\\
\Xi_a &=1-a^2 g^2\,,\quad \Xi_b = 1-b^2 g^2\ ,\ \ \qquad
f= 2 m \rho^2 - q^2 + 2 a b q g^2 \rho^2\ .
\end{align}
In the gauge field  we have introduced an arbitrary constant $\alpha$ which parameterizes a gauge choice  that will be important later.
The coordinates $\phi,\psi$ are taken $2\pi$ periodic, while $\theta \in [0,\pi/2]$, so that together these parameterize a three-sphere $S^3$ (seen as a torus fibration over the interval parameterized by $\theta$).

The solution depends on the four parameters $a, b,m, q$, with the first two satisfying $a^2g^2<1, b^2g^2<1$. Correspondingly, it carries four independent conserved charges: the energy $E$ (associated with translations generated by the Killing vector $\frac{\partial}{\partial t}$), two angular momenta $J_1, J_2\,$ (associated with rotations generated by the Killing vectors $\frac{\partial}{\partial\phi}$ and $\frac{\partial}{\partial\psi}$, respectively) and the electric charge $Q$. These read:
\begin{align}\label{CCLPcharges}
E &= \frac{m\pi (2\Xi_a +2\Xi_b - \Xi_a\,\Xi_b) +2\pi qabg^2(\Xi_a+\Xi_b)}{4\Xi_a^2\,\Xi_b^2}\ ,\qquad Q = \frac{\pi q}{2g \Xi_a\, \Xi_b}\ ,\nn\\[2mm]
&\ \  J_1 = \frac{\pi[2am + qb(1+a^2 g^2) ]}{4 \Xi_a^2\, \Xi_b}\ ,\qquad
J_2 = \frac{\pi[2bm + qa(1+b^2 g^2) ]}{4 \Xi_b^2\, \Xi_a}\ .
\end{align}

The outer event horizon of the solution is identified by the largest positive root of $\Delta_r=0$,  denoted by $r=r_+$. This is a Killing horizon,  
 generated by the Killing vector field
\be\label{generator_horizon}
V = \frac{\partial}{\partial t} + \Omega_1 \frac{\partial}{\partial \phi} +  \Omega_2 \frac{\partial}{\partial \psi} \ ,
\ee
where
\be\label{angular_velocities_CCLP}
\Omega_1 = \frac{a(r_+^2+ b^2)(1+g^2 r_+^2) + b q}{
               (r_+^2+a^2)(r_+^2+b^2)  + ab q}\ ,\qquad
\Omega_2 = \frac{b(r_+^2+ a^2)(1+g^2 r_+^2) + a q}{
               (r_+^2+a^2)(r_+^2+b^2)  + ab q}\ 
\ee
are the angular velocities on the horizon measured in a non-rotating frame at infinity.
Evaluating the surface gravity gives the Hawking temperature
\be\label{temperature_CCLP}
T \equiv\beta^{-1} = \frac{r_+^4[(1+ g^2(2r_+^2 + a^2+b^2)] -(ab + q)^2}{2\pi\,
         r_+\, [(r_+^2+a^2)(r_+^2+b^2) + abq]}\ ,
\ee
while the electrostatic potential on the horizon is defined as
\be\label{def_Phi}
\Phi \,=\, \iota_V A |_{r_+} -  \iota_V A |_{\infty}\ 
\ee
and reads \cite{Chen:2005zj}
\be\label{electrostatic_pot_CCLP}
\Phi = \frac{3\,g\, q \,r_+^2}{2\left((r_+^2 + a^2)(r_+^2 + b^2)+abq\right)}\ .
\ee
Finally, the entropy $S = \frac{\rm Area}{4}$ is given by
\be\label{entropyCCLP}
S=\frac{\pi^2 [(r_+^2 +a^2)(r_+^2 + b^2) +a b q]}{2\Xi_a \Xi_b r_+}
\ .
\ee
These quantities satisfy the first law of thermodynamics,
\be
\diff E = T \diff S + \Omega_1\, \diff J_1+ \Omega_2 \,\diff J_2 + \Phi\, \diff Q\ .
\ee

The intensive variables $\beta,\Omega_1, \Omega_2,\Phi$, that we will denote collectively as ``chemical potentials'', are conjugate in a thermodynamical sense to the charges $E,J_1,J_2,Q$, respectively. 
Adopting the approach of \cite{Gibbons:1976ue}, this is demonstrated by computing the on-shell action. The latter should be evaluated in a regular Euclidean section of the solution, that is a section in the complexified solution where the metric is real and positive definite. One then argues that the action will take the same form on any other section of the complexified solution which is homologous to the Euclidean section, even though the induced metric on this other section may be complex.
 For the solution considered in this paper, the on-shell action was computed in \cite{Chen:2005zj} (see also \cite{Kunduri:2005zg} for the case with $a=b$) by performing the analytic continuation $t\to -i \tau$, $a\to ia$, $b\to ib$ which yields a real Euclidean metric. Regularity of the Euclidean metric leads to identify  $\beta$ with the circumference of the Euclidean time circle that shrinks as $r\to r_+$.\footnote{Regularity of the Euclidean section will be discussed further later. A general discussion of the thermodynamics of asymptotically locally AdS spaces can be found in \cite{Papadimitriou:2005ii}.} Moreover the long-distance divergences arising from the action integral were regularized in \cite{Chen:2005zj,Kunduri:2005zg} using the background subtraction method. Eventually the parameters $a,b$ are continued back to their original value. The result thus found for the on-shell action is:
\be\label{OnShAction_CCLP}
I = \frac{\pi\beta}{4\Xi_a\Xi_b}
\Big[m - g^2 (r_+^2 + a^2)(r_+^2 + b^2) -
\frac{q^2 r_+^2}{(r_+^2 + a^2)(r_+^2 + b^2)+abq}\Big]\ ,
\ee
and one can verify that the quantum statistical relation is satisfied:
\be\label{QSR_CCLP}
I = \beta E - S - \beta \Omega_1J_1 - \beta\Omega_2J_2 - \beta\Phi Q\ .
\ee
The on-shell action, seen as a function of the chemical potentials, $I=I(\beta,\Omega_1,\Omega_2,\Phi)$, is interpreted as minus the logarithm of the grand-canonical partition function. 
 One can check that the black hole charges are indeed conjugate to the chemical potentials, that is they satisfy:
\be\label{charges_from_I}
E = \frac{\partial I}{\partial\beta}\ ,\qquad J_1 = -\frac{1}{\beta}\frac{\partial I}{\partial\Omega_1}\ ,\qquad J_2 = -\frac{1}{\beta}\frac{\partial I}{\partial \Omega_2} \ ,\qquad Q = -\frac{1}{\beta}\frac{\partial I}{\partial\Phi}\ .
\ee
The entropy \eqref{entropyCCLP} is thus the logarithm of the microcanonical partition function which is obtained as the Legendre transform of the logarithm of the grand-canonical partition function $I$ with respect to all chemical potentials.

The solution admits two notable and a priori distinct limits, namely the one to supersymmetry and the one to extremality. 
The solution is supersymmetric if the parameters are related as:
\be\label{susyCCLP}
q = \frac{m}{1+ag+bg}\ .
\ee
For later purposes we record the non-spacelike Killing vector arising as a bilinear of the spinor parameter solving the Killing spinor equation:
\be\label{susy_vector}
K = \frac{\partial}{\partial t} + g \frac{\partial}{\partial \phi} +  g \frac{\partial}{\partial \psi}\ .
\ee
One can check that the supersymmetry condition \eqref{susyCCLP} does not imply extremality since the temperature does not vanish in general.

Extremality instead corresponds to a double root of $\Delta_r$. In order to show this, we write the cubic polynomial $r^2\Delta_r$ as
\be
g^{-2}r^2\Delta_r = (r^2-r_+^2)(r^2-r_0^2)(r^2-r_-^2)\ ,
\ee
where by definition $r_+^2 \geq r_0^2 \geq r_-^2$.  Comparing this expression for $\Delta_r$ with the one in \eqref{CCLPfunctions}, we find:
\begin{align}\label{r_+r_0r_-relations}
r_+^2 +r_0^2 + r_-^2 &= -(a^2+b^2+ g^{-2})\ ,\nn\\
r_+^2r_0^2+r_0^2r_-^2+ r_-^2r_+^2 &= g^{-2}(a^2+b^2+a^2b^2g^2-2m)\ ,\nn\\
 r_+^2r_0^2r_-^2 &= -g^{-2}(ab+q)^2\ .
\end{align}
Using these relations, the numerator in the expression \eqref{temperature_CCLP} for the temperature can be rewritten as 
$r_+^2(r_+^2-r_0^2)(r_+^2-r_-^2)$ \cite{Choi:2008he}.
Hence the extremality condition corresponds to $r_+^2= r_0^2$, meaning that $r^2\Delta_r$ has a double root indeed. This condition does not imply the supersymmetry relation \eqref{susyCCLP}, in fact one can check that the solution of \cite{Chong:2005hr} contains  causally well-behaved
black holes that are extremal but non-supersymmetric. 

On the other hand, for the black hole satisfying the supersymmetry condition \eqref{susyCCLP} to be free of causal pathologies such as closed timelike curves (CTC's), one needs to further restrict the parameters as
\be\label{NoCTCs1}
m = \frac{1}{g}(a+b)(1+ag)(1+bg)(1+ag+bg)\ .
\ee
One also has to require $a + b + abg > 0$.
After imposing \eqref{NoCTCs1} in addition to \eqref{susyCCLP}, the solution becomes extremal. 
So although the generic supersymmetric solution depends on three parameters, 
 the supersymmetric and causally well-behaved solution is also extremal and depends on two parameters only.

Throughout the paper, we will carefully distinguish between supersymmetry and extremality. Moreover, a quantity evaluated after taking {\it both} the supersymmetry and the extremality limits will be called ``BPS'' and denoted by a $*$ symbol in the formulae.

In this BPS limit, the double root $\rBPS^2 = r_+^2 = r_0^2$ reads
\be\label{r_BPS}
\rBPS =  \sqrt{\frac{1}{g}(a+b+abg)} \ ,
\ee
while the chemical potentials above satisfy:
\be\label{leadingterms_chempot}
\beta \to \infty\ ,\qquad \Omega_1\to \OmBPS_1 = g\ ,\qquad \Omega_2\to \OmBPS_2 = g\ ,\qquad \Phi \to \PhiBPS = \frac{3}{2}\,g\ .
\ee
Note that the null generator of the horizon given in \eqref{generator_horizon} then coincides with the supersymmetric Killing vector \eqref{susy_vector}. The BPS values of the charges are:
\begin{align}\label{BPScharges}
\JBPS_1 &=  \frac{\pi(a+b)(2a+b+abg)}{4g(1-ag)^2(1-bg)}    \ , \qquad
\JBPS_2 = \frac{\pi(a+b)(a+2b+abg)}{4g(1-ag)(1-bg)^2} \ , \nn\\[1mm]
\QBPS & =  \frac{\pi (a+b)}{2g^2(1-ag)(1-bg)}\ ,
\end{align} 
with the energy being given by:
\be\label{susyrel_BPS}
\EBPS \= \OmBPS_1 \, \JBPS_1 + \OmBPS_2 \, \JBPS_2 + \PhiBPS\, \QBPS \ .
\ee
Relation \eqref{susyrel_BPS} is a consequence of supersymmetry. 
Since two parameters have been tuned to reach both supersymmetry and extremality, the charges satisfy an additional relation, which reads:
\be\label{nonlinear_rel_BPScharges}
\left(\QBPS\right)^3 + \frac{2\pi}{g^3}\JBPS_1\JBPS_2 = \left( 3 \QBPS  + \frac{\pi}{2g^3} \right)\left( 3 \left(\QBPS \right)^2 -  \frac{\pi}{g^3}(\JBPS_1+\JBPS_2)\right)\ .
\ee
The BPS entropy reads 
\be
\SBPS = \frac{\pi^2 (a+b)\rBPS}{2g(1-ag)(1-bg)}
\label{SBPSab}
\ee
and it can be written in terms of the charges as \cite{Kim:2006he}:
\be\label{S_BPS}
\SBPS = \pi \sqrt{3\left(\QBPS\right)^2 -\frac{\pi}{g^3} \big(\JBPS_1+\JBPS_2\big)}\ .
\ee  
Reproducing this expression for the BPS entropy from a field theory computation will be our final goal.
 
By setting the rotational parameters equal, $a=b$, the solution reduces to the one of \cite{Cvetic:2004hs}, where the two angular momenta are equal to each other. In the BPS limit, this reduces to the one-parameter supersymmetric black hole of \cite{Gutowski:2004ez}.

\section{BPS limit of black hole thermodynamics}\label{sec:BPSlimit}

We are interested in studying a BPS limit of the black hole thermodynamics described above. In particular, we want to express the quantum statistical relation \eqref{QSR_CCLP}, relating the entropy to the on-shell action, after both the supersymmetry and the extremality limits have been taken. Since one has to tune two parameters, there are multiple ways to reach the BPS black hole. In particular, the order of limits matters. 
We start by describing the consequences of imposing supersymmetry first, followed by extremality. 
We will then comment on other possible limits.

\subsection{Supersymmetry}\label{impose_susy}

From now on we will set $g=1$ for simplicity (this can easily be reinstated by dimensional analysis). Moreover instead of using $(a,b,m,q)$ as independent parameters, we will find it convenient to describe the solution in terms of $(a,b,r_+,q)$. 
From the expression of $\Delta_r$ in \eqref{CCLPfunctions} we see that $m$ is then determined as:
\be
m = \frac{(r_+^2+a^2)(r_+^2+b^2)(1+ r_+^2) + q^2 +2ab q}{2r_+^2} \ .
\ee

In terms of the variables $(a,b,r_+,q)$, the supersymmetry condition \eqref{susyCCLP} 
becomes
\be\label{susyq_moregeneral}
q = -ab + (1+a+b)\,r_+^2\pm  \,\sqrt{-r_+^2(r_+^2-\rBPS^2)^2}\ ,
\ee
where we recall that $\rBPS$ is given by \eqref{r_BPS}.
If we require $q$ to be real (and $r_+^2 >0$) then we need to take $r_+ = \rBPS$, that is we are immediately forced to impose extremality in addition to supersymmetry. This is consistent with the fact that the causally well-behaved, 
supersymmetric black hole is also extremal.
If instead we insist on imposing supersymmetry while keeping $r_+$ generic (but real), then $q$ must be complex and may be written as
\be\label{susy_with_r+}
q = -(a- i r_+)(b- i r_+)(1- i r_+)\ .
\ee
where we have fixed the sign in \eqref{susyq_moregeneral} for definiteness. The other sign choice corresponds to sending $i\to -i$ in this expression; this change straightforwardly propagates in the expressions  for the chemical potentials, the action, the entropy and the charges given below.

A complex value of the parameters would not be allowed in the Lorentzian solution. However our aim is to study an analytically continued solution which satisfies the requirement of preserving supersymmetry. For this purpose, it is legitimate to take $q$ and possibly other parameters complex: since the Killing spinor equation is analytic in the supergravity fields, it will still admit a solution in the complexified background.

Using expression \eqref{susy_with_r+} for $q$, the chemical potentials \eqref{angular_velocities_CCLP}, \eqref{temperature_CCLP}, \eqref{electrostatic_pot_CCLP} become
\begin{align}
\beta &=  \frac{-2\pi(a-i r_+)(b-i r_+)(\rBPS^2 + i r_+)}{(\rBPS^2-r_+^2)\left[2(1+a+b)r_++i(\rBPS^2 - 3 r_+^2)\right]}\ ,\nn\\[1mm]
\Omega_1 &= \frac{(\rBPS^2+ ia r_+)(1-ir_+)}{(\rBPS^2+ir_+)(a-ir_+)}\ ,\qquad
\Omega_2 = \frac{(\rBPS^2+ ib r_+)(1-ir_+)}{(\rBPS^2+ir_+)(b-ir_+)}\ ,\nn\\[1mm]
\Phi &=  \frac{3r_+(r_++i)}{2(\rBPS^2 + ir_+)} \ ,\label{susy_chempot}
\end{align}
so they are all complex: it is only when $r_+ \to \rBPS$ (while keeping $a,b$ fixed and real) that their imaginary part vanishes and they reach the BPS values \eqref{leadingterms_chempot}. Since we have imposed the condition \eqref{susy_with_r+} on the parameters, the chemical potentials are no more independent. It is readily checked that they now satisfy the constraint:
\be\label{constraint_uppercase}
\beta \left(1+ \Omega_1 +\Omega_2 -2 \Phi \right) =  2\pi i\ .
\ee
 Note that this condition could not be satisfied if the chemical potentials were real.

The constraint just derived has a simple interpretation: it corresponds to a global regularity condition for the spinor $\epsilon$ 
solving the Killing spinor equation~\eqref{KillingSpEq} and parameterizing the supersymmetry of the solution. 
In order to see this, let us discuss the action of the Killing vectors on $\epsilon$. 
The explicit form of the Killing spinor $\epsilon$  is given in Appendix~\ref{app:KillingSpinor}. Although this depends on the frame chosen, the spinorial Lie derivative $\mathcal{L}$, which is covariant under local Lorentz transformations, can be evaluated in any frame. 
 As discussed in detail in Appendix~\ref{app:KillingSpinor}, we find that the Killing spinor satisfies:
\be\label{Lie_der_spinor}
\mathcal{L}_{\frac{\partial}{\partial t}}\epsilon =  \tfrac{i}{2}(1+2\alpha)\,g \,\epsilon \ ,\qquad
\mathcal{L}_{\frac{\partial}{\partial\phi}} \epsilon = \tfrac{i}{2} \, \epsilon \ ,\qquad
\mathcal{L}_{\frac{\partial}{\partial\psi}} \epsilon = \tfrac{i}{2} \, \epsilon \ ,
\ee
where we recall that $\alpha$ controls the flat connection term in the gauge field \eqref{gaugepot}.
The last two equations are consistent with the fact that $\phi,\psi$ parameterize circles of length $2\pi$ inside $S^3$, which shrink smoothly to zero size at $\theta=0$ and $\theta=\pi/2$, respectively. When transported one full time around either one of these shrinking circles, the spinor must be antiperiodic, as it inherits the spin structure of flat space; indeed exponentiating the action of the Lie derivative one correctly finds that $\rme^{2\pi \mathcal{L}_{{\partial}/{\partial \phi}}}\,\epsilon = \rme^{2\pi \mathcal{L}_{{\partial}/{\partial \psi}}}\,\epsilon= - \epsilon$.

A similar condition must apply to the Killing generator $V$ of the horizon, which becomes null at $r=r_+$. In the Euclidean section of the solution this vector generates translations along the time circle, which has period $\beta$ and which shrinks to zero size as $r\to r_+$. Recalling \eqref{generator_horizon} and using \eqref{Lie_der_spinor}, we see that the action of $V$ on the Killing spinor yields:
\be
\mathcal{L}_V \, \epsilon \, =\,  \tfrac{i}{2}\,\big( 1 + \Omega_1  + \Omega_2 - 2\Phi + 2\,\iota_V A|_{r_+} \big)\,\epsilon \ ,
\ee
where we also noted that from the definition \eqref{def_Phi} of the electrostatic potential one has $\alpha \equiv \iota_V A|_\infty= \iota_V A|_{r_+} - \Phi$. The constraint \eqref{constraint_uppercase} then leaves us with
\be
\mathcal{L}_V \, \epsilon = \left( \mp\, \tfrac{\pi}{\beta}+ i\,\iota_V A|_{r_+} \right) \,\epsilon\ .
\ee
So transporting the spinor around the circle generated by $V$ one full time yields:
\be\label{periodicity_KillingSp_general}
\rme^{i\,\beta\,\mathcal{L}_V  }\,\epsilon \,=\, - \,\rme^{- \beta\,\iota_V A|_{r_+}}\,\epsilon \ .
\ee
For the gauge field to be well-defined, its component along the direction that shrinks must vanish, i.e.\ we should require $\iota_V A|_{r_+} = 0$. This is achieved by fixing the gauge choice in $A$ as $\alpha = - \Phi$. In this gauge we have: 
\be\label{periodicity_KillingSp_general}
\rme^{i\,\beta\,\mathcal{L}_V  }\,\epsilon \,=\, - \,\epsilon   \qquad \text{for} \qquad \iota_V A|_{r_+} =0\ ,
\ee
namely the Killing spinor is {\it antiperiodic} when transported one full time around the circle generated by $V$. As for the circles generated by $\frac{\partial}{\partial\phi}$ and $\frac{\partial}{\partial\psi}$, this reflects regularity of the spinor, since in the Euclidean section the thermal circle and the radial coordinate parameterize a space which looks like $\mathbb{R}^2$ for $r\to r_+$. Here we postulate that this condition should be satisfied also in the complexified, supersymmetric solution. 
Therefore {\it the constraint \eqref{constraint_uppercase} can be seen as a regularity condition ensuring the correct periodicity of the Killing spinor. In particular, it ensures that the spinor is {\it antiperiodic} in the regular gauge such that $\iota_V A=0$ at the horizon.}

In order to proceed further, it will be useful to redefine the chemical potentials $\Omega_1,\Omega_2,\Phi$ by introducing the new variables:
\begin{align}\label{def_lowercase_chempot}
\omega_1 = \beta(\Omega_1- \OmBPS_1)\ ,\qquad \omega_2 = \beta(\Omega_2-\OmBPS_2)\ ,\qquad \elpot = \beta(\Phi-\PhiBPS)\ ,
\end{align}
where 
\be
\OmBPS_1=\OmBPS_2=1\ , \qquad \PhiBPS=\frac{3}{2}\ ,
\ee 
in agreement with \eqref{leadingterms_chempot} (recall that we are taking $g=1$ in this section).
In terms of the new variables, the constraint \eqref{constraint_uppercase} reads:
\be\label{constraint_lowercase_hatted}
 \omega_1 +\omega_2 -2 \elpot =  2\pi i \ .
\ee
Explicitly, these read:
\begin{align}\label{ChPsusy1}
\omega_1 &=  \frac{2\pi(a-1)(b-ir_+)}{2(1+a+b)r_+ + i (\rBPS^2-3r_+^2)}\ ,\\[2mm]\label{ChPsusy2}
\omega_2 & =  \frac{2\pi(b-1)(a-ir_+)}{2(1+a+b)r_+ + i (\rBPS^2-3r_+^2)}\ ,\\[2mm] \label{ChPsusy3}
\elpot & =  \frac{3\pi(a-i r_+)(b-ir_+)}{2(1+a+b)r_+ + i (\rBPS^2-3r_+^2)}\ ,
\end{align}

Implementing the supersymmetry condition \eqref{susy_with_r+} and using the new variables, the on-shell action \eqref{OnShAction_CCLP} takes the nice form:
\be\label{Isusy_is_Esusy}
I = \frac{2\pi}{27g^3G_5}\,\frac{\elpot^3}{\omega_1\omega_2}\ .
\ee
Note that this is manifestly independent of $\beta$. 
In view of the comparison with the field theory result to be discussed in Section~\ref{field_theory_section}, 
in this formula we have reinstated the five-dimensional Newton's constant $G_5$, otherwise set to 1 in 
this paper, as well as the inverse AdS radius $g$.

Condition \eqref{susy_with_r+} implies that the entropy \eqref{entropyCCLP} takes the complex value
\begin{align}\label{complex_entropy}
S &= \frac{\pi^2(a-i r_+)(b-i r_+)(i\rBPS^2-r_+)}{2(1-a^2)(1-b^2)} \nn \\[2mm]
 &= \frac{\pi^2 \left[ r_+ (r_+^2 + a^2 +b^2 + ba^2 + ab^2 + ab)  + i a b (\rBPS^2 - r_+^2) \right]}{2(1-a^2)(1-b^2)} \, ,
 \end{align}
where notice that when $r_+=\rBPS$, the imaginary part vanishes and the real part becomes the BPS entropy \eqref{SBPSab}.
The charges are also complex, and take the following form 
\begin{align}
J_1 &=   \frac{\pi (b+ 2a + a b)}{4(1-a) (1-a^2)(1-b^2)}  \left[r_+^2 (1+ a+ b)- a b + i r_+ (\rBPS^2 - r_+^2) \right], \\[1mm]
J_2 &=    \frac{\pi (a+ 2b + a b)} {4(1-b) (1-a^2)(1-b^2)} \left[r_+^2 (1+ a+ b)- a b + i r_+ (\rBPS^2 - r_+^2) \right], \\[1mm]
Q &=   \frac{\pi} {2(1-a^2)(1-b^2)} \left[r_+^2 (1+ a+ b)- a b + i r_+ (\rBPS^2 - r_+^2) \right] \ .
\end{align}
This makes it manifest that  in the BPS limit, $r_+=\rBPS$, again the imaginary parts vanish, while the real parts coincide with the BPS values  in \eqref{BPScharges}. 
The energy $E$ is also complex and it satisfies the  relation
\be
 E - \OmBPS_1 J_1 - \OmBPS_2 J_2 - \PhiBPS  Q =0 \ ,
\ee
which as already mentioned is a consequence of supersymmetry, independently of extremality.

The quantum statistical relation \eqref{QSR_CCLP} is now satisfied in the form
\be
I = -S - \omega_1 J_1 - \omega_2 J_2- \elpot\, Q \ .
\ee
Notice that the term containing the energy has disappeared. This is because 
\begin{align}
I &= \beta E - S - \beta \Omega_1J_1 - \beta\Omega_2J_2 - \beta\Phi Q\nn \\
  &= \beta (E - \OmBPS_1 J_1 - \OmBPS_2 J_2 - \PhiBPS\, Q   ) - S - \omega_1 J_1 - \omega_2J_2 - \elpot\, Q\ ,
\end{align}
and the term proportional to $\beta$ vanishes as it is just the supersymmetry condition satisfied by the charges. This also means that $\omega_1,\omega_2,\elpot$ are the chemical potentials conjugate to the charges $J_1,J_2,Q$ when $\beta$ is regarded as the chemical potential conjugate to the charge associated with the supersymmetric Hamiltonian
\begin{align}\label{susy_Hamiltonian_old}
 \{ \mathcal{Q},\overline{\mathcal{Q}} \}&= E -  J_1 -  J_2 - \frac{3}{2} Q \ ,\nn \\[1mm]
& \equiv  E - \OmBPS_1 J_1 - \OmBPS_2J_2 - \PhiBPS Q\ ,
\end{align}
which acts via the Killing vector $K$.

\subsection{Reaching extremality}\label{reach_extremality}

Starting from the supersymmetric, complexified family of solutions defined above, we can now  take the limit to extremality by sending $r_+\to\rBPS$.
In this limit both the charges and the entropy become real and reach their BPS values given in Section \ref{sec:CCLPsolQSR}. 
Crucially, the only quantity diverging in this limit is $\beta$, while the redefined chemical potentials $\omega_1,\omega_2,\elpot$ remain finite.
Dubbing their limiting values as  
\be\label{BPschempot}
\omBPS_{1,2}= \lim_{r_+\to\, \rBPS} \omega_{1,2} \ ,\qquad \elpotBPS = \lim_{r_+\to \,\rBPS} \elpot \ ,
\ee
we obtain
\begin{align}\label{BPS_chem_pots}
\omBPS_1 &= \frac{\pi\,(1-a)}{1+a^2+b^2+3\rBPS^2}\left( - \frac{a+2b+2ab+b^2}{\rBPS}+ i\,(1+a)\right)\ , \nn\\[1mm]
\omBPS_2 &=  \frac{\pi\,(1-b)}{1+a^2+b^2+3\rBPS^2}\left( - \frac{b+2a+2ab+a^2}{\rBPS}+ i\,(1+b)\right) \ ,\nn\\[1mm]
\elpotBPS &= -\frac{3\pi \,(a+b)}{2(1+a^2+b^2+3\rBPS^2)}\left( \frac{1-ab}{\rBPS} + i\,(2+a+b) \right)\ .
\end{align}
Note that contrarily to the charges, these remain complex even after taking the BPS limit.
Since the limit is smooth, we still have the constraint
\be\label{constraint_lowercase}
 \omBPS_1 +\omBPS_2 - 2\elpotBPS =  2\pi i \ 
\ee
and the BPS on-shell action is:
\be\label{IBPS}
\IBPS = \frac{2\pi}{27}\frac{(\elpotBPS)^{3}}{\omBPS_1\omBPS_2}\ .
\ee
For later purposes, we observe that in the physical range $a^2<1, b^2<1$ and $\rBPS^2 \equiv a+b+a b>0$, the real parts of the angular chemical potentials satisfy either ${\rm Re}\,\omega_1 <0$, ${\rm Re}\,\omega_2 <0$, or ${\rm Re}\,\omega_1 \cdot {\rm Re}\,\omega_2 <0$, while they are never both positive.
We recall that these are not the leading terms \eqref{leadingterms_chempot} of the chemical potentials in the BPS limit. They are instead the subleading terms in the expansion $\Omega_i = \OmBPS_i + \frac{1}{\beta}\, \omBPS_i + \ldots$ (and similarly for $\Phi = \PhiBPS + \frac{1}{\beta}\, \elpotBPS + \ldots$). 
We argue that these quantities are to be identified with the BPS chemical potentials and BPS grand-canonical partition function in the BPS limit of black hole thermodynamics. 
 This is demonstrated starting from the BPS Quantum Statistical relation:
\be\label{BPS_QSR}
\IBPS = -\SBPS - \omBPS_1 \JBPS_1 - \omBPS_2 \JBPS_2- \elpotBPS\, \QBPS -\Lambda (\omBPS_1 + \omBPS_2 - 2\elpotBPS - 2\pi i)\ , 
\ee
where $\Lambda$ is a Lagrange multiplier enforcing the constraint \eqref{constraint_lowercase} between the chemical potentials. Although this BPS-quantum statistical relation is satisfied in the solution under study with the term multiplied by $\Lambda$ evaluating to zero, it is useful to introduce the latter in order to recall that the chemical potentials cannot be varied independently, as they must satisfy the constraint. The conjugacy relation between the BPS charges and the chemical potentials subject to the constraint is then expressed as:
\be
  -\frac{\partial I}{\partial\omBPS_1} =  \JBPS_1 + \Lambda\ ,\qquad -\frac{\partial I}{\partial \omBPS_2} = \JBPS_2 + \Lambda\ ,\qquad -\frac{\partial I}{\partial\elpotBPS} = \QBPS - 2\Lambda \ ,
\ee
where one of these equations should be regarded as the equation which determines the Lagrange multiplier.
Using these relations, the entropy is understood as the Legendre transform of the supergravity on-shell action \eqref{IBPS}, trading the chemical potentials $\omBPS_1,\omBPS_2,\elpotBPS$ for the BPS charges $\JBPS_1,\JBPS_2,\QBPS$. This reproduces precisely the BPS entropy in the form \eqref{S_BPS}.
We describe the extremization leading from the on-shell action to the entropy in a slightly more general setup in Appendix~\ref{sec:revisiting_extremization}.  There we also show that demanding reality of the entropy and of the charges implies a specific relation between the charges that in the case under study is precisely eq.~\eqref{nonlinear_rel_BPScharges}.

We observe that {\it the BPS limit of black hole thermodynamics defined above provides a  
derivation of the extremization principle proposed in \cite{Hosseini:2017mds} from the principles of Euclidean quantum gravity}. In particular, we have shown that the function of the chemical potentials considered in \cite{Hosseini:2017mds} is in fact a supersymmetric on-shell supergravity action, which can be interpreted as minus the log of the grand-canonical partition function.  Then the extremization principle proposed in \cite{Hosseini:2017mds} can be seen as a Legendre transformation sending the log of the grand-canonical partition function into the log of the microcanonical partition function (i.e.\ the black hole entropy).
This result is non-trivial as the on-shell action must be defined by taking an appropriate limit in which supersymmetry is imposed before reaching extremality. This prescription also identifies the complex chemical potentials satisfying the constraint \eqref{constraint_lowercase}. The latter played a central role in~\cite{Hosseini:2017mds} and is understood here as a specific regularity condition for the Killing spinor that arises by imposing supersymmetry of the non-extremal family of solutions. It should be clear from the discussion of Appendix~\ref{sec:revisiting_extremization} that the value taken by the constraint and the expression for the on-shell action are equally crucial for the entropy to be retrieved correctly.

\subsection{Different BPS limits}

Above we defined the BPS limit in two steps: first we imposed supersymmetry and afterwards we reached extremality. This limiting procedure is however not unique, and we now compare it to other possible trajectories in the parameter space of the solution leading to the BPS locus.

We saw in Section \ref{sec:CCLPsolQSR} that the BPS solution is obtained by imposing the conditions  \eqref{susyCCLP}, \eqref{NoCTCs1} on the parameters $a,b,m,q$. Using \eqref{r_+r_0r_-relations} to trade $m,q$ for the horizon radii $r_0,r_+$, the same BPS conditions can be expressed as:
\be
r_0^2 = r_+^2 = a + b +ab \equiv \rBPS^2\ .
\ee
We then consider the following deformation away from the BPS values:
\begin{align}\label{expand_r+r0}
r_+ &= \rBPS + \epsilon\ , \nn
\\
r_0 &= \rBPS + \epsilon \, r_{0,1} + \epsilon^2 \, r_{0,2} + \epsilon^3 \, r_{0,3} + \ldots
\ ,
\end{align}
where $r_{0,k}$ are parameters independent of $\epsilon$. From \eqref{r_+r_0r_-relations} one can see that this implies:
\begin{align}
m&= (a+b)(1+a)(1+b)(1+a+b) + (\,1+a^2+b^2+3\rBPS^2\,)\rBPS(1+r_{0,1})\,\epsilon + \mathcal{O}(\epsilon^2)\ ,\nn\\[1mm]
q &= (a+b)(1+a)(1+b) + \frac{ (\,1+a^2+b^2+3\rBPS^2\,)\rBPS (1+r_{0,1})}{1+a+b}\,\epsilon + \mathcal{O}(\epsilon^2) \label{expansion_mq}\ .
\end{align}
The supersymmetry condition \eqref{susyCCLP} is still satisfied at first order in $\epsilon$, while generically it is violated at second order. On the other hand, we have $r_+-r_0 = (1-r_{0,1})\epsilon + \mathcal{O}(\epsilon^2)$ and unless $r_{0,1}=1$ the deformation takes the solution out of extremality already at first order.

Plugging the near-BPS values \eqref{expand_r+r0}, \eqref{expansion_mq} of the parameters in, we find that the chemical potentials now read:
 \begin{align}\label{susy_chem_pot_udep}
\omBPS_1 &\equiv \lim_{\epsilon\to 0}\beta(\Omega_1-\OmBPS_1) = \frac{\pi\,(1-a)}{1+a^2+b^2+3\rBPS^2}\left( - \frac{a+2b+2ab+b^2}{\rBPS}+ (1+a)u\right)\ , \nn\\[1mm]
\omBPS_2 &\equiv \lim_{\epsilon\to 0}\beta(\Omega_2-\OmBPS_2) = \frac{\pi\,(1-b)}{1+a^2+b^2+3\rBPS^2}\left( - \frac{b+2a+2ab+a^2}{\rBPS}+ (1+b)u\right) \ ,\nn\\[1mm]
\elpotBPS &\equiv \lim_{\epsilon\to 0}\beta(\Phi-\PhiBPS) = -\frac{3\pi \,(a+b)}{2(1+a^2+b^2+3\rBPS^2)}\left( \frac{1-ab}{\rBPS} + (2+a+b)u \right)\ ,
\end{align}
where $u$ is defined in terms of $r_{0,1}$ as:
\be\label{defu}
u = \frac{\rBPS}{1+a+b}\,\frac{r_{0,1}+1}{r_{0,1}-1}\ .
\ee
These satisfy the constraint:
\be\label{sum_susy_chempot_udep}
\omBPS_1 + \omBPS_2 - 2\elpotBPS \,=\, 2\pi u\ .
\ee
We see that these chemical potentials depend---via $u$---on the way the BPS locus is attained. The 
 limiting value of the on-shell action \eqref{OnShAction_CCLP} also depends on $u$ and reads:
\be\label{IBPS_udep}
\IBPS
=\frac{\pi^2(a+b)^2\left[1-a^2(1+b)-b^2(1+a)-3ab-2\rBPS^2+ (3+a+b-ab)\rBPS\,u \right]}{4(1-a)(1-b)\rBPS(1+a^2 +b^2+3\rBPS^2)}\,.
\ee
One can check that the quantum statistical relation is satisfied for any value of $u$.

Comparing the expressions for $\omBPS_1,\omBPS_2,\elpotBPS$, we clearly see that the limiting procedure  described in Sections \ref{impose_susy}, \ref{reach_extremality} corresponds to choosing $u=i$. The alternate sign choice in \eqref{susyq_moregeneral} corresponds to taking $u=-i$ here. The peculiarity of these two choices is that fixing $r_{0,1}$ so that $u=\pm i$ and setting to zero all other $r_{0,k}$ coefficients in \eqref{expand_r+r0} corresponds to requiring that the supersymmetry condition \eqref{susyCCLP} is preserved at finite~$\epsilon$. Moreover, it is only after choosing $u=\pm i$ that the BPS on-shell action \eqref{IBPS_udep} can be written in the nice form \eqref{IBPS}.

A different choice of $u$ was previously discussed in \cite{Silva:2006xv}. In that work, the BPS limit was defined by setting: 
\be\label{q_Silva}
q = (a+b)(1+a)(1+b)\ ,
\ee
\be
m = (a+b)(1+a)(1+b)(1+a+b) + \mu\ ,
\ee
and then sending $\mu \to 0$. Clearly both BPS conditions \eqref{susyCCLP} and \eqref{NoCTCs1} are fulfilled in the limit.
 Note that here $q$ is kept fixed together with $a$ and $b$.\footnote{We remark that \eqref{q_Silva} is different from the extremality condition, it implies extremality only after imposing the supersymmetry condition \eqref{susyCCLP}.
}
In terms of our general parameterization of the BPS limit, this corresponds to choosing $r_{0,1}= -1$, that is, $u=0$. From \eqref{expansion_mq} we see that in this case $m$ and $q$ only receive corrections at order $\mathcal{O}(\epsilon^2)$ (then the other $r_{0,k}$ coefficients can be tuned so that $q$ maintains its value \eqref{q_Silva} also at finite $\epsilon$). 
The BPS chemical potentials given in \cite{Silva:2006xv} are consistent with those obtained by setting $u=0$ in \eqref{susy_chem_pot_udep} and \eqref{sum_susy_chempot_udep}.

The general analysis above shows that there are different ways to approach the BPS solution, which lead to different expressions 
for the chemical potentials $\omBPS_1,\omBPS_2,\elpotBPS$ and the on-shell action, all of which lead to the same physical 
black hole entropy.  
The one presented in Section \ref{impose_susy}, \ref{reach_extremality}, which first focuses on a non-extremal but supersymmetric 
family of solutions and in a second step reaches extremality, is singled out as  the unique 
manifestly supersymmetric BPS limit. As we shall now see, this naturally leads to a calculation of the black hole entropy in 
the dual rigid supersymmetric background at the boundary.  

\subsection{SCFT background from regularity in the bulk}

Having identified a BPS on-shell action and its precise relation with the black hole entropy, we would like to define a holographically dual SCFT quantity that matches it.
A central statement of the AdS/CFT correspondence is that the gravitational on-shell action matches the generating functional of connected correlators, $ -\log Z$, of a dual SCFT at large $N$. The on-shell action is a functional of the non-normalizable modes in an asymptotic expansion of the supergravity fields, which are identified with the background fields entering in the SCFT partition function $Z$ and sourcing the dual SCFT operators.
The gravitational on-shell action $I$ for the solution of \cite{Chong:2005hr} depends on  the chemical potentials $\beta,\Phi,\Omega_1,\Omega_2$, therefore we expect these to also appear in the SCFT background.
In the following we illustrate how indeed the asymptotic supergravity metric and gauge field contain information about the chemical potentials after imposing regularity of the Euclidean solution in the bulk. 
In this way we will identify the relevant SCFT background. In the next section we will evaluate the SCFT partition function on such background and match it with the supergravity on-shell action.

We start our analysis by discussing the boundary data for the solution reviewed in Section \ref{sec:CCLPsolQSR}, independently of supersymmetry or extremality. Sending $r \to \infty$, we find that the asymptotic metric is
\be\label{FGexpansion}
\diff s^2 = \frac{\diff r^2}{r^2} + r^2 \, \diff s^2_{\rm bdry} + \ldots \ ,
\ee
where the metric on the conformal boundary reads
\be\label{bdrymetrCCLP}
\diff s^2_{\rm bdry} \ =\ -\frac{ \Delta_\theta}{\Xi_a \Xi_b}\diff t^2+  \frac{\diff \theta^2}{\Delta_\theta }+ \frac{\sin^2 \theta}{\Xi_a}\diff \phi^2  +  \frac{\cos ^2\theta}{\Xi_b}\diff \psi^2   \ ,
\ee
and the boundary gauge field is 
\be
A_{\rm bdry} = \alpha\, \diff t\ .
\ee
The spacelike part of $\diff s^2_{\rm bdry}$ is a regular, conformally-flat metric on $S^3$. The boundary metric acquires a simpler form if before taking the large-$r$ limit we perform a  change of coordinates in the bulk transforming $(r,\theta)$ into $(\hat r,\hat\theta)$ as in~\cite{Hawking:1998kw}:
\be\label{change_r_theta}
\Xi_a \,\hat r^2 \sin^2\hat\theta = (r^2 + a^2)\,\sin^2\theta\ ,\qquad \Xi_b\, \hat r^2 \cos^2\hat\theta = (r^2 + b^2)\,\cos^2\theta\ .
\ee
Asymptotically, this induces a (regular) Weyl transformation
\be\label{conf_rescaling}
\diff s^2_{\rm bdry} = \frac{\Delta_\theta}{\Xi_a\Xi_b}\,\diff \hat s^2_{\rm bdry}\ ,
\ee 
such that the new metric on the conformal boundary reads
\be\label{canonical_bdry_metric}
\diff \hat s^2_{\rm bdry} =  -\diff t^2 +  \diff\hat\theta^2 + \sin^2 \hat{\theta}\, \diff\phi^2 + \cos^2 \hat{\theta} \,\diff\psi^2 \ .
\ee
We have thus obtained the canonical metric on the direct product of $\mathbb{R}$ with an $S^3$ of unit radius.

From these computations we conclude that locally the conformal boundary does not contain any information about the chemical potentials. These instead arise from a global analysis, as we now demonstrate  by studying regularity of the Euclidean section.

The Lorentzian metric \eqref{5met} can be analytically continued to an Euclidean metric by introducing the Euclidean time $\tau=i\,t$
and taking $a,b$ purely imaginary, while $m,q$ remain real.
Recalling their expression \eqref{angular_velocities_CCLP}, \eqref{temperature_CCLP}, \eqref{electrostatic_pot_CCLP}, it is easy to see that after this analytic continuation the quantities $\beta$, $i\Omega_1$, $i\Omega_2$, $\Phi$ are real.
Introducing a shifted radial coordinate $R^2 = r - r_+$,
 one can see that at leading order as $R\to 0$, the metric takes the form
\begin{align}\label{metric_nearplus}
\diff s^2 &=  h_{RR}\left(\diff R^2   +  R^2\big(\tfrac{2\pi}{\beta}\diff \tau\big)^2 \right) +h_{\theta\theta}\, \diff \theta^2 + h_{\phi\phi} (\diff\phi + i \Omega_1 \diff \tau)^2 + h_{\psi\psi} (\diff\psi +i \Omega_2 \diff \tau)^2 \nn\\[1mm]
& \quad + 2 h_{\phi\psi} (\diff\phi +i \Omega_1 \diff \tau)(\diff\psi + i\Omega_2 \diff \tau)\ ,
\end{align}
where $h_{RR},h_{\theta\theta},h_{\phi\phi},h_{\psi\psi},h_{\phi\psi}$ are specific functions of the coordinate $\theta$ and of the parameters $a,b,r_+,q$, whose explicit expression will not be needed here.
The metric \eqref{metric_nearplus} describes a warped fibration of the deformed $S^3$ parameterized by $(\theta,\phi,\psi)$ over the $\mathbb R^2$ parameterized by polar coordinates $(R,\tau)$. 
Absence of conical singularities at $R=0$ requires the twisted identification 
\be\label{twisted_identification}
(\tau \,  ,\, \phi \, , \, \psi \,) \; \sim \; 
(\tau+\beta\ ,\ \phi -i \Omega_1\beta \ , \ \psi -i\Omega_2\beta \, )
\ee
when one goes around the Euclidean time circle one full time \cite{Gibbons:1976ue,Hawking:1998kw}.
Equivalently, one can notice that the (real) one-forms $i \Omega_1 \diff \tau$ and $i \Omega_2 \diff \tau$ describing the fibration of $S^3$ over the Euclidean time circle in \eqref{metric_nearplus} are not well-defined as $R\to0$, where the latter shrinks. The offensive terms are removed by the change of angular coordinates
\be\label{hatted_coords}
\tau = \hat\tau\ , \qquad \phi = \phi_1 -i \Omega_1 \hat\tau\ ,\qquad
 \psi =\phi_2 - i\Omega_2 \hat\tau\ ,
\ee
where it should be noted that $\phi_1,\phi_2$ are still $2\pi$-periodic. 
Now one can impose the standard identifications 
\be\label{untwisted_identification}
(\hat\tau \, ,\, \phi_1  \, ,\, \phi_2 \, ) \; \sim \;
( \hat\tau+\beta\, ,\,  \phi_1 \, ,\,  \phi_2 \,)\ .
\ee
This identification is clearly equivalent to the twisted identification \eqref{twisted_identification} of the old coordinates.
We observe that in the new coordinates $(\hat\tau,\theta,\phi_1,\phi_2,r)$, the Killing vector becoming null at the horizon reads simply $V=  i \frac{\partial}{\partial \hat\tau} $. 

We can now turn back to the boundary geometry. After also performing the change of coordinates \eqref{change_r_theta}, the boundary metric reads
\be\label{coord_transf_metric}
\diff \hat s^2_{\rm bdry} =\,  \diff \hat\tau^2 + \frac{1}{g^2}\left( \diff\hat\theta^2 + \sin^2 \hat{\theta}\, (\diff \phi_1 -i\Omega_1 \diff\hat\tau\,)^2 + \cos^2 \hat{\theta} \,(\diff \phi_2 -i\Omega_2 \diff \hat\tau\,)^2\right) \ .
\ee
Although this is locally equivalent to the metric \eqref{canonical_bdry_metric}, 
the identification \eqref{untwisted_identification} implies that it is globally different from the geometry described 
by \eqref{canonical_bdry_metric} with the identifications ($t$, $\phi$, $\psi$) $\sim$ ($t - i \beta$, $\phi$, $\psi$). 
It is instead equivalent to the twisted background described by \eqref{canonical_bdry_metric} if one assumes the 
identifications ($t$, $\phi$, $\psi$) $\sim$ ($t-i\beta$, $\phi -i \Omega_1\beta$, $\psi -i\Omega_2\beta $).

Let us also look at the gauge field. As already discussed, regularity close to the origin of the space requires $\iota_V A|_{r=r_+} = 0$, which is achieved by fixing the gauge choice in $A$ as $\alpha = -\Phi$.
Hence the boundary gauge field is
\be\label{Abdry}
A_{\rm bdry} = i\, \Phi\,\diff\hat\tau\ .
\ee
Since $\Phi$ is real in the Euclidean section, the gauge field is pure imaginary.

We have thus explained how the chemical potentials $\beta, \Omega_1,\Omega_2,\Phi$ affect the boundary geometry and gauge field, as a consequence of regularity of the Euclideanized bulk solution. The chemical potentials essentially correspond to holonomies around the Euclidean time circle at the boundary. As such, they  map directly into the corresponding quantities in a dual field theory background.

We also observe that the energy $\wh E$ associated with translations along the new time coordinate $\hat\tau$ reads, 
in terms of the charges defined previously, 
\be
\wh E = E - \Omega_1 J_1 - \Omega_2 J_2\ .
\ee
Taking into account the gauge transformation leading to \eqref{Abdry}, the superalgebra \eqref{susy_Hamiltonian_old} now takes the form
\begin{align}\label{new_superalgebra}
\{\mathcal{Q},\overline{\mathcal{Q}}\} &=  \wh E + (\Omega_1-\OmBPS_1) J_1  + (\Omega_2-\OmBPS_2) J_2 +(\Phi - \PhiBPS)Q \nn\\[1mm]
&= \wh E + \frac{1}{\beta}(\omega_1 J_1  + \omega_2 J_2 + \elpot\, Q )\ .
\end{align}
This will also be the field theory superalgebra and will play an important role in the following.

At this point we can analytically continue the parameters of the supergravity solution allowing them to take more general complex values. In particular, we can impose the supersymmetry condition and the constraint between the chemical potentials described previously. Recall that under this constraint the Killing spinor is antiperiodic; this should be regarded as a further boundary condition which is inherited by the SCFT spinor fields.
We regard the four-dimensional background identified above as the one on which the SCFT should be defined in order to describe the holographic dual of the black hole. 
In the following section we will use the metric \eqref{coord_transf_metric} and gauge field \eqref{Abdry}, but we drop the hats in the notation. 
This should not be confused with our unhatted variables used earlier in this section.

\section{Calculation of the field theory partition function}\label{field_theory_section}

We now discuss the dual field theory calculation of the BPS black hole entropy. 
The black hole solutions that we discussed in the previous sections are solutions to five-dimensional minimal gauged supergravity,
whose duals are four-dimensional $\mathcal{N}=1$ SCFTs. In this section we consider a generic four-dimensional
$\mathcal{N}=1$ SQFT  with a chiral~$U(1)_R$ symmetry, which include~$\mathcal{N}=1$ SCFTs. 
In particular, we consider theories which admit a Lagrangian formulation whose field 
content consists of vector and chiral multiplets. For such theories we compute the exact partition function
using supersymmetric localization, 
and  discuss the relation to the entropy  of the black hole. 
We also briefly discuss the Hamiltonian interpretation of our calculation.

\subsection{The definition and the set up of the functional integral}
\label{sec:PathIntegral}

We consider the background metric 
\be\label{background_metric}
\diff s^2 \=  \diff \tau^2 +  \diff\theta^2 + \sin^2 \theta\, (\diff \phi_1- i \,\Omega_1 \diff \tau)^2 + \cos^2 \theta \,(\diff \phi_2 - i \, \Omega_2 \diff  \tau)^2  \,,
\ee
with the three independent coordinate identifications~$\tau\sim\tau+\beta$, $\phi_1 \sim \phi_1 + 2 \pi$, and $\phi_2 \sim \phi_2 + 2 \pi$.  This describes a fibration of a round $S^3$ of unit radius over $S^1$.
In accordance with the bulk analysis, we allow the quantities~$\Omega_{1},\Omega_{2}, \beta$ to be complex.
In order to study supersymmetric field theories on this geometry, we couple the theory to off-shell supergravity 
and consider its rigid limit~\cite{Festuccia:2011ws}. 
The asymptotic analysis of the bulk supergravity shows that the dual SCFT couples to the four-dimensional background 
conformal supergravity which is defined at the boundary (see e.g.~\cite{Klare:2012gn,Cassani:2012ri}). The boundary 
value of the bulk gauge field defines the gauge field~$A^{\rm cs}$ in the conformal supergravity multiplet, which couples 
canonically to the SCFT $R$-current. In agreement with \eqref{Abdry}, we take this to be the flat connection $A^{\rm cs}=  i \Phi\, \diff\tau$. 
For the moment we can forget about the bulk origin of $\Phi$ and regard it just as a complex parameter. 
The supersymmetric coupling of the field theory to the background described above can also be formulated in terms of 
new minimal supergravity~\cite{Sohnius:1981tp, Sohnius:1982fw}.
This allows us to work more generally and consider an $\mathcal N=1$  SQFT (not necessarily superconformal) with a~$U(1)_R$ symmetry. 
The bosonic content of the new minimal supergravity multiplet consists of the vielbein, a gauge field~$A^{\rm nm}_\mu$ for~$U(1)_R$ symmetry, 
and a vector field~$V^{\rm nm}_\mu$ which is conserved, i.e.~$\nabla^\mu V^{\rm nm}_\mu=0$. The relation between the conformal 
and the new minimal supergravity gauge fields is $A^{\rm cs}= A^{\rm nm}- \frac{3}{2}V^{\rm nm}$. A consistent choice for the 
new minimal background fields is\footnote{The new minimal supergravity fields $A^{\rm nm}$, $V^{\rm nm}$ are not uniquely defined by the conformal supergravity field $A^{\rm cs}$. As shown in \cite{Dumitrescu:2012ha}, they can be shifted by a term proportional to $K_\mu \diff x^\mu$, where $K^\mu \partial_\mu$ is the supersymmetric Killing vector obtained as a bilinear of the Killing spinors, which in our case coincides with \eqref{susy_vector}. We fix this ambiguity by demanding that $A^{\rm nm}$, $V^{\rm nm}$ only have components along $\diff \tau$. We also note that our $A^{\rm nm}$ describes the fluctuation of the electric chemical potential $\Phi$ over the particular value $\Phi^*=\frac{3}{2}$.}
\be \label{AVnm}
A^{\rm nm}  = i \Big(\Phi - \frac{3}{2} \Big)\diff\tau \ , \qquad
V^{\rm nm} =  -i \,\diff\tau \ .
\ee

Before starting the actual computation we spell out the conventions used in this section. We use 4-component (Dirac) 
spinors~$\psi$ which can be split into their left- and right-chiral parts~$\psi^{L,R} = \frac12 (1\pm \gamma^5) \psi$. 
In the Euclidean theory the left and right parts are taken to be independent with a choice of real slice to eventually be made in the functional 
integral.
We use~$\nabla_\mu$ for the derivative covariant 
with respect to the minimal coupling to gravity, so that for integer spin fields it is defined using the Christoffel connection, and for 
a spinor~$\psi$ we use the spin connection, i.e.,
\be 
\nabla _{\mu }\psi \,= \, \partial _{\mu }\psi +\tfrac{1}{4}\,  \omega _{\mu }^{ab}\gamma _{ab}\, \psi \,.
\ee
The flat indices~$a,b$ are with respect to vielbeins~$e^a$, $a=1,\cdots,4$, that are presented in Equations~\eqref{SpinFrame}.
The basis for $\gamma$ matrices is written in Equation~\eqref{gammamatrices}. 
Here~$\gamma _{ab} =\frac{1}{2}[\gamma _{a},\gamma _{b}]$.
In the following we also use~$D_\mu$ for the derivative covariant with respect to the coupling to all the gauge fields (as well as gravity). 
In addition to gravity and the~$U(1)_R$ gauge field mentioned above, we will also have dynamical gauge fields~$\mathcal{A}_\mu$, so 
that the generic form of the covariant derivative is
\be
D_\mu \, := \, \nabla _{\mu } -iA^{\rm nm}_{\mu }\gamma _{5} -i \mathcal{A}_{\mu } \,,
\ee 
where the~$\g_5$ acts as usual on spinors and is taken to be unity on integer-spin fields.  
Our convention is that the gauge field acts according to the representation of the corresponding 
field under the gauge symmetry, so that e.g.~for an adjoint scalar~$\phi$ we have~$\mathcal{A}_\mu \phi =[\mathcal{A}_\mu,\phi]$.
In particular, we will not explicitly write the $R$-charge in the covariant derivative.

The Killing spinor equation of new minimal supergravity in the four-component language is \cite{Sohnius:1981tp}
\be \label{KSeqn}
D _{\mu }\varepsilon +i V^{\rm nm}_{\mu }\gamma _{5} \, \varepsilon 
-\frac{i}{2} (V^{\rm nm})^{\nu } \gamma _{\mu \nu }\gamma _{5} \,\varepsilon \=0 \,.
\ee
We want to solve this equation with the given metric~\eqref{background_metric} and background vector fields~\eqref{AVnm}.  
The set of solutions that we are interested in is
\begin{eqnarray}
\label{KSSol}
\left(
\begin{array}{c}
u_1 \, \rme^{ \frac{\tau}{2} \left(1 - 2\Phi   +  \Omega_1 +  \Omega_2\right)} \\
u_2 \, \rme^{ \frac{\tau}{2} \left( 1- 2\Phi   - \Omega_1- \Omega_2\right)} \\
v_1 \, \rme^{ \frac{\tau}{2} \left(-1 + 2\Phi   + \Omega_1+ \Omega_2\right)} \\
 v_2 \, \rme^{\frac{\tau}{2} \left(-1 + 2\Phi  - \Omega_1 - \Omega_2\right)} \\
\end{array}
\right) \,,
\end{eqnarray} 
for arbitrary constants~$u_i, v_i$, $i=1,2$, that we will constrain below with the help of global conditions. 
We note that the term linear in~$A^{\rm nm}_\mu$ in the covariant derivative contains~$\gamma_5$, and so 
the R-charge of a given mode is correlated with its chirality. 
In our basis the chirality matrix reads $\gamma_5 = \text{diag}(1,1,-1,-1)$, and therefore we can interpret the modes~$u_i$, $i=1,2$, as left-chiral with~$\gamma_5=+1$ with~$R$-charge $+1$, 
and the modes~$v_i$, $i=1,2$, as right-chiral with $\gamma_5=-1$ with~$R$-charge $-1$.

The globally well-defined Killing spinor is either periodic or antiperiodic around the time cycle. 
This enforces the phases in~\eqref{KSSol} to be a specific multiple of $\pi$ times $\tau$. 
Now, the fact that the relative sign between~$(1-2\Phi)$ and~$ (\Omega_1+\Omega_2)$ is one value ($+1$) for the modes~$u_1$ and~$v_2$, 
and another~($-1$) for~$u_2$ and~$v_1$ implies that a well-defined Killing spinor must have either~$u_2=v_1=0$, or~$u_1=v_2=0$. 
These two set of solutions are related by the charge conjugation matrix~$C$ defined in Appendix~\ref{GammaMat}. 
We will make the first choice in the following.
Then we choose one supercharge which is a combination of a left- and a right-chiral supercharge. 
In our four-component language, this is implemented by imposing the condition~\eqref{spinorcondition}. 
Demanding the norm of the spinor,~$\ve^* \ve$, to be one gives us our Killing spinor 
\begin{eqnarray}
\label{KSSol2}
\varepsilon \=\left(
\begin{array}{c}
\rme^{ \frac{\tau}{2} \left(1 - 2\Phi   +  \Omega_1 +  \Omega_2\right)} \\
0 \\
0 \\
\rme^{-\frac{\tau}{2} \left(1 - 2\Phi  + \Omega_1 + \Omega_2\right)} \\
\end{array}
\right) \,.
\end{eqnarray} 
The two non-zero modes in~$\ve$ correspond to the left- and right-chiral parts of the Killing spinor.
The periodicity condition~$\varepsilon(\tau+\beta) = \pm \varepsilon(\tau)$ gives
\be \label{period1}
\b \bigl( 1 + \Omega_1+\Omega_2 -2 \Phi\bigr)  \= 2\pi i n  \,, \quad n \in \mathbb{Z} \,.
\ee
Recalling the definition \eqref{def_lowercase_chempot} of the supersymmetric chemical potentials, we see that the new minimal supergravity gauge field can be rewritten as 
\be \label{period2}
\beta A^{\rm nm}  \= i\,\beta  \Big(\Phi - \frac{3}{2} \Big)\diff\tau \=   i \,  \elpot \,\diff\tau  \,.
\ee
and the periodicity constraint~\eqref{period1} takes the form
\be \label{ConstraintFT}
 \, \omega_1+\, \omega_2 - 2 \elpot \= 2\pi i n \,, \quad n \in \mathbb{Z} \,.
\ee
The constraint~\eqref{ConstraintFT} was derived here as an independent computation in the field theory 
by demanding supersymmetry. Of course the same analysis can be performed---and the same conclusion can be
drawn---by looking at the boundary (UV) values of the bulk Killing spinor. In the bulk we actually say more---supersymmetry 
of the family of solutions reviewed in Section~\ref{sec:CCLPsolQSR} and regularity of the solution in the deep IR fixes~$n=\pm 1$.

As we have already mentioned, this constraint can be only satisfied by complex values of the chemical potential.
It is not clear to us at the moment what, if any, is the physical meaning of the other values of~$n$, apart from~$n=0$
which could be interpreted as the Euclidean pure~AdS$_5$ solution.

We now want to compute the partition function of the field theory in the above background for which 
we use localization with respect to the supercharge generated by~$\varepsilon$. 
We write the partition function schematically as a path integral over all the fields~$\phi$ of the theory, 
\be
Z \= \int [D\phi] \, \exp\bigl(-S_\text{phys}(\phi) \bigr) \,.
\ee
Here the physical action of the theory is supersymmetric, i.e.~$\d_\ve S_\text{phys} = 0$. 
The localization computation of the supersymmetric partition function proceeds 
by deforming the action by a~$\d_\ve$-exact term~$S_\text{phys}(\phi) \to S_\text{phys}(\phi) + t \, \d_\ve \CV$. 
One then shows that corresponding deformed path integral~$Z(t)$ is independent of~$t$, and therefore~$Z \equiv Z(0) = Z(\infty)$. 
In the~$t \to \infty$ limit the integral is dominated by the BPS configurations~$\d_\ve \psi =0$, so that we 
obtain\footnote{Here we assume that one can define a supersymmetric scheme. Starting from a holographic analysis, it has been argued in \cite{Papadimitriou:2017kzw} (see also \cite{An:2017ihs}) that four-dimensional $\mathcal{N}=1$ SCFTs have an anomaly in the supersymmetry variation of the supercurrent which modifies the superalgebra in curved space. Along the lines of \cite{Genolini:2016sxe,Genolini:2016ecx}, here we are assuming that there exists a local counterterm which removes the anomaly from the variation of the supercurrent in the background of interest (at the expense of breaking another symmetry of the background).}
\be \label{locresult}
Z = \int_{\mathcal{M}_\text{loc}} [D\phi] \, \exp\bigl(-S_\text{phys}(\phi) \bigr)\, Z_\text{1-loop}(\d_\ve\CV)  \,, \qquad 
\mathcal{M}_\text{loc} := \{\d_\ve \psi =0 \} \,.
\ee

This technique has been applied to perform similar computations of the partition function of~$\CN=1$ SQFTs on 
various manifolds diffeomorphic to~$S^1 \times S^3$~\cite{Nawata:2011un,Benini:2011nc,Closset:2013sxa,Assel:2014paa}\footnote{The 
computations of~\cite{Nawata:2011un,Benini:2011nc} rely on a nice technique presented in~\cite{Kim:2009wb} which 
identifies a path integral representation of twisted generalizations of Witten's index. } 
including twisted $S^1\times S^3$~\cite{Hosomichi:2014hja}.
In all these treatments, the Killing spinor is time independent (and in particular periodic) in the circle direction, i.e.~$n=0$,
while in our case the gravity solution imposes~$n= \pm 1$, which leads to \emph{antiperiodic} spinors. 
A related technical point is that one often solves for one of the potentials, say~$\varphi$, in terms of the others ($\omega_1,\omega_2$) before going through the computation. 
Here we consider all the potentials separately and obtain a simple-looking expression in terms of these three potentials, on which
we impose the constraint at the end.
As we shall see, the localization computation goes through as usual, and therefore the final result for the partition function, 
expressed as a function of~$ \omega_{1}, \omega_{2}, \elpot$,
is essentially independent of~$n$ except in that the constraint~\eqref{ConstraintFT} is obeyed. 

Given this new feature, we will revisit the localization computation. 
Before doing so we comment on an important conceptual point.
The fact of having an antiperiodic spinor does not mean supersymmetry is broken. The periodicity~$n$
affects both the spinor periodicities as well as the background chemical potentials under which the spinor is charged, 
so that the Killing spinor equation is satisfied for any value of~$n$. 
Assuming as usual that the bosonic fields are periodic, anti-periodicity of the 
supercharge implies that all the fermionic fields are also anti-periodic in the supersymmetric functional integral.
In particular, the partition function that we compute is not the physical thermal partition function, but rather 
a supersymmetric partition function with an unusual constraint. As a consequence, the partition function 
is independent of~$\beta$ as is usual in supersymmetric situations. We will comment more on this in the following.

\subsection{The localization computation}

We now begin the localization computation by considering the chiral
multiplet $(\phi, \psi, F)$ coupled to the background supergravity and in a representation
R with weights~$\{\rho\}$ of the 
dynamical vector multiplet~$(\mathcal{A}_\mu,\lambda, \overline\lambda, D)$.  The fields~$\psi$, $\lambda$ are left-chiral fields. 
The corresponding $R$-charges of the fields in the chiral multiplet are~$(r,r-1,r-2)$.
The variations of the chiral fields are \cite{Sohnius:1982fw} (with~$*$ denoting complex conjugation)
\begin{eqnarray} \label{Chiral}
\delta_{\ve} \phi&=&\sqrt{2} \, i  \bareR   \psi \,, \nn \\
\delta_{\ve} \psi &=& \sqrt{2} \, F \ve^L+i \sqrt{2}\gamma^\mu \ve^R  D_\mu \phi \,, \\
\delta_{\ve}F&=&- \sqrt{2} \,\bareL  \gamma^\mu \Bigl(D_\mu \psi -\frac{i}{2} V^{\rm nm}_\mu \psi \Bigr)+2   \bareL   \overline{\lambda} \, \phi  
 \,. \nn
\end{eqnarray}
The anti-chiral fields~$(\overline\phi, \overline\psi, \overline F)$ are in the representation $\bar R$ with weights~$\{-\rho\}$,  
and have $R$-charges~$(-r,-r+1,-r+2)$. Their supersymmetry variations are 
\begin{eqnarray}\label{eq53}
\delta_{\ve} \overline{\phi}&=&\sqrt{2}\, i  \bareL \bar{\psi} \,, \\
\delta_{\ve} \overline{\psi} &=&\sqrt{2} \, \overline{F} \ve^R+i \sqrt{2}\gamma^\mu \ve^L  D_\mu \overline{\phi} \,, \\
\delta_{\ve} \bar{F}&=&- \sqrt{2}\, \bareR  
\gamma^\mu \Bigl( D_\mu \overline{\psi}+ \frac{i}{2}V^{\rm nm}_\mu \overline{\psi} \Bigr)+2   \bareR   \lambda \, \overline{\phi}\,. 
\end{eqnarray}

For further use we define the following operators
\begin{eqnarray}
P^R_\mu &:=&- \rme^{\, 2\pi i\, n \,\frac{\tau}{\beta}}\, \rme^{i (\phi_1+\phi_2)}\,\ve^{L*}\,\gamma_\mu \, \ve^{R} \,, \\ 
P^L_\mu &:=&- \rme^{-2\pi i\, n \,\frac{\tau}{\beta}}\, \rme^{-i (\phi_1+\phi_2)}\,\ve^{R*}\,\gamma_\mu \, \ve^{L} \,, \\
K_\mu &:=&\ve^{*}\, \gamma_1 \, \gamma_\mu \, \ve \,.
\end{eqnarray}
Denoting a generic vector built from the components~$v_\mu$ as~$v=v^\mu \partial_\mu$, we have
\bea
P^R&=&  
\p_\theta -i \cot{\theta}\, \p_{\phi_1} + i \tan{\theta}\, \p_{\phi_2} \,, \\
P^L&=&
\p_\theta + i \cot{\theta}\, \p_{\phi_1}- i \tan{\theta}\, \p_{\phi_2}  \,, \\
K & =& {2 } \, \Bigl(   \partial_\tau+\frac{i \, \omega_1}{\beta} \partial_{\phi_1} + \frac{i \, \omega_2}{\beta}  \partial_{\phi_2} \Bigr) \,. \label{VectY2}
\eea
The operators~$P^L$ and~$P^R$ have $R$-charges $+2$ and $-2$, respectively, and~$K$ has $R$-charge~$0$.  
Note that (up to a trivial rescaling), $K$ is the same supersymmetric Killing vector \eqref{susy_vector} of the bulk solution, 
just expressed in different coordinates defined in~\eqref{hatted_coords}.\footnote{This is equivalent to the prescription of \cite{Kim:2009wb} 
which exchanges twisted boundary conditions along time circle for dynamical fields for periodic boundary conditions supplemented by 
a twist of the time component of the new covariant derivative.}
The action of this vector, covariantized by the gauge field \eqref{period2}, represents on the physical fields the 
superalgebra \eqref{new_superalgebra}, where $Q$ is now identified with the $R$-charge and the chemical potentials 
satisfy the constraint \eqref{ConstraintFT}.

It is convenient to define the twisted or cohomological variables $\TwistV$ with $R$-charges $(r,\,r-2,\,r,\, r-2)$ as follows,
\bea
\st&: = & \ve^*\psi \,, \\
\sh &: =&\ve^*\gamma_1 \psi \,,  \\
F_\phi &: =&F+i \, \ve^{L*}\,\gamma^\mu \, \ve^{R}
 D_\mu \phi \,. 
\eea
In terms of these variables, the supersymmetry transformations read
\bea\label{TwChir}
&& \delta_{\ve} \phi \= i  \sqrt{2} \, \sh \,, \qquad \delta_{\ve} \sh \= i \sqrt{2} \,H \phi \,, \\
&& \delta_{\ve} \st \=\sqrt{2} \, F_\phi \,, \qquad  \delta_{\ve} F_\phi \= -\sqrt{2} \, H \st \,.
\eea
Acting on the (scalar) twisted variables, the previous supersymmetry variations obey the algebra
\be \label{BPSconditionLoc}
\delta_{\ve}^2 \=  -K^\mu D_\mu \, \equiv \, - 2 H \,.
\ee
Note that the operators~$P^L, P^R$ commute with~$H$:
\be \label{PLPRcomm}
[P^R,H] \= [P^L,H] \= 0 \,.
\ee

We group the cohomological variables in terms of ``elementary" bosons and fermions and their superpartners as follows,
\bea
&& {X}^B  =  \{\XelB\} \,,  \qquad \quad \d_\ve X^B \= \{\QXelB\} \,, \\
&& {X}^F  \=  \{\XelF\} \,,  \qquad \d_\ve X^F \= \{\QXelF\} \,. 
\eea
The deformation term that we add to the original Lagrangian of the theory is~$\d_\ve \, \CV$ 
with the following choice of~$\CV$ which is fairly standard\footnote{The positivity of the action of the fluctuations of this deformation action
needs to be studied with care. We do not enter this discussion here, because our main intention is to show the key differences
with respect to earlier treatments, like for instance the one presented in Section 4  of \cite{Hosomichi:2014hja}. Such differences really come from the new constraint~\eqref{period2}.} 
\bea
\CV \=-\frac{1}{2}\Lv{\d_\ve X^B}{X^F} \, \dmat \, \Rv{X^B}{\d_\ve X^F} \,,
\eea 
where 
\bea
D_{00}&=&\m{0}{H}{H}{0},\qquad D_{01}=\m{0}{0}{0}{0} \,, \\
D_{10}&=&\m{0}{P^L}{P^R}{0}, \qquad D_{11}=\m{0}{-1}{-1}{0} \,,
\eea
so that we have 
\bea \label{deltaV}
\d_\ve \CV &=&\Lv{{ X^B }}{\d_\ve {X^{F} }} \, \m{H}{0}{0}{1}\, \dmat \, \Rv{{ X^B }}{\d_\ve{X^{F} }} \nn \\
&&\qquad\qquad-\Lv{\d_\ve X^B}{ { X^{F} }} \, \dmat \, \m{1}{0}{0}{H}\Rv{\d_\ve X^B}{ { X^{F} }} \,.
\eea

With this deformation term we need to solve for the localization locus and the one-loop determinant of quadratic fluctuations of this 
deformation action evaluated on the space of off-shell fluctuations around this locus.  
We assume\footnote{One should revisit this more with a more rigorous and detailed analysis for the localizing term that we choose above.} 
here, based on a similar analysis in~\cite{Assel:2014paa}, that the localization locus is given by the conditions that 
all the fields of the chiral multiplet vanish, and the only-non-zero fields in the vector multiplet is a constant Wilson line, with holonomy $u$ 
around the time-circle
\bea
\mathcal{A}_\tau=\frac{u}{\beta}\ .
\eea

The one-loop determinant~$Z_\text{1-loop}(\d_\ve \CV)$ that we want to compute is the square root of 
the ratio of determinants of the fermionic and bosonic kinetic operators~$K_f$ and~$K_b$ in the expression~\eqref{deltaV}. 
From linear algebra we have that this ratio of determinants reduces to a ratio defined only on the elementary fields, 
\be \label{detrat}
\frac{\det \;K_f}{\det\; K_b}\=\frac{\det_{X^F} H}{\det_{X^B} H} \,.
\ee 
Further, using the commutation relations~\eqref{PLPRcomm}, given a mode of~$\phi$
with a certain~$H$-eigenvalue, the operator~$P^R$ produces a mode of~$\st$ with the 
same eigenvalue, and similarly~$P^L$ pairs up the fluctuations of~$\overline \phi$ with those of~$\sh$.
This pairing produces a cancellation of eigenvalues in~\eqref{detrat} except if 
any of these modes vanish. The bosonic modes contributing to the ratio of determinants are those obeying 
\be  \label{Ker}
\phi: P^R \, \phi\=0 \,, \qquad \overline{\phi}: P^{L} \, \overline{\phi} \=0 \, ,
\ee
 i.e.~those in the kernel of $D_{10}$.
 Similarly the fermionic modes contributing to the determinant are those 
in the cokernel of $D_{10}$, i.e.,
\bea \label{CoKer}
\st: {P^L} \, \st\=0 \,, \qquad P^R \, \stb\=0 \,.
\eea
The ratio of determinants thus reduces to 
\be
\frac{\det_{X^F} H}{\det_{X^B} H} \=\frac{\det_{\text{Coker}(D_{10})} H}{\det_{\text{Ker}(D_{10})} H} \,.
\ee

\vspace{0.2cm}

The determinant that we seek to compute is thus a ratio of the product of the eigenvalues of~$H$ in the cokernel of~$D_{10}$
and the corresponding product in the kernel. 
Since we are using twisted field variables, the kernel and cokernel conditions are simply 
linear first-order equations acting on scalars which can be solved easily. 
In particular, both the bosons as well as the fermions (which are bilinears in~$\ve$ and~$\psi$) are periodic around 
the Euclidean time and angular circles. 
Therefore we can use basis functions of the form 
$f(\theta) \, \rme^{i \frac{2 \pi}{\beta}n_0 \tau} \rme^{i  n_1 {\phi_1}} \rme^{i   n_2 \phi_2} $, $n_0, n_1, n_2 \in \mathbb{Z}$
for all the twisted variables. Basis functions of this form are in the kernel of $P_L$ and $P_R$ if and only if, respectively,
\be
\begin{split}
f_L(\theta)&\,\sim\, \cos ^{n_2}(\theta ) \,\sin ^{n_1}(\theta )\, , \\
f_R(\theta)&\,\sim \, \cos ^{-n_2}(\theta )\, \sin ^{-n_1}(\theta )\, .
\end{split}
\ee
We immediately see that the modes~$f_L$ 
are regular if and only if 
\be
n_1 \geq 0 \,, \qquad n_2 \geq 0 \,.
\ee
Similarly, the modes~$f_R$ 
are regular if and only if 
\be
n_1 \leq 0 \,, \qquad n_2 \leq 0 \,.
\ee
The eigenvalues of these modes under~$-\frac{i}2 K^\mu \p_\mu$, which we denote by~$\lambda$, are 
\be \begin{split} \label{EMRegularity}
\lambda_{\st}&\= \evpar \,, \qquad \regL\ ,\\
{\lambda}_{\stb}&\=\evpar \,,\qquad \regR \ ,\\
\lambda_\phi&\=\evpar \,,  \qquad \regR \ , \\
{\lambda}_{\overline{\phi}}&\=\evpar \,,  \qquad   \regL\ .
\end{split} \ee

To obtain the final result for the 1-loop determinant, we have to use the full covariant derivative in 
the operator~$H = \frac12 K^\mu D_\mu$. This includes terms proportional to the Wilson line~$\mathcal{A}_\tau=\frac{u}{\beta}$ of 
the dynamical gauge field as well as the $R$-symmetry gauge field~\footnote{Notice that we have normalized the holonomy 
of the color Wilson line to be $u$ and independent of $\beta$.}. 
Putting the above results together, we have
\be  \label{chiral1loop}
Z^{\text{chiral},\rho}_\text{1-loop}(u) \= \prod_{n_0\in \mathbb{Z}}\,\prod_{n_1, n_2 \geq 0} \frac{2\pi n_0 
+ \rho \cdot u - i \left(r-2\right)    \elpot + i \, n_1\,  \omega_1  
+ i\,  n_2 \,  \omega_2  }{2\pi n_0 + \rho \cdot u 
- i r  \elpot  - i\, n_1 \,  \omega_1  -  i \,  n_2\, \omega_2 } \ . 
\ee
Note that this expression is independent of~$\b$. The result for the vector multiplet is related to that of a chiral multiplet with R-charge $r=2$.
Denoting the roots by~$\alpha$, we get
\be
\label{vector1loop}
Z^{\text{vector},\a}_\text{1-loop}(u) \= \prod_{n_0\in \mathbb{Z}}\,\prod_{n_1, n_2 \geq 0} \frac{2\pi n_0 
+  \alpha \cdot u + i  \, n_1\,  \omega_1 + i\, n_2 \, \omega_2  }{2\pi n_0 + \alpha \cdot u - 2i\, \elpot 
- i \,n_1\, \omega_1  - i \, n_2\, \omega_2 } \,,
\ee
where we are omitting a factor that eventually cancels out with the integration measure and the gauge-fixing term \cite{Assel:2014paa}.

\subsection{Regularization and the result for the partition function \label{Sec:regulate}}

The result for the one-loop partition function of a chiral multiplet in a representation~R of gauge group~$G$ with~$R$-charge~$r$ 
is recovered by collecting the corresponding result~\eqref{chiral1loop} for each weight $\rho$. We perform this 
computation in Appendix \ref{RegulatorEll}. As explained there in some detail, 
there are two choices of regulator parameterized by~$s=\pm 1$. The answer can be summarized in terms of the 
elliptic gamma function \eqref{GammaeDef} as follows, 
\be  
\exp \Bigl( -i \pi \Psi \bigl(w, \a_1, \a_2 \bigr) \Bigr) \, \Gamma_e (w+\gamma;\a_1,\a_2 ) \,,
\ee
where the arguments are given by
\bea 
&& \a_1 \=  s \frac{i {\omega}_1}{2\pi}, \qquad 
\a_2 \= s \frac{i {\omega}_2}{2\pi}, \qquad \gamma\= s \frac{i {\elpot} }{ 2\pi } \,, \nn \\ 
&& w(\rho,r) \= {-} s \frac{1}{2\pi} \left(\rho \cdot u-(r-1)i {\elpot}\right) \label{D8} \,.
\eea
The infinite product expression for the elliptic gamma function is convergent for~$\a_{1,2}$ in the upper half plane. 
In order to compare with the BPS black hole solutions discussed in Section~\ref{sec:BPSlimit} which 
have~$\text{Re}(\omega_{1,2})<0$, it is natural to choose the regularization corresponding 
to~$s=-1$.\footnote{\label{signofs} In v1, v2 of this paper, the regularization~$s=+1$ was chosen. 
In this case, one could still make sense of the elliptic gamma funtion in the black hole region~$\text{Re}(\omega_{1,2})<0$
region using an analytic continuation from the region of convergence as described in~\cite{Felder}. 
This procedure leads to an expression for the index as in~\eqref{indexasintegral}, but with the R-charges shifted 
with respect to the physical values, which makes this choice less appealing. 
The consequence of choosing this regulator would be that the result for the leading prefactor~$\Psi^{(0)}$ has 
the opposite sign as for the~$s=-1$ case. This sign then propagates on to the prefactor~$\CF$ in~\eqref{Fgen} below.
}

The result can be written as follows\footnote{\label{thefoot}We note that we have used the ``one-step" regularization in reaching this result 
as in \cite{Closset:2013sxa,Assel:2014paa}. In the case $n=0$, the expression \eqref{PSI20} for $\Psi^{(0)}_2$ obtained by 
using this regularization differs from the one obtained by using the improved ``two-step'' regularization of \cite{Assel:2015nca} (see also \cite{Ardehali:2015hya}). 
However, the results of \cite{Assel:2014paa} and~\cite{Assel:2015nca} agree at  leading order in~$N$,  which is what we focus 
on in this paper. In the~$n=0$ situation analyzed in these papers, it was important for consistency with the~$\beta \to 0$ limit  \cite{DiPietro:2014bca}
that the factor~$\Psi^{(0)}_2$ does not contain the second and third terms, and indeed the two-step regularization was consistent 
with that requirement. 
At present it is not clear  to us at a technical level how to do the two-step regularization with the~$n\neq 0$ constraint. 
In Section \ref{F_from_single_letter} we will show that a naive application of the regularization prescription discussed in \cite{Martelli:2015kuk} suggests that the correct  result may be
$ \Psi^{(0)}_2 = \frac{i \,  \elpot }{12\pi}\Bigl(1 - 2  \pi i n \Bigl(\, \frac{1}{{\omega}_1}+\frac{1}{{\omega}_2} \, \Bigr)  -  
\frac{8 \pi ^2  n^2 }{{\omega} _1 \, {\omega}_2} \Bigr)  $, 
which is of course consistent with the $n=0$ case; but we intend to return to this problem in the future. 
As it will become clear momentarily, this issue does not arise in theories with ${\bf a}={\bf c}$, such as $\mathcal{N}=4$ SYM.}
\be \label{Zchiral}
Z^{\text{chiral}, \text{R}, r}_\text{1-loop}(u) \= \rme^{- i \,\pi \, \text{dim(R)}\, \Psi^{(0)}(r) } \rme^{- i \pi \Psi^{(1)} (u,r)} \prod_{\rho \in \text{R}} \,
\Gamma_e \Bigl(v(\rho,r);\frac{{\omega}_1 }{2\pi i},\frac{{\omega}_2 }{2\pi i} \Bigr) \,,
\ee
where $\Gamma_e$ is the elliptic Gamma function defined in \eqref{GammaeDef} and 
\bea
v(\rho,r) &\=&  \frac{1}{2\pi} \bigl(\rho \cdot u- i\,r\,{\elpot} \, \bigr),\\
\Psi^{(0)}(r)&\=&  i \bigl( (r-1)^3-(r-1)\bigr) \frac{1}{6 \pi}  \, \frac{\elpot^3}{ \omega_1  \omega_2} + (r-1)\Psi^{(0)}_2  \,, \\
 \Psi^{(0)}_2 &\=& \frac{i \,  \elpot }{12\pi}\biggl(1 - 2  \pi i n \Bigl(\, \frac{1}{{\omega}_1}+\frac{1}{{\omega}_2} \, \Bigr) 
 -  \frac{4 \pi ^2 \left(2 n^2+3 n+1\right)}{{\omega} _1 \, {\omega}_2} \biggr)  \,. \label{PSI20}
\eea   
Here the expression~$\Psi^{(1)}$, defined in \eqref{psi1}, does not contribute to the final answer if the 
theory is anomaly-free \cite{Assel:2014paa}, which we assume is the case. 
For a vector multiplet we specialize to~$r=2$, and R = Adj to obtain 
\bea \label{Zvector}
Z^{\text{vector}}_\text{1-loop}(u)&=& \rme^{- i \,\pi\, \text{dim(G)}\,\Psi^{(0)}(2) } \,
\rme^{- i \pi \, \Psi^{(1)} (2)} \prADJ\Gamma_e \Big(v(\rho, 2) ;\frac{{\omega}_1 }{2\pi i},\frac{{\omega}_2 }{2\pi i}  \Big) \,.
\eea

We can now write down the answer for the localized partition function~\eqref{locresult} of an $\CN=1$ supersymmetric gauge theory 
with $n_C$ chiral multiplets $I=1, \ldots, n_C$, with $R$-charges $r_I$ and in representations R$_I$ of the gauge group~$G$. 
We recall that the localization locus is labelled by the Wilson lines of the dynamical gauge field. 
We thus obtain the result for the partition function as a~${\text{rank}}(G)$-dimensional integral,
\be \label{Zfull}
Z \= \frac{1}{|\mathcal{W}|}\int \,\prod_{i=1}^{{\rm rk}(G)} \frac{\diff u_i}{2 \pi} \,  Z^{\text{vector}}_\text{1-loop}(u) \, \prod_{I=1}^{n_C} \, Z^{\text{chiral}, \text{R}_I, r_I}_\text{1-loop}(u) \,,
\ee
where $|\mathcal{W}|$ is the order of the Weyl group and the $u_i$, $i=1, \ldots, {\rm rk}(G)$ are real angle variables ranging from $-\pi$ to $\pi$, parameterizing the maximal torus of $G$.

Putting together the expressions~\eqref{Zchiral}, \eqref{Zvector}, \eqref{Zfull}, we obtain 
\be \label{ZFIsplit}
Z (\omega_1,\omega_2,{\elpot})\= \rme^{-\mathcal{F}(\omega_1,\omega_2,{\elpot})} \,
\mathcal{I}(\omega_1,\omega_2,{\elpot}) \,,
\ee
where the prefactor  
\be \label{Fdef}
\begin{split}
  \mathcal{F}(\omega_1,\omega_2,{\elpot})
&\= - \sum_{I} \text{dim(R$_I$)} \, \uVartwJ \, \frac{ \elpot^3}{ \omega_1  \omega_2}  \\
& \qquad\qquad+\Bigl(\text{dim}(G) +\sum_{I} \text{dim(R$_I$)} (r_I -1) \Bigr) i\, \pi \,\Psi^{(0)}_2 \,,
\end{split}
\ee
is independent of the Wilson line $u$,\footnote{In reaching the right-hand side of \eqref{Fdef} we have used that~$\Psi^{(0)}(r=2) = \Psi^{(0)}_2$.} and 
\be \label{indexasintegral}
\mathcal I(\omega_1,\omega_2,{\elpot}) \=   
\frac{1}{|\mathcal{W}|}\int  \prod_{i=1}^{{\rm rk}(G)}\frac{\diff u_i}{2 \pi} \!\!\!  
\prADJ\!\!\!\!\Gamma_e \Bigl(v(\rho,2);\frac{{\omega}_1 }{2\pi i},\frac{{\omega}_2 }{2\pi i} \Bigr) 
\prod_{I} \prod_{\rho \in \text{R}_I} \!
\Gamma_e \Bigl(v(\rho,r_I);\frac{{\omega}_1 }{2\pi i},\frac{{\omega}_2 }{2\pi i}  \Bigr)  .
\ee
The elliptic Gamma functions  associated with the vanishing roots of the gauge group $G$ have 
a zero mode, these are understood to be removed from the above expression. 
After this is done, these Gamma functions can be expressed in terms of Pochhammer symbols (while those associated with 
non-vanishing roots may be rewritten in terms of Jacobi theta functions).
When~$n=0$, we immediately see that $\mathcal{I}$ reduces to the usual expression of the Hamiltonian index \cite{Dolan:2008qi}. 
As we discuss in the next subsection, the quantity~$\mathcal I$, for arbitrary~$n$, is essentially the Hamiltonian index.

From now on we focus on the prefactor~$\mathcal{F}$ and leave the analysis of the large $N$ asymptotic behaviour of the index to the future.  Using the definition of the central charges that appear in the Weyl anomaly, 
\be
\label{acdef}
\begin{split}
{\bf a}&\=\frac{3}{32} \Bigl( 2 \,\text{dim($G$)} + \Rca \Bigr) \,,\\
{\bf c}&\=\frac{1}{32} \Bigl(4 \, \text{dim($G$)}+\Rcc \Bigr) \,,
\end{split}
\ee
we can rewrite \eqref{Fdef} as\footnote{The sign of~$\CF$ here is correlated to the choice of regulator~$s$, see Footnote \ref{signofs} above. 
}
\be \label{Fgen}
\mathcal{F} (\omega_1,\omega_2,{\elpot})
\= - (3 {\bf c}-2 {\bf a}) \frac{16}{27} \, \frac{ \elpot^3}{ \omega_1  \omega_2} + ({\bf a}-{\bf c}) 16 \pi i \, \Psi^{(0)}_2 \,.
\ee
Recall that using the constraint~\eqref{ConstraintFT}, the chemical potential $\elpot$ can be eliminated in favour of $\omega_1,\omega_2$ and $n$. 
The prefactor $\mathcal{F}$ can be seen as a generalization of the 
supersymmetric Casimir energy~\cite{Assel:2014paa,Assel:2015nca}. The latter corresponds 
to real fugacities $ \omega_1  = - |b_1|$, $\omega_2=-|b_2|$ and $n=0$, leading to zero entropy upon a Legendre 
transform. 

The inclusion of chemical potentials for flavor symmetries is quite straightforward. We simply introduce background gauge fields
that couple to the flavor symmetries and identify the chemical potentials with the Wilson lines of these background gauge fields. 
The flavor chemical potentials thus enter the covariant derivative expressions in the standard manner, and all our calculations 
go through as before. The product expression~\eqref{chiral1loop} for the one-loop determinant now contains a term proportional 
to the flavor chemical potentials, this is presented in Appendix~\ref{RegulatorEll}.
In the following we assume the flavor symmetries to be abelian for convenience of presentation.

As an illustration we can now apply these considerations to the case of~$\CN=4$ SYM with gauge group~$SU(N)$.
In the~$\CN=1$ language the theory consists of three chiral multiplets of R-charge $r=\frac23$, 
and one vector multiplet with~$r=2$. Besides the~$\CN=1$ $R$-symmetry, we now have two other independent 
flavor symmetries whose charges we denote~$F_1$, $F_2$. We denote the three corresponding chemical potentials 
as~$\v$, $\wt \v_1$, $\wt \v_2$, respectively. We can choose the flavor symmetries to 
be the ones that rotate the first and the third chiral multiplet, respectively, so that they have charges~$(F_1,F_2)=(1,0)$
and~$(0,1)$, respectively. The second chiral then has flavor charges~$(-1,-1)$. 
The relation of these charges to the~$SU(4)$ $R$-symmetry of~$\CN=4$ SYM is as follows.\footnote{In Appendix~\ref{RegulatorEll},
we work out the expression for~$\CF$ in the language of the~$SU(4)$ $R$-symmetry.}
Denoting the Cartan generators of the~$SU(4)$ as~$T_i$, $i=1,2,3$, we have
\be
Q \= T_1+\frac23 \, T_2 + \frac13 \, T_3 \,, \qquad F_1 \=  T_2 \,, \qquad F_2 \=  -T_3 \,,
\ee
where~$Q$ is the $R$-charge operator. 
Putting these facts together, we obtain   
\be  
\label{symsym}
\begin{split}
-  \mathcal{F}_{\mathcal{N}=4}  & \= \frac{N^2-1}{6 \,\omega_1  \omega_2} \, \Bigl( \bigl(-\tfrac13\v+ \wt \v_1 \bigr)^3 + \bigl(-\tfrac13\v+  \wt \v_2 \bigr)^3 
+\bigl(-\tfrac13\v-  \wt \v_1 - \wt \v_2\bigr)^3 +  \v^{\;3} \Bigr) \\
&  \= \frac{(N^2-1)}{2} \, \frac{\v_1  \v_2  \v_3}{ \omega_1  \omega_2} \, ,
\end{split}
\ee
where, in going to the second line, we have defined the linear combinations
\be
\v_1 \=  \frac13 (3 \wt \v_1 + 2 \v ) \,, \qquad  \v_2 \= \frac13 ( 3 \wt \v_2 + 2 \v) \,, \qquad  \v_3 \=  \frac13 (- 3 \wt \v_1 -  3 \wt \v_2 + 2 \v )\,.
\ee
We note that there is no term proportional to~${\bf a}-{\bf c}$ because the sum of each of the charges~$F_1$, $F_2$, and~$r-1$ over the four 
multiplets vanish.
The earlier case when only the~$R$-symmetry chemical potential is turned on is recovered by setting~$\wt \v_1= \wt \v_2=0$.

We now go back to our general~$\CN=1$ field theory result~\eqref{Fgen} and compare with the corresponding supergravity result. 
In the  large-$N$ limit we have that  ${\bf a}={\bf c}$ and    converting  to gravity units, we obtain
\be\label{ZlargeN}
-\CF \xrightarrow[N\to \infty]{}  ~   \frac{2\pi}{27 g^3 G_5} \, \frac{ \elpot^3}{ \omega_1  \omega_2} = I \, .
\ee

Upon setting $n=\pm 1$ in the constraint~\eqref{ConstraintFT} between the chemical potentials, this matches precisely the result \eqref{Isusy_is_Esusy} for the supergravity on-shell action and therefore the entropy function. Notice that we have in fact matched the on-shell action 
of the whole set of complexified supersymmetric solutions discussed in Section~\ref{impose_susy}, which contains 
the physical, extremal black hole as a particular case.  
We can now evaluate the constrained Legendre transform of 
\eqref{ZlargeN} 
as described in detail in Appendix~\ref{sec:revisiting_extremization}. Upon imposing that the charges as well as 
the entropy are real, and again setting $n=\pm 1$, this gives precisely the BPS black hole entropy~\eqref{S_BPS}.
For~$\CN=4$ SYM we can also turn on the other two flavor chemical potentials as discussed above and, 
by the same constrained Legendre transform of Appendix~\ref{sec:revisiting_extremization}, we 
can derive the entropy of the supersymmetric~AdS$_5$ black holes with multiple electric charges and an uplift to type IIB supergravity on $S^5$ presented in~\cite{Gutowski:2004yv,Chong:2005da,Kunduri:2006ek}.

\subsection{Relation to the Hamiltonian index}

In this section we discuss the factor~$\CI$ in the split \eqref{ZFIsplit}. We will show that~$\CI(\omega_1,\omega_2,\elpot)$ is, 
in a very precise sense, related to the Hamiltonian index computation of~\cite{Kinney:2005ej,Romelsberger:2005eg}. 
We will first discuss a specific computation in the context of a chiral multiplet as a way to illustrate this relation, and then use 
this to make a more general comment about~\eqref{ZFIsplit}.

When the supercharge is independent of~$\tau$ ($n=0$), it is known that after subtraction of zero point energies the supersymmetric 
path integral on a twisted $S^1\times S^3$ is equivalent to the Hamiltonian index in~$\CN=4$ SYM~\cite{Kinney:2005ej,Aharony:2003sx,Sundborg:1999ue}, 
and more generally in $\mathcal{N}=1$ theories~\cite{Romelsberger:2005eg}. 
Here we revisit this and check that the corresponding analysis holds for the 
more general case of $n \neq 0$. 

Let us start by focusing on the complex chiral multiplet. In the twisted notation that we introduced in the previous section,
the elementary fields are
\be \label{elLetters}
\{\Xel\}. 
\ee   
We recall that all these four twisted variables are scalars and, in particular, both the bosonic as well as the fermionic components 
are periodic. We will need the charge assignments under the charges ($Q,Q_\text{gauge},F,J_1,J_2$) 
where~$Q$ is the R-charge, $Q_\text{gauge}$ is the charge 
vector under Cartan generators of the gauge group $G$,  $F$ is the fermion number, and~$J_{1,2}$ are the angular momenta on the sphere.
The fields $\phi$ and $\overline{\phi}$ have charges $(r,\rho,0,0,0)$ and $(-r,-\rho,0,0,0)$ respectively, 
and the fermions $\st$ and $\stb$ have charges $(r-2,\rho,1,0,0)$ and $(-r+2,-\rho,1,0,0)$ respectively. 

In terms of the Wilson line $u$ and the chemical potentials ($ \omega_1,\,  \omega_2, \,  \varphi$) defined in the previous subsection, 
we define 
\be\label{rapidities}
s \= \rme^{\,i\,u} \,,\qquad t \=\rme^{\,\varphi} \, , \qquad p \=\rme^{\, \omega_1} \, , \qquad q \= \rme^{\, \omega_2} \,,
\ee
dual to the charges $\rho$, $Q$, $J_1$, and $J_2$, respectively.
By a reduction on $S^3$, the Hamiltonian index is equivalent to a counting problem in quantum mechanics. In this quantum mechanics, we are instructed to first compute the trace
\be \label{Itrace}
\mathcal{I}^{\rho,\, r}_{\rm chiral}(u) \, := \, {\rm Tr}\, (-1)^F t^{ \,Q} s^{\, Q_{gauge}} p^{\, J_1} q^{\, J_2} \,.
\ee 
In order to obtain the final answer for the complete index, one multiplies contributions from every multiplet 
in the theory and integrates over the color holonomies~$u$ as in the previous subsection. 
Thus we see that the result of this computation is very close to the localization
computation that we performed in the previous subsection, up to the subtlety about zero-point energy. 

The trace~\eqref{Itrace} is computed in two steps. First we compute the trace 
over the set of elementary ``letters" and their descendants\footnote{The descendants are obtained by spacetime derivative 
action $\partial_{\phi_1}$ and $\partial_{\phi_2}$ on the elementary letters. 
The  quantum numbers of $\partial_{\phi_1}$ and $\partial_{\phi_2}$ are $(0,0,0,1,0)$ and $(0,0,0,0,1)$, respectively. 
The charges ($Q,Q_{gauge},F,J_1,J_2$) commute with our supercharge~$\delta_\ve$.}.
The reality conditions and regularity of the wave functions~\eqref{EMRegularity}
only allow a subset of the Kaluza-Klein modes on the~$S^3$. In the quantum mechanics computation, 
this can be thought of as effectively reducing the number of single letters to two complex letters 
and all their descendants. We choose these two letters to have the quantum numbers of the zero modes of~$\{{\phi},~ \stb \}$. 
The calculation of the single letter index is a very simple adaptation of the  analogous $n=0$ calculation in~\cite{Dolan:2008qi}. 
We find that the answer  for a chiral multiplet of $R$-charge $Q=r$ and gauge group weight~$\rho$ is 
\be \label{pippo}
f^{\rho,r}_\text{chiral}(s,t,p,q) \, := \, \frac{\, t^{r} \, s^{\rho}}{(1-p)(1-q)}-\frac{t^{2-r}s^{-\rho}}{(1-p)(1-q)} \,.
\ee

The second step is to calculate the multiparticle contribution, which is captured in an elegant manner by the plethystic exponential
\bea
\mathcal{I}^{\rho,\, r}_\text{chiral}(u)  \= \exp \Biggl( \sum\limits_{n=1}^{\infty }\frac{1}{n}%
f^{\rho,r}_\text{chiral}(s^{n}, t^{n},p^{n},q^{n}) \Biggr). \,
\eea
Using the following identity (which is a particular case of (5.6) in \cite{Benvenuti:2006qr}), 
\be 
\pexp{p}{q}{\chi \upsilon}\=\prod_{k,\, l=0}^{\infty}\frac{1}{\left(1-\chi\, \upsilon \,p^k\, q^l\right)} \,,
\ee 
we obtain
\bea\label{FinalEquality}
\mathcal{I}^{\rho,\, r}_\text{chiral}(u) \= \prod_{k,\, l=0}^{\infty}\frac{\left(1-s^{-\rho}\, t^{2-r} \,p^k\, q^l\right)}{\left(1-s^{\rho}\, t^{r} \,p^k\, q^l\right)}\=\Gamma_e
\Big(\frac{1}{2\pi} \left(\rho \cdot u-r \,i \,{\varphi}\right);\frac{{\omega}_1}{2\pi i}, \frac{{\omega}_2}{2\pi i}\Big)\,.
\eea
The definition of elliptic gamma function $\Gamma_e$ appearing
on the right-hand side is the same one appearing in \eqref{Zchiral}.
In proving the second equality in \eqref{FinalEquality} we have used the constraint \eqref{ConstraintFT} and the fact that $n \in \mathbb{Z}$.

We can also write down the analogous expression~$\mathcal{I}_\text{vector}(u)$ for the vector multiplet by 
using the fact that the vector multiplet in these calculations behaves like a chiral one with $R$-charge~$2$. 
To obtain the final result for the index, we are instructed to project to gauge invariant states by integrating over~$u$.
Doing so, we land precisely on the expression~\eqref{indexasintegral}. 
We have thus shown that the factor~$\CI$ in Equation~\eqref{ZFIsplit} is a Hamiltonian index. 
This supports our interpretation of the prefactor~$\CF$ as the supersymmetric Casimir energy for arbitrary~$n$.

We now make the relation between the index computed above (valid for arbitrary~$n$) and the index computed 
in~\cite{Kinney:2005ej,Romelsberger:2005eg} (valid for~$n=0)$. 
For the purposes of presentation of this argument, it is enough to focus our attention to $s=1$ ($u=0$).
Our calculation above corresponds to the following trace,
\be\label{Irho}
\mathcal{I}^{\rho,\,r}_\text{chiral}(u=0) \= \, {\rm Tr}_{\CH_\text{tw}}\,  (-1)^F \rme^{\,\omega_1 \, J_1\, + \, {\omega}_2 \, J_2\, +\, {\varphi} \,  Q  } \, .
\ee 
where the twisted Hilbert space~$\CH_\text{tw}$ consists of all elements of the Hilbert space generated by the letters \eqref{elLetters}.
Substituting the constraint \eqref{ConstraintFT} in \eqref{Irho}, we obtain
\bea\label{step1}
\mathcal{I}^{\rho,r}_\text{chiral}(u=0; n)&=& \, {\rm Tr}_{\CH_\text{tw}}\,  (-1)^F \rme^{\pi i n Q} \,\rme^{\, \omega_1  \left( \, J_1+\frac{Q}{2} \right)\, +\,  {\omega}_2 \, \left(J_2+\frac{Q}{2} \right)\,  }.
\eea
Adding and subtracting a factor of $2\pi i n J_1$ in the exponent, we  
obtain\footnote{In this argument we could have equally well chosen~$J_2$ and shifted~$\omega_2$ instead of~$\omega_1$.\label{foot:spinstatistics}}
\be 
\mathcal{I}^{\rho,\,r}_\text{chiral}(u=0) \= \, {\rm Tr}_{\CH_\text{tw}}\,  (-1)^F  \rme^{2 \pi i n J_1} \,\rme^{\, ( \omega_1 - 2 \pi i n) \left( J_1+\frac{Q}{2} \right)\, + \,  {\omega}_2 \, \left(J_2+\frac{Q}{2} \right)\,  }. 
\ee
At this point we note that
\be
\rme^{2 \pi i n J_1} \=1 \quad \text{on} \quad \CH_\text{tw} \,.
\ee
This can be confirmed from the charge assignments given in the beginning of this subsection. 
This statement is a simple consequence of the fact that the twisted bosons as well as the twisted fermions 
are spacetime scalars.
Of course the physical fermionic field still obeys the spin-statistics relation. 
We thus obtain
\bea\label{RelIndicesN}
\mathcal{I}^{\rho,\,r}_\text{chiral}(u=0) &=&  \, {\rm Tr}_{\CH_\text{tw}}\, (-1)^F\,\rme^{\, ( \omega_1 - 2 \pi i n) 
\left( J_1+\frac{Q}{2} \right)\, + \,  {\omega}_2 \, \left(J_2+\frac{Q}{2} \right)\,  }.
\eea

In the discussion in the previous subsection we treated the index~$\CI$ as a function of the three chemical potentials, implicitly
assuming the constraint between them. After having solved for~$\elpot$ we can also write~$\CI$ as a function of the 
independent potentials~$\omega_1, \omega_2$, and~$n$. Making this notation explicit, we obtain:
\be \label{Ichir1}
\mathcal{I}^{\rho,\,r}_\text{chiral}(u=0,\omega_1,\omega_2;n) \= \mathcal{I}^{\rho,\,r}_\text{chiral}(u=0,\omega_1- 2 \pi i n,\omega_2;0) \,.
\ee
Indeed this equality can be checked explicitly by writing~$\elpot = \frac12(\omega_1+ \omega_2) - \pi i n$ in~\eqref{FinalEquality}, and using 
invariance properties of the elliptic gamma function. 
Plugging the expression~\eqref{Ichir1} into the expression for the full index (after integrating over~$u$), we obtain the relation 
that we wanted between the index~$\CI$ at arbitrary~$n$ and the one of~\cite{Kinney:2005ej,Romelsberger:2005eg} which has~$n=0$.

\vspace{0.2cm}

We can equivalently rephrase this above argument as a trace over the Hilbert space of physical fields in which all the fermions are spinors and 
are antiperiodic around the time circle. Once we have understood the quantity~$\CF$ in~\eqref{ZFIsplit} as a generalization of the supersymmetric 
Casimir energy, we can make the identification of the quantity~$\CI$ in~\eqref{ZFIsplit} with the Hamiltonian trace over the 
physical Hilbert space~$\CH_\text{phys}$,
\be
\label{IIndef}
\CI (\omega_1,\omega_2,{\elpot}) 
\=  \, {\rm Tr}_{\CH_\text{phys}}\,  \rme^{\pi i (n+1) F} \, \rme^{-\beta \{\mathcal{Q},\overline{\mathcal{Q}}\}+
\omega_1  J_1 + \omega_2  J_2 + \elpot \, Q } \,,
\ee
where the three potentials are constrained by~\eqref{ConstraintFT}.\footnote{The superalgebra \eqref{new_superalgebra} implies 
that this trace can also be written as ${\rm Tr}\, \rme^{\pi i (n+1) F} \, \rme^{-\beta \wh E} \,$. We understand the index \eqref{IIndef} as a trace where this energy operator $\wh E$ is normal ordered, and the path integral $Z = \rme^{-\mathcal{F}}\mathcal{I}$ as the same trace, albeit with a Weyl ordered operator. The prefactor $\mathcal{F}$ thus essentially corresponds to the expectation value of $\wh E$.} 
Here~$F$ is the fermion number operator, and we have an insertion of~$\rme^{\pi i (n+1) F}$ because for even (odd)~$n$,
the fermions are periodic (anti-periodic) when we translate the functional integral into the Hamiltonian formalism.
The dependence of~$\CI$ on~$n$ appears implicitly as usual through the constraint~\eqref{ConstraintFT} that relates the three chemical potentials, and 
also explicitly through the insertion of~$\rme^{\pi i (n+1) F}$.

Solving the constraint for~$\elpot$ as above, we obtain this trace as a function of two independent chemical potentials and~$n$, 
\be
\label{ourhatI}
\CI(\omega_1,\omega_2;n) \= \, {\rm Tr}_{\CH_\text{phys}}\,  \rme^{\pi i (n+1) F-\pi i n Q} \,\rme^{-\beta \{\mathcal{Q},\overline{\mathcal{Q}}\} + 
 \omega_1 (J_1+\frac{1}{2}Q) + \omega_2 (J_2+\frac{1}{2}Q) } \,.
\ee
In this case we insert in the above equation, a factor of unity in the following form given by the spin-statistics theorem 
valid for the physical fermions,
\be
\rme^{2 \pi i n (J_1 + \frac12 F)} \= 1  \quad \text{on the physical Hilbert space} \,,
\ee 
in order to obtain
\be
\CI(\omega_1,\omega_2;n) \= \, {\rm Tr}_{\CH_\text{phys}}\,  \rme^{\pi i F} \,\rme^{-\beta \{\mathcal{Q},\overline{\mathcal{Q}}\} + 
( \omega_1 - 2 \pi i n)(J_1+\frac{1}{2}Q) + \omega_2 (J_2+\frac{1}{2}Q) } \,.
\ee

In other words, the~$n$-dependence of the quantity~$\CI$ can be completely absorbed in a shift of 
one\footnote{Mirroring footnote~\ref{foot:spinstatistics}, here we could have 
equally well chosen~$ \omega_2$ instead of~$\omega_1$ in writing the spin-statistics theorem.}
of the chemical potentials, i.e.,
\be\label{RelInd0}
\CI(\omega_1,\omega_2;n) \= \CI(\omega_1- 2 \pi i n,\omega_2;0) \,,
\ee
where the right-hand side is the familiar index
\be 
\CI(\omega_1,\omega_2;0) 
\=  \, {\rm Tr}_{\CH_\text{phys}}\,  (-1)^F \rme^{-\beta \{\mathcal{Q},\overline{\mathcal{Q}}\}  +\omega_1 (J_1+\frac{1}{2}Q)+\omega_2 (J_2+\frac{1}{2}Q)} \,.
\ee

We would like to make a comment here about the relevance of the constraint~\eqref{ConstraintFT}  in the field theory calculation. 
Naively one may say that taking the spinor antiperiodic and $n=1$ is not particularly meaningful, because the same~$n=1$ appears 
in the background gauge field that is proportional to $\varphi$ entering the constraint equation, and therefore one could simply
``gauge-shift it away" to reach the usual situation with~$n=0$ and periodic spinors. Indeed the calculations in the present subsection 
of the Hamiltonian index and, in particular, Equation~\eqref{RelInd0} bear this out. On the other hand, as we saw in Section~\ref{Sec:regulate},
the quantity~$\CF$ does depend on~$n$, and indeed this is what 
leads to a non-vanishing entropy.
This is reminiscent of an anomaly mechanism, which we think is one of the interesting 
questions that should be addressed in the future.

\subsection{$\mathcal{F}$ from single letter index}\label{F_from_single_letter}

In this subsection we provide an alternative derivation for the generalization of the supersymmetric Casimir energy, including arbitrary 
flavor fugacities, following the discussion in~\cite{Martelli:2015kuk}. In this reference it was shown that the supersymmetric Casimir energy, defined as the vacuum expectation 
value of the supersymmetric Hamiltonian, may be computed from a limit of an ``index-character'' counting zero modes of a twisted Dirac operator, in the spirit of \cite{Martelli:2006yb}.  This in turn was 
shown to be closely related  to the single letter indices\footnote{The relevance of this limit for the computation of Casimir energies was anticipated in \cite{Kim:2012ava}.} and the limit was interpreted physically as  a heat-kernel regularization.  
We expect that simple modifications of the arguments of \cite{Martelli:2015kuk} will show that for $n\neq0$ the prefactor $\CF$ is still computed by the same limit of the modified  single letter index (\ref{FinalEquality}). Below we describe this limit, leaving 
a more rigorous justification for future work.

We have seen that the contribution of the one-loop determinant of the chiral multiplet may be written as the plethystic exponential of a single letter index as in 
\eqref{pippo}. This  generalizes the usual single letter index because it has three independent global fugacities 
$(t,p,q)$.  We start rewriting the above expression as
\be
f^{r}_\text{chiral}(\newt,p,q) \, = \, \frac{\, x^{r}  - x^{2-r} }{(1-p)(1-q)}   \,,
\label{fff}
\ee
where here we changed the notation for the fugacity $t$ as  $\newt
$ and from now on we suppress the  gauge fugacity $s$.
This extends the usual expression, which depends on two fugacities only, reflecting the constraint $n=0$.

We also noted  the fact that the one-loop determinant of the vector multiplet can be obtained from that of the chiral multiplet by replacing 
 $r\to 2$. On the other hand, it is known that this can also be expressed as the plethystic expenential of a single letter index, which is 
 not  equivalent to the single letter chiral index with $r\to 2$; however, we have the following relation
\bea
 f_\mathrm{vector}  (\newt ,p,q)  & =   &  1+  f_\mathrm{chiral}^{r=2}  (\newt,p,q) \nn\\
 & = &    1+  \frac{\, x^{2}  - 1 }{(1-p)(1-q)} ~.
 \label{newfchiral}
\eea

Before proceeding with the limit, let us note that ``flavor'' (namely global symmetries that commute with the supercharge) fugacities can be included straighforwardly in our computations. 
Denoting the fugacity as $h$, and the corresponding charge as $q_f$,  the  elliptic gamma function becomes 
\be
\mathcal{I}^{ r  ,q_f}_\text{chiral} (\newt,p,q,h)  = 
\prod_{k,\, l=0}^{\infty}\frac{1-h^{-q_f}\, \newt^{2-r} \,p^k\, q^l }{1-h^{q_f}\, \newt^{r} \,p^k\, q^l}~,
\ee
which in turn can be written as the plethystic expenential of a single letter index with the extra fugacity, namely 
\be
f^{r,q_f}_\text{chiral}(\newt,p,q,h) \, = \, \frac{ x^{r} h^{q_f} - x^{2-r} h^{-q_f}  }{(1-p)(1-q)}   \,,
\label{newfvector}
\ee
Here $h^{q_f}$ can be immediately generalized to 
\be
 h^{q_f} = \prod_i^{n_f} h_i^{q_f^i}
\ee
where $n_f$ is the number of $U(1)$ flavor symmetries, but for simplicity we will write formulas with only one flavor\footnote{We do not consider non-Abelian flavor symmetries here, but it is straighforward to include them too.}.  Armed with (\ref{newfchiral}) and (\ref{newfvector}) we can derive expressions for the generalized supersymmetric Casimir energy, including arbitrary flavor fugacities, following the limiting procedure explained in \cite{Martelli:2015kuk}.
For simplicity we present the computations in the case of an Abelian  vector multiplet plus one chiral multiplet with R-charge $r$ and flavor   $U(1)_f$ charge $q_f$.   
As we shall see, the extension to arbitrary gauge groups and chiral multiplets in general representations can be easily reconstructed from this simple case. 
We thus start with the total single letter index 
\be
 f_\mathrm{vector}  (\newt ,p,q)  + f^{r,q_f}_\text{chiral}(\newt,p,q,h)      =    1+  \frac{\, x^{2}  - 1 }{(1-p)(1-q)}  +     \frac{ x^{r} h^{q_f} - x^{2-r} h^{-q_f}  }{(1-p)(1-q)}  
\ee
we set
\be\label{rapidchange}
\newt =\rme^{ \ccc \varphi} \, , \qquad p =\rme^{\ccc \omega_1} \, , \qquad q = \rme^{\ccc \omega_2}  \, ,   \qquad h = \rme^{\ccc u_f } \,.
\ee
Taylor expanding  around $c\to 0$ and  picking up  the linear term in $-c$, after using the constraint
\be 
 \, \omega_1+\, \omega_2 - 2 \varphi = 2\pi i n \,, 
\ee
we get the following contribution from the vector single letter
\be
\mathcal{F}_\mathrm{vector}(\omega_1,\omega_2,{\elpot})   = \frac{\varphi (-\omega_1\omega_2 +  2 i \pi n (\omega_1+ \omega_2 )   + 8\pi^2 n^2)} {12\omega_1\omega_2}~.
\ee
For the chiral multiplet we get the contribution
\bea
\mathcal{F}_\mathrm{chiral}(\omega_1,\omega_2,{\elpot})    & = &  -\frac{1}{6\omega_1\omega_2}\left(  \varphi  (r-1)  +   u_f q_f  \right)^3  \nn\\
 &+&  \frac{-\omega_1\omega_2 +  2 i \pi n (\omega_1+ \omega_2)      + 8 \pi^2 n^2 + 2\varphi^2} {12\omega_1\omega_2}\left(   \varphi  (r-1)  +   u_f q_f \right) \,.\quad
\eea
Recalling the expressions for the anomalies  (\ref{acdef}),  and setting temporarily  $q_f=0$, we see that the total contribution can be rewritten as
\bea
\label{Ffromlimit}
\mathcal{F} (\omega_1,\omega_2,{\elpot})  & = & -(3{\bf c}-2 {\bf a})  \frac{16}{27}\frac{\varphi^3}{\omega_1\omega_2}   +   ({\bf a}- {\bf c})    16  \pi i \Psi_2^{(0)}~,
 \eea
which is precisely the expression that we derived with the path integral method, but with  
\begin{align}
\Psi_2^{(0)} &=      \frac{i\varphi }{12\pi } \frac{(\omega_1\omega_2 -  2 i \pi n (\omega_1+ \omega_2 )    - 8\pi^2 n^2) } {\omega_1\omega_2}  \nn\\[1mm]
&=  \frac{i\varphi }{12\pi } \frac{(  - 8\varphi^2    - 6 \varphi (\omega_1+ \omega_2 ) +  \omega_1^2    +  3 \omega_1\omega_2  + \omega_2^2  )  } {\omega_1\omega_2}  ~.
  \end{align}
The first line is the result anticipated in Footnote \ref{thefoot}. The second line, where we have eliminated $n$ by writing it in terms of the three fugacities, makes it manifest that $\mathcal{F}(\omega_1,\omega_2,\elpot)$ is homogeneous of degree one. This implies that when $n=0$ its Legendre transform, which is the entropy at order $\mathcal{O}(N)$, vanishes. As we have discussed, this is no more true for $n\neq0$.

After restoring the flavor fugacity, the general expression reads 
\bea
 \mathcal{F} (\omega_1,\omega_2,{\elpot},u_f)  & = &- \frac{\varphi^3 (1+(r-1)^3)}{6\omega_1\omega_2}- \frac{u_f^3q_f^3}{6\omega_1\omega_2} -  \frac{\varphi^2 u_f (r-1)^2 q_f}{2\omega_1\omega_2} -  \frac{\varphi u_f^2 (r-1) q_f^2}{2\omega_1\omega_2}  \nn\\[1mm]
& & +\,   \frac{\omega_1^2 + \omega_2^2 +  12\pi^2  n^2 }{24\omega_1\omega_2} \left(   \varphi ( 1+(r-1))  +   u_f q_f \right)   ~.
\eea
This extends to general $n$ the result (4.19) presented in   \cite{Bobev:2015kza}, that is recovered  setting $n=0$\footnote{To compare with  \cite{Bobev:2015kza} we have to make  the  identificatons 
$\varphi \to -\beta\sigma,\; \omega_i^{\rm here}=-\beta\omega_i^{\rm there},\; (r-1)^3+1\to k_{rrr},\;   (r-1)^2q_f \to k_{rrf},\;  u_f \to -\beta m_I~,$ etc.}.
This can be compactly written in terms of 't Hooft anomaly coeefficients as 
\bea
\mathcal{F} (\omega_1,\omega_2,{\elpot},u_f)   & = & -\frac{\varphi^3k_{rrr} + 3 \varphi^2 u_f k_{rrf}  +3  \varphi\, u_f^2 k_{rff} + u_f^3k_{fff}} {6\omega_1\omega_2} \nn \\[1mm]
&&+ \,  \frac{\omega_1^2 + \omega_2^2 +  12\pi^2 n^2 }{24\omega_1\omega_2} \left(  k_r  \varphi  +   k_f  u_f  \right) ~,
\eea
which exactly reduces to (4.19) of \cite{Bobev:2015kza} for $n=0$.

A slightly different perspective  on the result (\ref{Ffromlimit}) is given by thinking of this as arising from a regularised sum of eigenvalues of a supersymmetric Hamiltonian,  namely imagining that 
\bea
\mathcal{F} (\omega_1,\omega_2,{\elpot})   & = &\frac{1}{2} \sum_{n_1,n_2\geq 0} \lambda^\phi_{n_1,n_2} + \frac{1}{2}\sum_{n_1,n_2\geq 0} \lambda^B_{n_1,n_2} ~,
\label{Fassum}
\eea
where, in the notation of   \cite{Assel:2015nca}, $\lambda^\phi_{n_1,n_2}, \lambda^B_{n_1,n_2}$ are the eigenvalues of unpaired modes, which upon reduction on $S^3$ are  chiral and Fermi multiplets, respectively. Assuming that these eigenvalues can be read off simply from 
the one-loop determinants (\ref{chiral1loop}),   (\ref{vector1loop}), as in the $n=0$ case,  then for the four-dimensional chiral multiplet we have
\bea
 \lambda^\phi_{n_1,n_2} & = &      - n_1     \omega_1  -   n_2  \omega_2  -  r  \elpot ~,\nn\\
\lambda^B_{n_1,n_2}  & = &      n_1\,  \omega_1   +   n_2 \,  \omega_2 -   \left(r-2\right)  \elpot ~,
\eea
with similar expressions for the contributions of the vector multiplet.  If we regularise the two infinite sums \emph{separately}, using the insight of \cite{Assel:2015nca}, we conclude that the regularised sum is 
\bea
\mathcal{F} (\omega_1,\omega_2,{\elpot})   & = & \frac{1}{2} \zeta_2 (-1;-\omega_1,-\omega_2,-r\elpot)- \frac{1}{2} \zeta_2 (-1;-\omega_1,-\omega_2,-(2-r)\elpot)~,
\eea
which coincides with the chiral multiplet contribution to (\ref{Ffromlimit}). Indeed, as discussed in \cite{Martelli:2015kuk}, the two methods for computing $\mathcal{F}$  described here are simply two equivalent regularizations of (\ref{Fassum}).

\subsubsection*{The example of ${\cal N}=4$ SYM}

Let us finally return to the example  of ${\cal N}=4$ SYM, and reproduce the result \eqref{symsym}, applying the procedure discussed above. 
In principle we can use the expressions derived above,
specified to the case  of ${\cal N}=4$ SYM viewed as   of ${\cal N}=1$ theory with three chiral multiplets. It is however  more efficient 
 to derive the desired result from a limit of a single letter index that generalizes that presented in  \cite{Kinney:2005ej}. Let us therefore consider the following single letter index
\bea
     f_{{\cal N}=4} (\newt,p,q,v,w)   & = &        1+  \frac{\, x^{2}  - 1 }{(1-p)(1-q)}  \nn\\[1mm]
     &&\,+ \frac{  x^{r_1}  \tfrac{1}{w}   -   x^{2-r_1}w   + x^{r_2}  \frac{w}{v}   -   x^{2-r_2} \tfrac{v}{w}   + x^{r_3} v   -   x^{2-r_3} \frac{1}{v}}  {(1-p)(1-q)}   ~.
\eea
This is obtained as the sum of generalized vector multiplet single letter index and three chiral multiplets, with two flavor fugacities $v,w$ for the charges commuting with one supercharge. 
Setting $r_1=r_2=r_3=2/3$, $\newt=t^3$, and further making the change of variables
\be
p = t^3 y ,\qquad q= \frac{t^3}{y}~,
\ee
this reduces exactly to the index written in  equation (4.2) of  \cite{Kinney:2005ej}, depending on four fugacities. 
Setting $r_1=r_2=r_3=2/3$ and  further changing variables as follows 
\be
u_1 = \frac{\newt^{2/3}}{w}~, \qquad u_2 = \frac{\newt^{2/3}w}{v}~,\qquad u_3 = \newt^{2/3}v~,
\ee
we get the expression
\be
   f_{{\cal N}=4} (p,q,u_1,u_2,u_3)  \=    1-\frac{(1-u_1) (1-u_2) (1-u_3)}{(1-p) (1-q)} \,.
 \ee
To implement the limit we now set 
\be
p =\rme^{\ccc \omega_1} \, , \qquad q = \rme^{\ccc \omega_2}  \, ,   \qquad   u_1 =\rme^{ \ccc \varphi_1 } \, , \qquad u_2 =\rme^{ \ccc \varphi_2} \, , \qquad 
u_3 =\rme^{ \ccc \varphi_3 } \, , 
\ee
expand near to $c\to 0$ and pick up the linear term in $-c$, getting 
\be
\mathcal{F}_{{\cal N}=4} = -\frac{1}{2}(N^2-1)\frac{\varphi_1\varphi_2\varphi_3}{\omega_1\omega_2}
\ee
where we have reintroduced the dimension of the gauge group $SU(N)$. Now we have that
\be
\rme^{c ( \varphi_1+\varphi_2+\varphi_3) } =  u_1 u_2 u_3  = \newt^2 =  \rme^{2c\varphi}
\ee
implying that 
\be
\varphi_1+\varphi_2+\varphi_3 = 2 \varphi =    \omega_1 +  \omega_2 -  2\pi i n   ~,
 \ee
 which agrees with our Lagrangian analysis in~\eqref{symsym},  and   for $n=0$ reproduces the formula presented in~\cite{Bobev:2015kza}. 
 In the special case that $\varphi_1=\varphi_2=\varphi_3=\frac{2}{3}\varphi$, this reduces to 
\be
\mathcal{F}_{{\cal N}=4} = -\frac{4(N^2-1)}{27}\frac{\varphi^3}{\omega_1\omega_2}~,
\ee
 in agreement with our general formula  \eqref{Fgen}.

\section{Summary and discussion}\label{sec:discussion}

In this paper we presented a holographic computation of the entropy of supersymmetric AdS$_5$ black holes.
In particular, we reproduced the grand-canonical partition function whose constrained Legendre transform yields the Bekenstein-Hawking entropy both on the supergravity and on the SCFT side.   
 In this manner, we clarified the extremization principle proposed in \cite{Hosseini:2017mds}. One of our main results is the identification of the holographic dictionary, in particular the precise match of the SCFT chemical potentials (namely, the background fields entering in the SCFT partition function) with the chemical potentials arising in the BPS limit of black hole thermodynamics and appearing in the supergravity on-shell action. We defined the BPS limit by starting from a complexified solution that is supersymmetric but non-extremal, and taking the limit to extremality. We have explained how the resulting chemical potentials are different from those obtained previously in \cite{Silva:2006xv} with a similar approach but another definition of the BPS limit. In particular, the supersymmetric chemical potentials that we defined are complex, while those defined in 
 \cite{Silva:2006xv} are real, and therefore cannot satisfy the key relation \eqref{crucialconstraint}.
We have noted that this relation enforces antiperiodic boundary conditions for the spinors around the thermal circle in the black hole geometry.

Our gravitational analysis determines the correct functional integral in the holographic field theory. In the large-$N$ limit, we found that the degeneracy of states in the boundary SCFT carrying the same charges as the black hole
is captured by a generalization of the supersymmetric Casimir energy, which can essentially 
be viewed as the vacuum energy. This vacuum energy is completely controlled, in this limit, by one anomaly 
coefficient, namely the central charge~{\bf c} (={\bf a}) of the four-dimensional theory. 
This phenomenon is of course very reminiscent of the Cardy formula in two-dimensional CFTs. It would be very interesting if this line of thought can be made more precise as it would then 
lead to predictions about the subleading quantum properties of supersymmetric 
black holes similar to the case of black holes in asymptotically flat space, which were governed by the  
modular transformation of two-dimensional SCFTs, or their refinements~\cite{Sen:2007qy, Sen:2008yk, Dabholkar:2010uh, Dabholkar:2011ec, Dabholkar:2012nd}. 
In order to do this for four-dimensional SCFTs, we need to understand the details of the 
modular-like properties of the elliptic gamma functions that appear in the boundary partition function~\cite{Spiridonov:2012ww},
and how exactly they are related to our physical computation. A duality of the type proposed in \cite{Shaghoulian:2016gol} may also shed some light on this problem.

Another point worth noting is that we studied the full asymptotically AdS$_5$ black hole solution in the context 
of the~AdS$_5$/CFT$_4$ correspondence, rather than the near-horizon  AdS$_2$/CFT$_1$ correspondence 
as in Sen's entropy function approach \cite{Sen:2005wa}. 
In fact, the entropy of BPS AdS$_5$ black holes has been studied in~\cite{Morales:2006gm,Dias:2007dj,Suryanarayana:2007rk} 
in the entropy function formalism. It was found in~\cite{Dias:2007dj,Suryanarayana:2007rk} that 
the supersymmetric chemical potentials obtained in this 
limit can be mapped to the near horizon field strengths in the~AdS$_2$ region, and using this dictionary the BPS Euclidean on-shell action 
coincides with Sen's entropy function. 
It would be interesting to study the relation between our BPS limit and the entropy function formalism. Perhaps this could 
teach us about the details of the~CFT$_1$ relevant to the black hole entropy through an embedding into a better 
understood~CFT$_4$, such as the~$\CN=4$ SYM theory.

 The strategy of this paper may also be employed in other contexts, for instance it could clarify further the holographic dictionary for the static supersymmetric AdS$_4$ black holes studied in \cite{Benini:2015eyy,Benini:2016rke,Cabo-Bizet:2017xdr}, and possibly for their rotating generalizations. In particular, it would be interesting to see if a regularity condition in the bulk can explain on the gravity side the constraint $\sum_a \Delta_a = 2\pi$, where $\Delta_a$ are background gauge field holonomies, that arises when evaluating the large $N$ limit of the topologically twisted index for the ABJM theory \cite{Benini:2015eyy,Hosseini:2016tor}.

The results we derived in Section \ref{field_theory_section} have an independent interest from the main thread of the paper, 
and are valid for arbitrary (Lagrangian) four dimensional ${\cal N}=1$ field theories with an $R$-symmetry, not necessarily 
having any gravity dual. In particular, using localization, we have computed a supersymmetric partition function on 
twisted $S^1\times S^3$ manifolds, generalizing  the results of~\cite{Assel:2014paa}, allowing for an additional discrete 
chemical potential labeled by an integer $n$. These are  exact results, valid for general gauge groups, and without involving 
any large $N$ limit. 
We have shown that this partition function factorizes into a trace part, which is the supersymmetric 
index~\cite{Romelsberger:2005eg}, and a prefactor that is the~$n$-dependent version of the supersymmetric Casimir energy,
that we calculate using a regularization procedure for the one-loop determinant arising from localization. 
We have also calculated this prefactor by a limiting procedure on the single-letter index along the lines of~\cite{Martelli:2015kuk}. 
It will be nice to complement our analysis with the study of the supersymmetric quantum mechanics 
obtained by reduction on the~$S^3$ extending~\cite{Assel:2015nca,Martelli:2015kuk,Lorenzen:2014pna}. 

We anticipate that it will be possible to expand the results of this paper in several  directions. 
Here we derived the microscopic entropy of the multi-charge BPS black holes of~\cite{Gutowski:2004yv,Chong:2005da,Kunduri:2006ek}. 
These multi-charge solutions feature running scalar fields, and it will be interesting to investigate whether the extremization of the on-shell action is related to a five-dimensional attractor mechanism for the scalars as suggested in \cite{Hosseini:2017mds}. It will also be worth considering the recently constructed asymptotically locally AdS$_5$ black holes with a squashed boundary \cite{Blazquez-Salcedo:2017ghg,Cassani:2018mlh}. The squashing deformations studied in these papers should not affect the SCFT partition function, so one could expect the story to work similarly to the case with no squashing; however the subtleties related to the non-trivial gauge field at the boundary noted in \cite{Cassani:2018mlh} deserve a careful study.  In another direction, one may wonder what field theory computation reproduces the entropy of the hairy black holes found in \cite{Markeviciute:2018yal,Markeviciute:2018cqs}.
It will also be interesting to extend our findings to AdS$_7$ black holes and thus provide a physical explanation for the corresponding extremization principle proposed in~\cite{Hosseini:2018dob}.

\vskip 5mm 

\noindent {\bf Note added in v3:} Some of the statements in earlier versions of this paper were based on the assumption  
that at  large $N$ the index $\mathcal I(\omega_1,\omega_2,{\elpot})$  scales as ${\cal O}(1)$, as originally argued in \cite{Kinney:2005ej}. 
In particular,  this led us to conclude that the field theory supersymmetric partition function
\be
Z (\omega_1,\omega_2,{\elpot})\= \rme^{-\mathcal{F}(\omega_1,\omega_2,{\elpot})} \,
\mathcal{I}(\omega_1,\omega_2,{\elpot}) \,,
\ee
behaves as $-\log Z (\omega_1,\omega_2,{\elpot}) \to  \mathcal{F}(\omega_1,\omega_2,{\elpot}) $ in this limit. 
However, it has been later pointed out in   \cite{Benini:2018ywd}
that at least for ${\cal N}=4$ SYM, for a suitable domain of \emph{complex} fugacities,  
the index itself scales as 
\be
-  \log \mathcal{I}_{\mathcal{N}=4}  \,  \xrightarrow[N\to \infty]{}  \, \frac{N^2}{2} \, \frac{\v_1  \v_2  \v_3}{ \omega_1  \omega_2}  
\,  =  \,  -   \mathcal{F}_{\mathcal{N}=4} ~,
 \ee
thus matching precisely the gravitational on-shell action of the black hole, $I$.
This, together with our findings, suggests that for generic ${\cal N}=1$ field theories (at least those with a type IIB gravity dual), 
the large $N$ asymptotics of the index will be governed by the generalization of the  
supersymmetric Casimir energy, that we computed in this paper, and therefore  generically the on-shell action should 
be identified with the large $N$ limit of  
\be
-  \log \mathcal{I}(\omega_1,\omega_2,{\elpot}) \,  \xrightarrow[N\to \infty]{} \, -\mathcal{F}(\omega_1,\omega_2,{\elpot})   
\= I   (\omega_1,\omega_2,{\elpot})   ~.
\ee
It also suggests that the background subtraction renormalization scheme that we employed to compute the on-shell action
corresponds in the field theory to a scheme in which the supersymmetric Casimir energy is subtracted from the path integral 
calculation of the partition function, so that in this scheme $Z=\mathcal{I}$.

While on the one hand this  supports the expectation that the index  $ \mathcal{I} $ carries information about the IR physics, 
corresponding to the horizon of the black hole, on the other hand it leads to the intriguing 
prediction that the large $N$ asymptotics of the index (as well as the Cardy-like limit \cite{Choi:2018hmj,Honda:2019cio}) are 
controlled by the generalization of the supersymmetric Casimir energy, which is by definition 
related to the ``small temperature'' limit of the full partition function.   As we already noticed, this phenomenon is reminiscent 
of the Cardy behaviour and modular invariance in two-dimensional CFTs, and it will be very interesting 
to explore this in depth.

\section*{Acknowledgments}

We would like to thank Rajesh Gupta for useful discussions and initial collaboration on this project. 
We also thank  Francesco Benini,  Morteza Hosseini,   Juan Maldacena, and Shiraz Minwalla  for  discussions and 
comments on an earlier version of this paper. The research of A.~C.~B.~and S.~M.~is supported by the 
ERC Consolidator Grant N.~681908, ``Quantum black holes:~A microscopic window into the microstructure of gravity''. 
D.~M.~is  supported by the ERC Starting Grant N. 304806,  
``The gauge/gravity duality and geometry in string theory''. The research of S.~M.~is also supported by 
the STFC grant ST/P000258/1. 

\appendix

\section{The bulk Killing spinor}\label{app:KillingSpinor}

In this appendix we explicitly solve the Killing spinor equation in the background given by the metric and gauge field of Section~\ref{sec:CCLPsolQSR}. We work in Lorentzian signature and comment on the analytic continuation at the end.

In general, a bosonic solution to five-dimensional minimal gauged supergravity is supersymmetric if there exists a non-zero Dirac spinor $\epsilon$ solving the Killing spinor equation which arises from setting to zero the supersymmetry variation of the gravitino. In the conventions of \cite{Chong:2005hr} (recalling our footnote \ref{foot:rescaleA}), the Killing spinor equation reads: 
\be\label{KillingSpEq}
\left[ \nabla_\mu - \frac{i}{12g}\left( \Gamma_{\mu}{}^{\nu\kappa} - 4\delta^\nu_\mu\Gamma^\kappa\right) F_{\nu\kappa} - \frac{g}{2} \,  \Gamma_\mu - \,i A_\mu \right]\epsilon  \ =\ 0\ ,
\ee
where the gamma-matrices obey the Clifford algebra $\{\Gamma_\mu,\Gamma_\nu\}=2g_{\mu\nu}$ and in this appendix $\mu,\nu$ are five-dimensional indices. One can see that the integrability condition of this equation is satisfied in the background of \cite{Chong:2005hr} if the parameters $a,b,m,q$ satisfy relation \eqref{susyCCLP}. We assume this condition is imposed and describe the supersymmetric solution using the three parameters $a,b,m$.

It is convenient to choose a set of coordinates and a frame adapted to the general structure of supersymmetric solutions first described in \cite{Gauntlett:2003fk}; this will make it easier to solve the equation. For the three-parameter supersymmetric solution of \cite{Chong:2005hr}, the form adapted to supersymmetry was given in \cite{Cassani:2015upa}. While we refer to that paper for  a  detailed discussion, here we just provide the information needed to solve \eqref{KillingSpEq}.

We start by introducing new ``orthotoric'' coordinates $(y,\eta,\xi,\Phi,\Psi)$, related to those of~\cite{Chong:2005hr} as:
\begin{align}\label{CCLPtoOrtho1}
t & =  y,\nn\\[1mm]
\theta &= \frac 12\arccos\eta, \nn \\[1mm]
r^2 &= \frac{1}{2}(a^2-b^2)\widetilde m\, \xi + \frac{1}{g}\left[(a+b)\widetilde m + a+b+abg \right] + \frac 12 (a+b)^2\widetilde m \ ,\quad \nn \\[1mm]
\phi &= g\,y -\frac{4(1-a^2g^2)}{(a^2-b^2)g^2\widetilde m}\,(\Phi-\Psi) \ ,\nn\\[1mm] 
\psi &= g\,y -\frac{4(1-b^2g^2)}{(a^2-b^2)g^2\widetilde m}\,(\Phi+\Psi)\ ,
\end{align}
where for convenience we traded the parameter $m$ for
\be\label{defmtilde}
\widetilde m \ =\ \frac{m \,g}{(a+b)(1+ag)(1+bg)(1+ag+bg)} -1\ .
\ee
In these coordinates, one can choose the relatively simple frame:
\begin{align}\label{ortho_frame}
E^0 & = f\left(\diff y + \omega \right) \ ,\nonumber\\[1mm]
E^1 &= \frac{1}{g f^{1/2}}\sqrt{\frac{\eta-\xi}{\mathcal{F}(\xi)}}\,\diff \xi ~,  \, \qquad \  E^2 =  \frac{1}{g f^{1/2}}\sqrt{\frac{\mathcal{F}(\xi)}{\eta-\xi}}\,(\diff \Phi + \eta\, \diff \Psi) ~, \nonumber\\[1mm]
E^3 &= -\frac{1}{g f^{1/2}}\sqrt{\frac{\eta-\xi}{\mathcal{G}(\eta)}}\, \diff \eta~, \ \ \quad E^4 =  \frac{1}{g f^{1/2}}\sqrt{\frac{\mathcal{G}(\eta)}{\eta-\xi}}\,(\diff \Phi + \xi\, \diff \Psi)  ~,
\end{align}
where $\mathcal{F}(\xi)$, $\mathcal{G}(\eta)$ are the cubic polynomials:
\begin{align}\label{CCLPtoOrtho2}
\mathcal{G}(\eta)&=-\frac{4}{(a^2-b^2)g^2\widetilde m}(1-\eta^2) \left[ (1-a^2g^2)(1+\eta) + (1-b^2g^2)(1-\eta) \right] \ ,\nn \\[1mm]
\mathcal{F}(\xi) &= - \mathcal{G}(\xi) - 4\,  \frac{1+\widetilde m}{\widetilde m}\left( \frac{2+ag+bg}{(a-b)g} + \xi \right)^3\ ,
\end{align}
while
\begin{align}
f & =  \frac{24(\eta-\xi)}{\mathcal{F}''(\xi) + \mathcal{G}''(\eta)}\ , \nn\\[2mm]
\omega &= \frac{\mathcal{F}'''+\mathcal{G}'''}{48g(\eta-\xi)^2}\bigg\{\left[ \mathcal{F}(\xi) + (\eta-\xi) \left(\tfrac 12 \mathcal{F}'(\xi) -\tfrac 14 \big(\tfrac{2+ag+bg}{(a-b)g}+\xi\big)^2\mathcal{F}'''\right) \right](\diff \Phi + \eta\, \diff \Psi) \nn \\[1mm] 
&+ \mathcal{G}(\eta) (\diff \Phi + \xi\, \diff \Psi)  \bigg\} 
\,-\, \frac{\mathcal{F}'''\mathcal{G}'''}{288g} \left[ (\eta+\xi)\diff \Phi + \eta\xi\, \diff \Psi \right]  -\frac{2}{g\widetilde m}\diff\Psi \ .
\end{align}

Then the five-dimensional metric \eqref{5met} --- with $q$ fixed as in \eqref{susyCCLP} --- is just
\be
\diff s^2 = - (E^0)^2 + (E^1)^2+ (E^2)^2+(E^3)^2+(E^4)^2 \ ,
\ee
while the graviphoton field \eqref{gaugepot} reads:
\be
A= -\frac{3}{2} \,\Big(gE^0  + \frac{1}{3} P   \Big) + \Big(\frac{3}{2}g+\alpha\Big)\diff y -2\,\frac{(1-a^2g^2)(\diff\Phi-\diff\Psi)+ (1-b^2g^2)(\diff\Phi+\diff\Psi)}{(a^2-b^2)g^2\widetilde m}\ ,
\ee
with
\be
P = - \frac{\mathcal{F}'(\xi)(\diff\Phi+\eta\,\diff \Psi) + \mathcal{G}'(\eta)(\diff\Phi + \xi\,\diff\Psi)}{2(\eta-\xi)}\ .
\ee

Within this framework it is not hard to find an explicit solution to the Killing spinor equation. Imposing the projection
\be\label{projection_spinor}
\frac{i}{2}\left(\Gamma^{12} - \Gamma^{34}\right) \epsilon \,=\, \epsilon\ ,
\ee
which also implies $i\Gamma^0 \epsilon = \epsilon$, the four complex degrees of freedom of the Dirac spinor reduce to a single complex function to be solved for (corresponding to the two supercharges preserved by the solution). 
We find the explicit solution:
\be
\epsilon = \rme^{\frac{i}{2}\left((3g+ 2 \alpha) y -\frac{4(1-a^2g^2)}{(a^2-b^2)g^2\widetilde m}(\Phi-\Psi) -\frac{4(1-b^2g^2)}{(a^2-b^2)g^2\widetilde m}(\Phi+\Psi)\right)}\!\! \left[\frac{\widetilde m(a-b)(\eta-\xi)}{\big(\tfrac{2}{g}+a+b\big)(1+\widetilde m)+ (a-b)(\eta + \widetilde m \xi)}  \right]^{1/2}\!\!\!\!\!\epsilon_0 \,,
\ee
where $\epsilon_0$ is a constant spinor satisfying the projection \eqref{projection_spinor}. 

We can also express the Killing spinor in the original coordinates $(t,\theta,\phi,\psi,r)$. We obtain:
\be\label{solKillSp}
\epsilon \, =\, \rme^{\frac{i}{2}\left((g+2 \alpha) t + \phi + \psi\right)} \left[\frac{g^{-1}(a+b+abg)(1+\widetilde m)+\widetilde m(a^2\cos^2\theta+b^2\sin^2\theta) - r^2}{a^2\cos^2\theta+b^2\sin^2\theta+r^2} \right]^{1/2}\epsilon_0\ .
\ee

One can check that choosing $\epsilon_0$ so that it has unit norm, the spinor bilinear
\be
K = \bar\epsilon\, \Gamma^\mu \epsilon\, \partial_\mu
\ee
is precisely the Killing vector given in \eqref{susy_vector}.

In the main text we will need the spinorial Lie derivative along the Killing vectors $\frac{\partial}{\partial t}, \frac{\partial}{\partial \phi}$ and $\frac{\partial}{\partial \psi}$.
The spinorial Lie derivative along a Killing vector $X$ is defined as
\be
\mathcal{L}_X \epsilon \, = \, X^\mu \nabla_\mu \epsilon - \tfrac{1}{4} \nabla_{\mu}X_{\nu} \Gamma^{\mu\nu} \epsilon\ . 
\ee
This has the nice property of being covariant under local Lorentz transformations (see e.g. \cite{Ortin:2002qb}).
An explicit computation shows that the spinorial Lie derivatives along the directions $\frac{\partial}{\partial t},\frac{\partial}{\partial \phi},\frac{\partial}{\partial \psi}$ simply reduce to partial derivatives, hence the Killing spinor \eqref{solKillSp} satisfies:
\be\label{Lie_der_spinor_app}
\mathcal{L}_{\frac{\partial}{\partial t}}\epsilon =  \tfrac{i}{2}(g+2\alpha) \,\epsilon \ ,\qquad
\mathcal{L}_{\frac{\partial}{\partial\phi}} \epsilon = \tfrac{i}{2} \, \epsilon \ ,\qquad
\mathcal{L}_{\frac{\partial}{\partial\psi}} \epsilon = \tfrac{i}{2} \, \epsilon \ .
\ee

We observe that even after coming back to the original coordinates $(t,\theta,\phi,\psi,r)$, the frame \eqref{ortho_frame} is not adapted to a Fefferman-Graham asymptotic expansion, where the metric takes the form \eqref{FGexpansion}. In order to achieve this, one should perform a frame rotation. In the rotated frame such that $E^5 = \frac{\diff r}{g r}$, the leading term of the Killing spinor is $\mathcal{O}(r^{1/2})$ and is annihilated by the projector $\frac{1}{2}(1-\Gamma^5)$. Since the spinorial Lie derivative is covariant under local Lorentz transformations, the properties \eqref{Lie_der_spinor_app} also hold in the new frame and are inherited by the boundary Killing spinor.

As a final step, the Killing spinor equation and its solution can be analytically continued by taking $t =-i\tau $ and $m  = -(1+ag+bg)(a\mp i r_+)(b\mp i r_+)(1\mp i g r_+)$. The resulting Killing spinor is the one associated with the complexified supersymmetric solution discussed in Section \ref{sec:BPSlimit}.
The Lorentzian charge conjugate spinor $\widetilde \epsilon$, which in the analytically continued solution should a priori be considered independent of $\epsilon$, satisfies an equation that is obtained from \eqref{KillingSpEq} by formally sending $\epsilon\to\widetilde\epsilon$ and $g \to -g$. It follows that the complexified solution admits one Killing spinor $\epsilon$ and  one Killing spinor $\widetilde\epsilon$.

\section{Revisiting the extremization principle}\label{sec:revisiting_extremization}

We revisit here the extremization principle proposed in \cite{Hosseini:2017mds}. We first review how it works mathematically in a slightly more general setup than the one discussed in the main text, then we  apply it to the black holes of \cite{Chong:2005hr}.

In general, a supersymmetric, asymptotically AdS$_5 \times S^5$ black hole solution to type IIB supergravity may carry two angular momenta $J_i$, $i=1,2$ (associated with the $U(1)^2\subset SO(4)$ symmetry of the asymptotic AdS$_5$) and three electric charges $Q_K$, $K=1,2,3$ (associated with the $U(1)^3\subset SO(6)$ symmetry of $S^5$).
We thus consider a set of chemical potentials $\omega_i$, conjugate to the two angular momenta, and $\Delta_K$, conjugate to the three electric charges.\footnote{In supergravity, the position of the index on the electric charge is usually taken as $Q_K$, thus the position of the index on the symplectically conjugate chemical potential should rather be $\Delta^K$ (same for the angular momenta). In this paper however we will not distinguish between upper and lower indices on the chemical potentials.} 
We assume that the chemical potentials satisfy the linear constraint
\be\label{constraint_general}
 \Delta_1 +\Delta_2 + \Delta_3 +\omega_1 + \omega_2  = 2\pi i n\ ,
\ee
where $n$ is a fixed number. 
Moreover we assume that the grand-canonical partition function ${Z}(\Delta,\omega)$ describing our system is given by:
\be\label{def_calE}
-\log{Z(\Delta,\omega)} \,\equiv\, {I}(\Delta,\omega) \,=\, \coeff\, \frac{\Delta_1\Delta_2\Delta_3}{\omega_1\omega_2}\ ,
\ee
where $\coeff$ is another fixed (real) number. We are interested in the Legendre transform
\be\label{def_Legendre_transf}
 {S}(Q,J) =  {\rm ext}_{\{\Delta,\omega,\Lambda\}}\big[- {I} - \Delta_K Q_K - \omega_i J_i - \Lambda(\Delta_1 +\Delta_2 + \Delta_3 +\omega_1 + \omega_2 - 2\pi i n )\big]\ ,
\ee
where $\Lambda$ is a Lagrange multiplier implementing the constraint \eqref{constraint_general}. The function of the charges $ {S}(Q,J)$ can be seen as the logarithm of the microcanonical partition function, that is the entropy.
Since we must work over the complex numbers, we defined the Legendre trasform above as a (complex) extremum of the function on the right hand side of \eqref{def_Legendre_transf}.
There are six extremization equations: the five equations
\be\label{extremiz_eqs}
-\frac{\partial {I}}{\partial\Delta_K} = Q_K + \Lambda \ ,\qquad\qquad
-\frac{\partial {I}}{\partial\omega_i} = J_i + \Lambda \ ,
\ee
that follow from varying \eqref{def_Legendre_transf} with respect to the chemical potentials, plus the constraint \eqref{constraint_general} that is obtained by varying with respect to the Lagrange multiplier $\Lambda$. It is straightforward to see that the equations in \eqref{extremiz_eqs} imply the following cubic equation for $\Lambda$:
\begin{align}\label{eq_for_Lambda}
0 &\,=\, (Q_1+\Lambda)(Q_2+\Lambda)(Q_3+\Lambda) + \coeff (J_1+\Lambda)(J_2+\Lambda) \nn\\[1mm]
&\,=\, p_0 + p_1 \Lambda + p_2 \Lambda^2 + \Lambda^3\ ,
\end{align}
where we defined
\begin{align}
p_0 &= Q_1Q_2Q_3 +\coeff J_1J_2\ ,\nn\\[1mm]
p_1 &=  Q_1Q_2 + Q_2 Q_3 +Q_3Q_1 +\coeff( J_1 + J_2) \ ,\nn\\
p_2 &= Q_1+Q_2+Q_3+\coeff\ .
\end{align}
We denote by ``${\rm Roots}$'' the set of three solutions to \eqref{eq_for_Lambda}.
The saddle point values of $(\Delta_K,\omega_i)$ solving the rest of the equations are:
\begin{align}\label{saddles_chemical}
&\Delta_1  =   \Xi\,\tilde Q_2 \tilde Q_3 \tilde J_1 \tilde J_2 \ , \qquad
 \Delta_2  = \Xi\,\tilde Q_1 \tilde Q_3 \tilde J_1 \tilde J_2\ , \qquad
 \Delta_3  = \Xi\,\tilde Q_1 \tilde Q_2 \tilde J_1 \tilde J_2\ , \nn\\[1mm]
& \omega_1  = - \Xi\,\tilde Q_1\tilde Q_2 \tilde Q_3 \tilde J_2 \ , \qquad
  \omega_2 = - \Xi\,\tilde Q_1\tilde Q_2 \tilde Q_3 \tilde J_1  \ ,
\end{align}
where we introduced
$\tilde Q_K = Q_K +\Lambda$, $\tilde J_i = J_i +\Lambda$, as well as
\be
\Xi = \frac{2\pi i n}{\tilde J_1 \tilde J_2  (\tilde Q_1 \tilde Q_2+ \tilde Q_2 \tilde Q_3 + \tilde Q_1 \tilde Q_3)- (\tilde J_1 + \tilde J_2) \tilde Q_1 \tilde Q_2 \tilde Q_3}\ ,
\ee
and it is understood that $\Lambda \in {\rm Roots}$.

Then without further work we can write down the Legendre transform. Indeed noting that the function $ {I}$ is homogeneous of degree one, by Euler's theorem we have that
\be
 \Delta_K\, \frac{\partial {I}}{\partial\Delta_K} +\omega_i \,\frac{\partial {I}}{\partial\omega_i} \,=\,  {I}\ ,
\ee
which implies that \eqref{def_Legendre_transf} reduces to:
\be\label{S_supLambda}
 {S} = {\rm ext}_{\Lambda\,\in\,{\rm Roots}} \,(2\pi in\,\Lambda)\ .
\ee
Note that if the constraint \eqref{constraint_general} is satisfied with $n=0$, then $ {S}$ necessarily vanishes.

In general the roots of the cubic equation \eqref{eq_for_Lambda} for $\Lambda$ may be real or complex. If we require reality of the entropy, as it should be in a physical black hole solution, then from \eqref{S_supLambda} we see that we need 
to pick a purely imaginary root. 
Since complex roots come in complex conjugate pairs, we must be in the situation that there are two complex conjugate purely imaginary roots.  
It is straightforward to see that, assuming the charges are all real, this is equivalent to the condition
\be\label{condition_p0p1p2}
p_0 = p_1 p_2 \ .
\ee
In this case the cubic equation \eqref{eq_for_Lambda} for $\Lambda$ factorizes as
\be
( \Lambda^2 + p_1 )( \Lambda + p_2 ) = 0\ ,
\ee
and the three roots are
\be
{\rm Roots}\,=\,\{ \, -p_2 \, ,  \,  \pm\, i\sqrt{p_1}\,\}\ .
\ee
Depending on the sign of $n$, we choose either $\Lambda = i\sqrt{p_1}$, or $\Lambda =-i\sqrt{p_1}$, so that 
we get a positive entropy. We conclude that, assuming both the entropy and the charges are real, the result of our extremization problem is
\be\label{entropy_p1}
 {S} = 2\pi| n |\,\sqrt{p_1}\ .
\ee
Plugging the chosen imaginary root $\Lambda$ in the saddle point solution \eqref{saddles_chemical} for the chemical potentials, one finds that these are complex functions of the charges.

The assumptions \eqref{constraint_general}, \eqref{def_calE} which were put forward in~\cite{Hosseini:2017mds},  
follow naturally from our analysis in the main text. 
Indeed the general discussion above applies to the black hole of \cite{Chong:2005hr} and the BPS limit defined in Section \ref{sec:BPSlimit}. We identify:
\be
\Delta_1=\Delta_2=\Delta_3 = -\frac{2}{3}\,\elpot \ ,\qquad \coeff = -\frac{\pi}{4}\ .
\ee
Then the linear constraint \eqref{constraint_general} between the chemical potentials matches the one obtained in the main text  by choosing $n =  1$ 
and the function \eqref{def_calE} coincides with the supersymmetric on-shell action \eqref{IBPS}. 
Upon identifying further
\be
Q^{\rm here}_1=Q^{\rm here}_2=Q^{\rm here}_3=-\frac{1}{2}\QBPS\ ,\quad J^{\rm here}_1 = \JBPS_1\ , \quad J^{\rm here}_2 = \JBPS_2\ ,
\ee 
 the BPS quantum statistical relation \eqref{BPS_QSR} agrees with \eqref{def_Legendre_transf}.
Moreover, one can check that the factorization condition \eqref{condition_p0p1p2} is  satisfied by the BPS charges \eqref{BPScharges}, as it had to be since both the charges and the entropy are real.
With these identifications, the result \eqref{entropy_p1} precisely reproduces the BPS black hole entropy \eqref{S_BPS}. Moreover the saddles \eqref{saddles_chemical}  agree with the BPS chemical potentials \eqref{BPS_chem_pots}.

We note that the analysis  presented here is valid 
even in the more general setting of supersymmetric, 
but non-extremal solution, where the charges and the entropy are \emph{complex}.
In particular,  we note that in the solutions discussed in the main text the factorization condition~\eqref{condition_p0p1p2} is not satisfied if one just uses the supersymmetry condition \eqref{susy_with_r+} in the expressions \eqref{CCLPcharges} for the charges, without imposing the extremality condition $r_+=\rBPS$. In fact, one can see that if the black hole parameters $a,b,r_+$ are assumed real and $a,b$ are independent, then \eqref{condition_p0p1p2} is satisfied if and only if $r_+=\rBPS$, that is  at extremality. Therefore the simple form \eqref{entropy_p1} of the entropy  will in general not hold if one only imposes \eqref{susy_with_r+}. 
Nevertheless, remarkably, it remains true that, for appropriate
choices of roots of  \eqref{eq_for_Lambda},  the expression \eqref{S_supLambda} reproduces the complex entropy  \eqref{complex_entropy}.

\section{Field theory conventions \label{GammaMat}}

We work with Euclidean signature~$(+,+,+,+)$. Our choice of representation for gamma matrices in Section \ref{sec:PathIntegral} is:
\bea
\label{gammamatrices}
\gamma_1&\=&+\sigma_2\otimes 1_{2\times 2} \,,\nn \\
\gamma_{2}&\=&-\sigma _1\otimes \sigma _{1} \, , \nn\\
\gamma_3&\=&+\sigma _1\otimes \sigma _{2} \,,\\
\gamma_4&\=&+\sigma _1\otimes \sigma _{3} \,. \nn
\eea
The charge conjugation matrix~$C =\gamma_2 \gamma_4$ obeys 
\be
C \=-C^T=-C^{-1}\= 1_{2\times 2}\otimes i \sigma_2  \,.
\ee

The Killing spinor denoted by $\ve$ in \eqref{KSSol2} obeys the following reality condition
\bea \label{spinorcondition}
i\, \ve^* \gamma_1\, \= \, \ve\, C^{-1},
\eea
provided the constraint \eqref{period1} is imposed (this implies that the two non vanishing components of $\ve$ are pure phases, complex conjugate to each other).

The frame used in Section \ref{sec:PathIntegral} to find angular independent form of Killing spinors is
\be\label{SpinFrame}
\begin{split}
e^1 \,&\= \,\diff \tau \,, \\
e^2\, &\= \, \text{d$\theta $} \cos (\phi_2 +\phi_1 ) \ + \, \frac{1}{2} \sin (2 \theta ) \sin (\phi_2 +\phi_1 )
   \left(i \text{d}\tau \left(\Omega _1-\Omega _2\right)+\text{d$\phi_2 $}-\text{d$\phi_1$}\right) \,,\\
   e^3\, &\=  -\text{d$\theta $} \sin (\phi_2 +\phi_1 )+\frac{1}{2} \sin (2 \theta ) \cos (\phi_2 +\phi_1 )
   \left(i \text{d}\tau \left(\Omega _1-\Omega _2\right)+\text{d$\phi_2 $}-\text{d$\phi_1$}\right) \,, \\
   e^4\,&\= \, \sin ^2\theta(\diff \phi_1- i \Omega _1  \diff\tau) +  \cos^2\theta(\diff\phi_2 - i  \Omega _2\diff\tau) \,.
\end{split}
\ee


\section{Regularization of 1-loop determinants \label{RegulatorEll}}

In this appendix we regularize the infinite product in the RHS of \eqref{chiral1loop}, i.e.,
\be \label{chiral1loopapp}
Z^{\text{chiral},\rho}_\text{1-loop}(u) \= \prod_{n_0\in \mathbb{Z}}\,\prod_{n_1, n_2 \geq 0} \frac{2\pi n_0 
+ \rho \cdot u - i \left(r-2\right)    \elpot + i \, n_1\,  \omega_1  
+ i\,  n_2 \,  \omega_2  }{2\pi n_0 + \rho \cdot u 
- i r  \elpot  - i\, n_1 \,  \omega_1  -  i \,  n_2\, \omega_2 } \ . 
\ee
We use the ``one-step" regularization  
as in \cite{Closset:2013sxa,Assel:2014paa} which uses the properties of triple gamma functions. 
The mathematical analysis of~\cite{Spiridonov:2010em, Felder} may be useful in performing 
the two-step regularization, as in~\cite{Assel:2015nca}, with the~$n\neq 0$ constraint, but at present it is not clear 
to us at a technical level how exactly this would work out. 

We follow the conventions and definitions of multiple gamma and zeta functions given in \cite{Friedman_2004}. 
The RHS of \eqref{chiral1loopapp} can be written as 
\bea \label{Shintp}
Z^+&=&\pr{0}{\infty} \frac{\frac{\rho \cdot u -i r \varphi + 2 i \elpot}{2 \pi} +n_0+\frac{i \omega_1 n_1+i \omega_2 n_2}{2 \pi}}{\frac{-\rho \cdot u +i r \varphi }{2 \pi} +n_0+\frac{i \omega_1 n_1+i \omega_2 n_2}{2 \pi}} \times \frac{\frac{\rho \cdot u -i r \varphi + 2 i \elpot}{2 \pi} -(1+n_0)+\frac{i \omega_1 n_1+i \omega_2 n_2}{2 \pi}}{\frac{-\rho \cdot u +i r \varphi }{2 \pi} -(1+n_0)+\frac{i \omega_1 n_1+i \omega_2 n_2}{2 \pi}} \nn\\
&=&\pr{0}{\infty} \frac{\frac{\rho \cdot u -i r \varphi + 2 i \elpot}{2 \pi} +n_0+\frac{i \omega_1 n_1+i \omega_2 n_2}{2 \pi}}{\frac{-\rho \cdot u +i r \varphi }{2 \pi} +n_0+\frac{i \omega_1 n_1+i \omega_2 n_2}{2 \pi}} \times \frac{1-\frac{\rho \cdot u -i r \varphi + 2 i \elpot}{2 \pi} +n_0-\frac{i \omega_1 n_1+i \omega_2 n_2}{2 \pi}}{1-\frac{-\rho \cdot u +i r \varphi }{2 \pi} +n_0-\frac{i \omega_1 n_1+i \omega_2 n_2}{2 \pi}} \,. \nn \\
\eea
Note that here we have made an implicit choice to place the term~$n_0=0$ in the first ratio of~$\G_3$ functions, but 
we could as well have placed it in the second ratio, in which case we would get:
\bea \label{Shintm}
Z^-&=&\pr{0}{\infty} \frac{\frac{\rho \cdot u -i r \varphi + 2 i \elpot}{2 \pi} +1+n_0+\frac{i \omega_1 n_1+i \omega_2 n_2}{2 \pi}}{\frac{-\rho \cdot u +i r \varphi }{2 \pi} +1+n_0+\frac{i \omega_1 n_1+i \omega_2 n_2}{2 \pi}} \times \frac{\frac{\rho \cdot u -i r \varphi + 2 i \elpot}{2 \pi} -n_0+\frac{i \omega_1 n_1+i \omega_2 n_2}{2 \pi}}{\frac{-\rho \cdot u +i r \varphi }{2 \pi} -n_0+\frac{i \omega_1 n_1+i \omega_2 n_2}{2 \pi}} \nn\\
&=&\pr{0}{\infty} \frac{\frac{\rho \cdot u -i r \varphi + 2 i \elpot}{2 \pi} +1+n_0+\frac{i \omega_1 n_1+i \omega_2 n_2}{2 \pi}}{\frac{-\rho \cdot u +i r \varphi }{2 \pi} +1+n_0+\frac{i \omega_1 n_1+i \omega_2 n_2}{2 \pi}} \times \frac{-\frac{\rho \cdot u -i r \varphi + 2 i \elpot}{2 \pi} +n_0-\frac{i \omega_1 n_1+i \omega_2 n_2}{2 \pi}}{-\frac{-\rho \cdot u +i r \varphi }{2 \pi} +n_0-\frac{i \omega_1 n_1+i \omega_2 n_2}{2 \pi}} \,. \nn \\
\eea
The above two expressions~(\ref{Shintp}, \ref{Shintm}) are formally the same infinite products, we have denoted them by~$Z^\pm$
in anticipation of the fact that there are two natural regulators which lead to two different expressions.

These infinite products can be summarized in terms of $\Gamma_3$ functions as  
\be \label{Shint}
Z^s \=\frac{\Gth{w+\gamma}{\a_1}{\a_2}}{\Gth{-w+\gamma}{\a_1}{\a_2}}  \; 
\frac{\Gth{1-w-\gamma}{-\a_1}{-\a_2}}{\Gth{1+w-\gamma}{-\a_1}{-\a_2} } \,, \qquad s \= \pm 1 \,,
\ee
where the parameters in the arguments are given by the following expressions:
\bea \label{awparameters}
&& \a_1 \=  s \frac{i {\omega}_1}{2\pi}, \qquad 
\a_2 \= s \frac{i {\omega}_2}{2\pi}, \qquad \gamma\= s \frac{i {\elpot} }{ 2\pi } \,, \nn \\ 
&& w(\rho,r) \= {-} s \frac{1}{2\pi} \left(\rho \cdot u-(r-1)i {\elpot}\right) \label{D8} \,.
\eea
In terms of these variables, our constraint~$-2 {\elpot}  + \omega_1+\omega_2 = 2 \pi i n$ translates to 
\be \label{gammaa12rel}
\g \= \frac{\a_1+\a_2+s\,n}{2} \,.
\ee

At this point we use the following identity~\cite{Friedman_2004} in both the numerator and the denominator of \eqref{Shint} which 
holds if Im($\a_{1,2}$)$>0$,
\be \label{IdentityGG}
{\Gththr{z}{1}{\a_1}{\a_2} \, \Gththr{1-z}{1}{-\a_1}{-\a_2}} \=
\rme^{- \pi i\, \zeta_3(0,z;1,\a_1,\a_2)} \prod^{(\infty,\infty)}_{\overset{\rightarrow}{m}=
(0,0)}\frac{1}{(1-\rme^{2 \pi i (z+\oarrow{m}\cdot \oarrow{\a})})}\,,
\ee
Here the zeta function of third order $\zeta_3$ is defined as (see \cite{Friedman_2004}): 
\bea
 \zetathr{z}{\a_1}{\a_2} \, &:=&\, -\frac{z^3}{6 \a_1 \a_2}+\frac{(\a_1+\a_2+1) z^2}{4 \a_1 \a_2}\nn\\&-& \frac{\left(\a_1^2 + \a_2^2+1 +  3\left(\a_1+\a_2+\a_1\a_2\right)\right) z}{12 \a_1 \a_2} 
 \\&+&\frac{(1+\a_1+\a_2)(\a_1+\a_2+\a_1 \a_2)}{24 \a_1 \a_2}\ . \nn
\eea

We can now transform the product of triple gamma functions as follows,
\be \begin{split}\label{RegFinal}
& \frac{\Gth{w+\gamma}{\a_1}{\a_2}}{\Gth{-w+\gamma}{\a_1}{\a_2}}  \; 
\frac{\Gth{1-w-\gamma}{-\a_1}{-\a_2}}{\Gth{1+w-\gamma}{-\a_1}{-\a_2} }\\
& \qquad \qquad \= \rme^{-i \pi \Psi(w,\a_1,\a_2)} \prID{-w+\gamma}{w+\gamma} \\
& \qquad \qquad \=  \rme^{-i \pi \Psi(w,\a_1,\a_2)} \, \Gamma_e(w+\gamma;\a_1,\a_2) \,.
\end{split} \ee
Here the elliptic Gamma function is defined as (for~$z \in \mathbb{C}$, $\text{Im}(\a_{1,2})  >0$)
\bea\label{GammaeDef}
\Gamma_e(z;\a_1,\a_2)\, := \, \Gell{z}{\a_1}{\a_2} \,.
 \eea
In reaching the second equality in \eqref{RegFinal} we have used the second equality in the last line of equation \eqref{D8}, together with the fact that $ n  \in \mathbb{Z}$.
 The prefactor in the exponential, that is denoted as $\Psi$, is defined as follows 
\bea
\Psi(w;\a_1,\a_2) &\; := \;&\zetathr{w+\gamma}{\a_1}{\a_2}-\zetathr{-w+\gamma}{\a_1}{\a_2}
\\
&\=&-\frac{w^3}{3 \a_1 \a_2} +A(\gamma;\a_1,\a_2) w,
\eea
where
\be
A(\gamma;\a_1,\a_2)\; := \;
-\frac{3 \a_1 \left(\a_2-2 \gamma+1\right)+\a_2 \left(3-6 \gamma \right)+\a_1^2+\a_2^2+6 \gamma^2-6 \gamma +1}{6 \a_1 \a_2} \,.
\ee
Upon substituting the relation~\eqref{gammaa12rel} we obtain
\be
A  \=\frac{\a_1^2+\a_2^2-2+6 s n-3 n^2}{12 \a_1 \a_2} \,.
\ee
The expression $A$, after summing over all the multiplets, gives rise to the term in the prefactor that is proportional to~$\Psi^{(0)}_2$,
which, as discussed in the main text, does not contribute to any theory with ${\bf a}= {\bf c}$, and therefore any holographic $\mathcal{N}=1$ SCFT in the large $N$ limit.

Putting everything together, the one loop determinant $Z$ takes the form
\be \label{ZchiralP}
Z^s \= \exp \Bigl( -i \pi \Psi \bigl(w, \a_1, \a_2 \bigr) \Bigr) \, 
\Gamma_e \left(w+\gamma;\a_1,\a_2 \right) \,,
\ee
where the arguments are given in~\eqref{awparameters}. 
The BPS black hole solutions discussed in Section~\ref{sec:BPSlimit} have~$\text{Re}(\omega_{1,2}) < 0$. 
In order to compare 
the field theory result to these black holes, it is natural to choose the regularization corresponding to~$s=-1$ in 
\eqref{Shint}, \eqref{awparameters}.\footnote{If we take~$s=1$ then the infinite product representation of the elliptic 
gamma function~\eqref{GammaeDef} would be divergent in the region~$\text{Re}(\omega_{1,2}) <0$, which is the 
region chosen by the BPS black hole chemical potentials. The elliptic gamma funtion could also be defined in this 
region using an analytic continuation from the region of convergence as described in~\cite{Felder}. In this case 
the result for the prefactor will have the opposite sign as for the~$s=-1$ as mentioned in the main text.}
This is indeed the expression quoted in~\eqref{Zchiral} with 
\bea \label{ZchiralPA}
w+\gamma\= \frac{1}{2\pi} \left(\rho \cdot u-r i  {\elpot}\right) \equiv v(\rho,r)  \,, 
\eea
and
\bea
\Psi^{(0)}&\; := \;&\Psi \Big|_{u\rightarrow0}\, , \label{psi0d}\\
\Psi^{(1)}&\; := \;&\Psi-\Psi^{(0)} \, . \label{psi1}
\eea
The contribution of the quantity $\Psi^{(1)}$ to the final partition function of a generic $\mathcal{N}=1$ SQFT theory 
can be easily checked to vanish if the corresponding matter content satisfies the anomaly vanishing conditions 
quoted in Section 5.1  of~\cite{Assel:2014paa}.

\subsubsection* {Turning on flavor fugacities and the case of~$\CN=4$ SYM}

The calculation with flavor chemical potentials goes through as above. We begin with the situation when we have a collection of~$U(1)$ 
flavor symmetries~$F^p$, $p=1,\cdots, M$ with corresponding potentials~$\v_p$. 
In the calculation of the determinant, the only change is that the covariant derivative now has factors proportional to~$\v_p$. 
As the result, choosing $s=-1$ we obtain Equation~\eqref{Shint} with
\be
w_I = w_I(\rho,\rho_I,r) \= \frac{1}{2\pi} \Bigl(\rho \cdot u-(r-1)i {\elpot} - 
i\, \sum_{p=1}^M F^p_I \,{\elpot_p} \Bigr) \,,
\ee 
where~$F_I^p$ is the flavor charge of the~$I^\text{th}$ multiplet in the theory.

The case of~$\CN=4$ SYM can be treated in this~$\CN=1$ language by 
embedding the~$U(1)$ $R$-symmetry into the~$SU(4)$ $R$-symmetry, and by treating the 
other two Cartan generators of the~$SU(4)$ as flavor rotations, as in the main text.
Alternatively, we can directly treat the  full non-abelian $R$-symmetry as follows.

We denote by $\varphi_p$, $p=1,2,3$, the chemical potentials for the three Cartan generators of the $SU(4)$. 
The three fermions of the chiral multiplets together with the gaugino fall into the fundamental of the~$SU(4)_R$. We label
these four fermions by~$I=1,\cdots, 4$, and denote their weights by~$\rho_I$, which in our conventions are
\bea
\rho_1&\=&(+1,0,0), \\
\rho_2&\=&(-1,+1,0), \\
\rho_3&\=&(0,-1,1), \\
\rho_4&\=&(0,0,-1).
\eea
For each fermion degree of freedom the quantity $w$ in Equation~\eqref{Shint}  is now refined to 
\be
w_I = w_I(\rho,\rho_I) \= \frac{1}{2\pi} \Bigl(\rho \cdot u  - 
i\, \sum_{p=1}^3 \rho^p_I \,{\wh \elpot_p} \Bigr) \,,
\ee 

The prefactor $\mathcal{F}$ is now
 \bea
- \frac{1}{6 \, \omega_1 \omega_2} \sum_{I=1}^4 
 \text{dim(R$_I$)}\sum_{p_1,\, p_2,\, p_3=1}^{3}\rho_I^{p_1}\, \rho_I^{p_2}  \, \rho_I^{p_3} \, \wh \elpot_{p_1}\, \wh \elpot_{p_2} \, \wh \elpot_{p_3} \,
 \eea
where $\text{R}_I=\text{Adj}$.
The weights of the fundamental obey the following identities, 
\bea
\sum_I   \rho^{p_1}_I \, \rho^{p_2}_I \, \rho^{p_3}_I & \= & C^{p_1\, p_2\, p_3} \,, \\
\sum_I   \rho^{p_1}_I \, \rho^{p_2}_I & \= & C^{p_1\, p_2} \,, \\
\sum_I   \rho^{p_1}_I & \= & 0 \,, \label{PropWeights}
\eea
where $C^{p_1\, p_2}$ is the Cartan matrix of $SU(4)$, and $C^{p_1\, p_2\, p_3}$  is such that  
\be
\sum_{p_1,\, p_2,\, p_3=1}^3 
C^{p_1\, p_2\, p_3} \, \wh \elpot_{p_1} \wh \elpot_{p_2}  \wh \elpot_{p_3} \=  3\, \wh \elpot_2  \, (\wh \elpot_1- \wh \elpot_3 ) \, (\wh \elpot_1- \wh \elpot_2+ \wh\elpot_3 ) \,.
\ee 
After the redefinition 
\be
 \v_1 \=  2 (\wh \elpot_1- \wh \elpot_3 ) \,, \qquad 
 \v_2 \= \frac23 \, \wh \v_2 \,, \qquad 
 \v_3 \= \frac23\, (\wh \elpot_1- \wh \elpot_2+ \wh \elpot_3 ) \,,  
\ee
one obtains, for the gauge group~$G$, 
\be
-\CF_{\CN=4} \=  \frac{4 \, \text{dim}(G)}{27} \, \frac{{\elpot}_1 {\elpot}_2 {\elpot}_3}{\omega_1 \omega_2} \,.
\ee
The contributions proportional to ${\bf c}-{\bf a}$ vanish due to property \eqref{PropWeights}.


\begin{thebibliography}{10}

\bibitem{Sen:1995in}
A.~Sen, {\it {Extremal black holes and elementary string states}},  {\em Mod.
  Phys. Lett.} {\bf A10} (1995) 2081--2094,
  [\href{http://arxiv.org/abs/hep-th/9504147}{{\tt hep-th/9504147}}].

\bibitem{Strominger:1996sh}
A.~Strominger and C.~Vafa, {\it {Microscopic origin of the Bekenstein-Hawking
  entropy}},  {\em Phys. Lett.} {\bf B379} (1996) 99--104,
  [\href{http://arxiv.org/abs/hep-th/9601029}{{\tt hep-th/9601029}}].

\bibitem{Benini:2015eyy}
F.~Benini, K.~Hristov, and A.~Zaffaroni, {\it {Black hole microstates in
  AdS$_{4}$ from supersymmetric localization}},  {\em JHEP} {\bf 05} (2016)
  054, [\href{http://arxiv.org/abs/1511.04085}{{\tt arXiv:1511.04085}}].

\bibitem{Benini:2015noa}
F.~Benini and A.~Zaffaroni, {\it {A topologically twisted index for
  three-dimensional supersymmetric theories}},  {\em JHEP} {\bf 07} (2015) 127,
  [\href{http://arxiv.org/abs/1504.03698}{{\tt arXiv:1504.03698}}].

\bibitem{Gutowski:2004ez}
J.~B. Gutowski and H.~S. Reall, {\it {Supersymmetric AdS$_5$ black holes}},
  {\em JHEP} {\bf 02} (2004) 006,
  [\href{http://arxiv.org/abs/hep-th/0401042}{{\tt hep-th/0401042}}].

\bibitem{Gutowski:2004yv}
J.~B. Gutowski and H.~S. Reall, {\it {General supersymmetric AdS$_5$ black
  holes}},  {\em JHEP} {\bf 04} (2004) 048,
  [\href{http://arxiv.org/abs/hep-th/0401129}{{\tt hep-th/0401129}}].

\bibitem{Chong:2005hr}
Z.~W. Chong, M.~Cvetic, H.~Lu, and C.~N. Pope, {\it {General non-extremal
  rotating black holes in minimal five-dimensional gauged supergravity}},  {\em
  Phys. Rev. Lett.} {\bf 95} (2005) 161301,
  [\href{http://arxiv.org/abs/hep-th/0506029}{{\tt hep-th/0506029}}].

\bibitem{Chong:2005da}
Z.~W. Chong, M.~Cvetic, H.~Lu, and C.~N. Pope, {\it {Five-dimensional gauged
  supergravity black holes with independent rotation parameters}},  {\em Phys.
  Rev.} {\bf D72} (2005) 041901,
  [\href{http://arxiv.org/abs/hep-th/0505112}{{\tt hep-th/0505112}}].

\bibitem{Kunduri:2006ek}
H.~K. Kunduri, J.~Lucietti, and H.~S. Reall, {\it {Supersymmetric multi-charge
  AdS$_5$ black holes}},  {\em JHEP} {\bf 04} (2006) 036,
  [\href{http://arxiv.org/abs/hep-th/0601156}{{\tt hep-th/0601156}}].

\bibitem{Kinney:2005ej}
J.~Kinney, J.~M. Maldacena, S.~Minwalla, and S.~Raju, {\it {An Index for 4
  dimensional super conformal theories}},  {\em Commun. Math. Phys.} {\bf 275}
  (2007) 209--254, [\href{http://arxiv.org/abs/hep-th/0510251}{{\tt
  hep-th/0510251}}].

\bibitem{Romelsberger:2005eg}
C.~Romelsberger, {\it {Counting chiral primaries in N = 1, d=4 superconformal
  field theories}},  {\em Nucl. Phys.} {\bf B747} (2006) 329--353,
  [\href{http://arxiv.org/abs/hep-th/0510060}{{\tt hep-th/0510060}}].

\bibitem{Janik:2007pm}
R.~A. Janik and M.~Trzetrzelewski, {\it {Supergravitons from one loop
  perturbative N=4 SYM}},  {\em Phys. Rev.} {\bf D77} (2008) 085024,
  [\href{http://arxiv.org/abs/0712.2714}{{\tt arXiv:0712.2714}}].

\bibitem{Grant:2008sk}
L.~Grant, P.~A. Grassi, S.~Kim, and S.~Minwalla, {\it {Comments on 1/16 BPS
  Quantum States and Classical Configurations}},  {\em JHEP} {\bf 05} (2008)
  049, [\href{http://arxiv.org/abs/0803.4183}{{\tt arXiv:0803.4183}}].

\bibitem{Chang:2013fba}
C.-M. Chang and X.~Yin, {\it {1/16 BPS states in $\mathcal N=$ 4
  super-Yang-Mills theory}},  {\em Phys. Rev.} {\bf D88} (2013), no.~10 106005,
  [\href{http://arxiv.org/abs/1305.6314}{{\tt arXiv:1305.6314}}].

\bibitem{Witten:1988ze}
E.~Witten, {\it {Topological Quantum Field Theory}},  {\em Commun. Math. Phys.}
  {\bf 117} (1988) 353.

\bibitem{Nekrasov:2002qd}
N.~A. Nekrasov, {\it {Seiberg-Witten prepotential from instanton counting}},
  {\em Adv. Theor. Math. Phys.} {\bf 7} (2003), no.~5 831--864,
  [\href{http://arxiv.org/abs/hep-th/0206161}{{\tt hep-th/0206161}}].

\bibitem{Pestun:2007rz}
V.~Pestun, {\it {Localization of gauge theory on a four-sphere and
  supersymmetric Wilson loops}},  {\em Commun. Math. Phys.} {\bf 313} (2012)
  71--129, [\href{http://arxiv.org/abs/0712.2824}{{\tt arXiv:0712.2824}}].

\bibitem{Assel:2014paa}
B.~Assel, D.~Cassani, and D.~Martelli, {\it {Localization on Hopf surfaces}},
  {\em JHEP} {\bf 08} (2014) 123, [\href{http://arxiv.org/abs/1405.5144}{{\tt
  arXiv:1405.5144}}].

\bibitem{Assel:2015nca}
B.~Assel, D.~Cassani, L.~Di~Pietro, Z.~Komargodski, J.~Lorenzen, and
  D.~Martelli, {\it {The Casimir Energy in Curved Space and its Supersymmetric
  Counterpart}},  {\em JHEP} {\bf 07} (2015) 043,
  [\href{http://arxiv.org/abs/1503.05537}{{\tt arXiv:1503.05537}}].

\bibitem{Witten:1998qj}
E.~Witten, {\it {Anti-de Sitter space and holography}},  {\em Adv. Theor. Math.
  Phys.} {\bf 2} (1998) 253--291,
  [\href{http://arxiv.org/abs/hep-th/9802150}{{\tt hep-th/9802150}}].

\bibitem{Sen:2008yk}
A.~Sen, {\it {Entropy Function and AdS(2) / CFT(1) Correspondence}},  {\em
  JHEP} {\bf 11} (2008) 075, [\href{http://arxiv.org/abs/0805.0095}{{\tt
  arXiv:0805.0095}}].

\bibitem{Silva:2006xv}
P.~J. Silva, {\it {Thermodynamics at the BPS bound for Black Holes in AdS}},
  {\em JHEP} {\bf 10} (2006) 022,
  [\href{http://arxiv.org/abs/hep-th/0607056}{{\tt hep-th/0607056}}].

\bibitem{Hosseini:2017mds}
S.~M. Hosseini, K.~Hristov, and A.~Zaffaroni, {\it {An extremization principle
  for the entropy of rotating BPS black holes in AdS$_{5}$}},  {\em JHEP} {\bf
  07} (2017) 106, [\href{http://arxiv.org/abs/1705.05383}{{\tt
  arXiv:1705.05383}}].

\bibitem{Benini:2016rke}
F.~Benini, K.~Hristov, and A.~Zaffaroni, {\it {Exact microstate counting for
  dyonic black holes in AdS4}},  {\em Phys. Lett.} {\bf B771} (2017) 462--466,
  [\href{http://arxiv.org/abs/1608.07294}{{\tt arXiv:1608.07294}}].

\bibitem{Azzurli:2017kxo}
F.~Azzurli, N.~Bobev, P.~M. Crichigno, V.~S. Min, and A.~Zaffaroni, {\it {A
  universal counting of black hole microstates in AdS$_{4}$}},  {\em JHEP} {\bf
  02} (2018) 054, [\href{http://arxiv.org/abs/1707.04257}{{\tt
  arXiv:1707.04257}}].

\bibitem{Halmagyi:2017hmw}
N.~Halmagyi and S.~Lal, {\it {On the on-shell: the action of AdS$_{4}$ black
  holes}},  {\em JHEP} {\bf 03} (2018) 146,
  [\href{http://arxiv.org/abs/1710.09580}{{\tt arXiv:1710.09580}}].

\bibitem{Cabo-Bizet:2017xdr}
A.~Cabo-Bizet, U.~Kol, L.~A. Pando~Zayas, I.~Papadimitriou, and V.~Rathee, {\it
  {Entropy functional and the holographic attractor mechanism}},  {\em JHEP}
  {\bf 05} (2018) 155, [\href{http://arxiv.org/abs/1712.01849}{{\tt
  arXiv:1712.01849}}].

\bibitem{Choi:2018hmj}
S.~Choi, J.~Kim, S.~Kim, and J.~Nahmgoong, {\it {Large AdS black holes from
  QFT}},  \href{http://arxiv.org/abs/1810.12067}{{\tt arXiv:1810.12067}}.

\bibitem{Honda:2019cio}
M.~Honda, {\it {Quantum Black Hole Entropy from 4d Supersymmetric Cardy
  formula}},  \href{http://arxiv.org/abs/1901.08091}{{\tt arXiv:1901.08091}}.

\bibitem{ArabiArdehali:2019tdm}
A.~Arabi~Ardehali, {\it {Cardy-like asymptotics of the 4d $\mathcal{N}=4$ index
  and AdS$_5$ blackholes}},  \href{http://arxiv.org/abs/1902.06619}{{\tt
  arXiv:1902.06619}}.

\bibitem{Choi:2018vbz}
S.~Choi, J.~Kim, S.~Kim, and J.~Nahmgoong, {\it {Comments on deconfinement in
  AdS/CFT}},  \href{http://arxiv.org/abs/1811.08646}{{\tt arXiv:1811.08646}}.

\bibitem{Benini:2018ywd}
F.~Benini and P.~Milan, {\it {Black holes in 4d $\mathcal{N}=4$
  Super-Yang-Mills}},  \href{http://arxiv.org/abs/1812.09613}{{\tt
  arXiv:1812.09613}}.

\bibitem{Buchel:2006gb}
A.~Buchel and J.~T. Liu, {\it {Gauged supergravity from type IIB string theory
  on $Y^{p,q}$ manifolds}},  {\em Nucl. Phys.} {\bf B771} (2007) 93--112,
  [\href{http://arxiv.org/abs/hep-th/0608002}{{\tt hep-th/0608002}}].

\bibitem{Gauntlett:2007ma}
J.~P. Gauntlett and O.~Varela, {\it {Consistent Kaluza-Klein reductions for
  general supersymmetric AdS solutions}},  {\em Phys. Rev.} {\bf D76} (2007)
  126007, [\href{http://arxiv.org/abs/0707.2315}{{\tt arXiv:0707.2315}}].

\bibitem{Chen:2005zj}
W.~Chen, H.~Lu, and C.~N. Pope, {\it {Mass of rotating black holes in gauged
  supergravities}},  {\em Phys. Rev.} {\bf D73} (2006) 104036,
  [\href{http://arxiv.org/abs/hep-th/0510081}{{\tt hep-th/0510081}}].

\bibitem{Gibbons:1976ue}
G.~W. Gibbons and S.~W. Hawking, {\it {Action Integrals and Partition Functions
  in Quantum Gravity}},  {\em Phys. Rev.} {\bf D15} (1977) 2752--2756.

\bibitem{Kunduri:2005zg}
H.~K. Kunduri and J.~Lucietti, {\it {Notes on non-extremal, charged, rotating
  black holes in minimal D=5 gauged supergravity}},  {\em Nucl. Phys.} {\bf
  B724} (2005) 343--356, [\href{http://arxiv.org/abs/hep-th/0504158}{{\tt
  hep-th/0504158}}].

\bibitem{Papadimitriou:2005ii}
I.~Papadimitriou and K.~Skenderis, {\it {Thermodynamics of asymptotically
  locally AdS spacetimes}},  {\em JHEP} {\bf 08} (2005) 004,
  [\href{http://arxiv.org/abs/hep-th/0505190}{{\tt hep-th/0505190}}].

\bibitem{Choi:2008he}
J.~Choi, S.~Lee, and S.~Lee, {\it {Near Horizon Analysis of Extremal AdS(5)
  Black Holes}},  {\em JHEP} {\bf 05} (2008) 002,
  [\href{http://arxiv.org/abs/0802.3330}{{\tt arXiv:0802.3330}}].

\bibitem{Kim:2006he}
S.~Kim and K.-M. Lee, {\it {1/16-BPS Black Holes and Giant Gravitons in the
  AdS$_5 \times S^5$ Space}},  {\em JHEP} {\bf 12} (2006) 077,
  [\href{http://arxiv.org/abs/hep-th/0607085}{{\tt hep-th/0607085}}].

\bibitem{Cvetic:2004hs}
M.~Cvetic, H.~Lu, and C.~N. Pope, {\it {Charged Kerr-de Sitter black holes in
  five dimensions}},  {\em Phys. Lett.} {\bf B598} (2004) 273--278,
  [\href{http://arxiv.org/abs/hep-th/0406196}{{\tt hep-th/0406196}}].

\bibitem{Hawking:1998kw}
S.~W. Hawking, C.~J. Hunter, and M.~Taylor, {\it {Rotation and the AdS / CFT
  correspondence}},  {\em Phys. Rev.} {\bf D59} (1999) 064005,
  [\href{http://arxiv.org/abs/hep-th/9811056}{{\tt hep-th/9811056}}].

\bibitem{Festuccia:2011ws}
G.~Festuccia and N.~Seiberg, {\it {Rigid Supersymmetric Theories in Curved
  Superspace}},  {\em JHEP} {\bf 06} (2011) 114,
  [\href{http://arxiv.org/abs/1105.0689}{{\tt arXiv:1105.0689}}].

\bibitem{Klare:2012gn}
C.~Klare, A.~Tomasiello, and A.~Zaffaroni, {\it {Supersymmetry on Curved Spaces
  and Holography}},  {\em JHEP} {\bf 08} (2012) 061,
  [\href{http://arxiv.org/abs/1205.1062}{{\tt arXiv:1205.1062}}].

\bibitem{Cassani:2012ri}
D.~Cassani, C.~Klare, D.~Martelli, A.~Tomasiello, and A.~Zaffaroni, {\it
  {Supersymmetry in Lorentzian Curved Spaces and Holography}},  {\em Commun.
  Math. Phys.} {\bf 327} (2014) 577--602,
  [\href{http://arxiv.org/abs/1207.2181}{{\tt arXiv:1207.2181}}].

\bibitem{Sohnius:1981tp}
M.~F. Sohnius and P.~C. West, {\it {An Alternative Minimal Off-Shell Version of
  N=1 Supergravity}},  {\em Phys. Lett.} {\bf 105B} (1981) 353--357.

\bibitem{Sohnius:1982fw}
M.~Sohnius and P.~C. West, {\it {The Tensor Calculus and Matter Coupling of the
  Alternative Minimal Auxiliary Field Formulation of $N=1$ Supergravity}},
  {\em Nucl. Phys.} {\bf B198} (1982) 493--507.

\bibitem{Dumitrescu:2012ha}
T.~T. Dumitrescu, G.~Festuccia, and N.~Seiberg, {\it {Exploring Curved
  Superspace}},  {\em JHEP} {\bf 08} (2012) 141,
  [\href{http://arxiv.org/abs/1205.1115}{{\tt arXiv:1205.1115}}].

\bibitem{Papadimitriou:2017kzw}
I.~Papadimitriou, {\it {Supercurrent anomalies in 4d SCFTs}},  {\em JHEP} {\bf
  07} (2017) 038, [\href{http://arxiv.org/abs/1703.04299}{{\tt
  arXiv:1703.04299}}].

\bibitem{An:2017ihs}
O.~S. An, {\it {Anomaly-corrected supersymmetry algebra and supersymmetric
  holographic renormalization}},  {\em JHEP} {\bf 12} (2017) 107,
  [\href{http://arxiv.org/abs/1703.09607}{{\tt arXiv:1703.09607}}].

\bibitem{Genolini:2016sxe}
P.~Benetti~Genolini, D.~Cassani, D.~Martelli, and J.~Sparks, {\it {The
  holographic supersymmetric Casimir energy}},  {\em Phys. Rev.} {\bf D95}
  (2017), no.~2 021902, [\href{http://arxiv.org/abs/1606.02724}{{\tt
  arXiv:1606.02724}}].

\bibitem{Genolini:2016ecx}
P.~Benetti~Genolini, D.~Cassani, D.~Martelli, and J.~Sparks, {\it {Holographic
  renormalization and supersymmetry}},  {\em JHEP} {\bf 02} (2017) 132,
  [\href{http://arxiv.org/abs/1612.06761}{{\tt arXiv:1612.06761}}].

\bibitem{Nawata:2011un}
S.~Nawata, {\it {Localization of N=4 Superconformal Field Theory on $S^1 \times
  S^3$ and Index}},  {\em JHEP} {\bf 11} (2011) 144,
  [\href{http://arxiv.org/abs/1104.4470}{{\tt arXiv:1104.4470}}].

\bibitem{Benini:2011nc}
F.~Benini, T.~Nishioka, and M.~Yamazaki, {\it {4d Index to 3d Index and 2d
  TQFT}},  {\em Phys. Rev.} {\bf D86} (2012) 065015,
  [\href{http://arxiv.org/abs/1109.0283}{{\tt arXiv:1109.0283}}].

\bibitem{Closset:2013sxa}
C.~Closset and I.~Shamir, {\it {The $\mathcal{N}=1$ Chiral Multiplet on
  $T^2\times S^2$ and Supersymmetric Localization}},  {\em JHEP} {\bf 03}
  (2014) 040, [\href{http://arxiv.org/abs/1311.2430}{{\tt arXiv:1311.2430}}].

\bibitem{Kim:2009wb}
S.~Kim, {\it {The Complete superconformal index for N=6 Chern-Simons theory}},
  {\em Nucl. Phys.} {\bf B821} (2009) 241--284,
  [\href{http://arxiv.org/abs/0903.4172}{{\tt arXiv:0903.4172}}]. [Erratum:
  Nucl. Phys.B864,884(2012)].

\bibitem{Hosomichi:2014hja}
K.~Hosomichi, {\it {A Review on SUSY Gauge Theories on $\mathbf{S^3}$}},  in
  {\em New Dualities of Supersymmetric Gauge Theories} (J.~Teschner, ed.),
  pp.~307--338.
\newblock 2016.
\newblock \href{http://arxiv.org/abs/1412.7128}{{\tt arXiv:1412.7128}}.

\bibitem{Felder}
G.~{Felder} and A.~{Varchenko}, {\it {The elliptic gamma function and
  $SL(3,Z)\ltimes Z^3$}},  {\em ArXiv Mathematics e-prints} (July, 1999)
  [\href{http://arxiv.org/abs/math/9907061}{{\tt math/9907061}}].

\bibitem{Ardehali:2015hya}
A.~Arabi~Ardehali, J.~T. Liu, and P.~Szepietowski, {\it {High-Temperature
  Expansion of Supersymmetric Partition Functions}},  {\em JHEP} {\bf 07}
  (2015) 113, [\href{http://arxiv.org/abs/1502.07737}{{\tt arXiv:1502.07737}}].

\bibitem{DiPietro:2014bca}
L.~Di~Pietro and Z.~Komargodski, {\it {Cardy formulae for SUSY theories in $d
  =$ 4 and $d =$ 6}},  {\em JHEP} {\bf 12} (2014) 031,
  [\href{http://arxiv.org/abs/1407.6061}{{\tt arXiv:1407.6061}}].

\bibitem{Martelli:2015kuk}
D.~Martelli and J.~Sparks, {\it {The character of the supersymmetric Casimir
  energy}},  {\em JHEP} {\bf 08} (2016) 117,
  [\href{http://arxiv.org/abs/1512.02521}{{\tt arXiv:1512.02521}}].

\bibitem{Dolan:2008qi}
F.~A. Dolan and H.~Osborn, {\it {Applications of the Superconformal Index for
  Protected Operators and q-Hypergeometric Identities to N=1 Dual Theories}},
  {\em Nucl. Phys.} {\bf B818} (2009) 137--178,
  [\href{http://arxiv.org/abs/0801.4947}{{\tt arXiv:0801.4947}}].

\bibitem{Aharony:2003sx}
O.~Aharony, J.~Marsano, S.~Minwalla, K.~Papadodimas, and M.~Van~Raamsdonk, {\it
  {The Hagedorn - deconfinement phase transition in weakly coupled large N
  gauge theories}},  {\em Adv. Theor. Math. Phys.} {\bf 8} (2004) 603--696,
  [\href{http://arxiv.org/abs/hep-th/0310285}{{\tt hep-th/0310285}}].
  [,161(2003)].

\bibitem{Sundborg:1999ue}
B.~Sundborg, {\it {The Hagedorn transition, deconfinement and N=4 SYM theory}},
   {\em Nucl. Phys.} {\bf B573} (2000) 349--363,
  [\href{http://arxiv.org/abs/hep-th/9908001}{{\tt hep-th/9908001}}].

\bibitem{Benvenuti:2006qr}
S.~Benvenuti, B.~Feng, A.~Hanany, and Y.-H. He, {\it {Counting BPS Operators in
  Gauge Theories: Quivers, Syzygies and Plethystics}},  {\em JHEP} {\bf 11}
  (2007) 050, [\href{http://arxiv.org/abs/hep-th/0608050}{{\tt
  hep-th/0608050}}].

\bibitem{Martelli:2006yb}
D.~Martelli, J.~Sparks, and S.-T. Yau, {\it {Sasaki-Einstein manifolds and
  volume minimisation}},  {\em Commun. Math. Phys.} {\bf 280} (2008) 611--673,
  [\href{http://arxiv.org/abs/hep-th/0603021}{{\tt hep-th/0603021}}].

\bibitem{Kim:2012ava}
H.-C. Kim and S.~Kim, {\it {M5-branes from gauge theories on the 5-sphere}},
  {\em JHEP} {\bf 05} (2013) 144, [\href{http://arxiv.org/abs/1206.6339}{{\tt
  arXiv:1206.6339}}].

\bibitem{Bobev:2015kza}
N.~Bobev, M.~Bullimore, and H.-C. Kim, {\it {Supersymmetric Casimir Energy and
  the Anomaly Polynomial}},  {\em JHEP} {\bf 09} (2015) 142,
  [\href{http://arxiv.org/abs/1507.08553}{{\tt arXiv:1507.08553}}].

\bibitem{Sen:2007qy}
A.~Sen, {\it {Black Hole Entropy Function, Attractors and Precision Counting of
  Microstates}},  {\em Gen. Rel. Grav.} {\bf 40} (2008) 2249--2431,
  [\href{http://arxiv.org/abs/0708.1270}{{\tt arXiv:0708.1270}}].

\bibitem{Dabholkar:2010uh}
A.~Dabholkar, J.~Gomes, and S.~Murthy, {\it {Quantum black holes, localization
  and the topological string}},  {\em JHEP} {\bf 06} (2011) 019,
  [\href{http://arxiv.org/abs/1012.0265}{{\tt arXiv:1012.0265}}].

\bibitem{Dabholkar:2011ec}
A.~Dabholkar, J.~Gomes, and S.~Murthy, {\it {Localization \& Exact
  Holography}},  {\em JHEP} {\bf 04} (2013) 062,
  [\href{http://arxiv.org/abs/1111.1161}{{\tt arXiv:1111.1161}}].

\bibitem{Dabholkar:2012nd}
A.~Dabholkar, S.~Murthy, and D.~Zagier, {\it {Quantum Black Holes, Wall
  Crossing, and Mock Modular Forms}},
  \href{http://arxiv.org/abs/1208.4074}{{\tt arXiv:1208.4074}}.

\bibitem{Spiridonov:2012ww}
V.~P. Spiridonov and G.~S. Vartanov, {\it {Elliptic hypergeometric integrals
  and 't Hooft anomaly matching conditions}},  {\em JHEP} {\bf 06} (2012) 016,
  [\href{http://arxiv.org/abs/1203.5677}{{\tt arXiv:1203.5677}}].

\bibitem{Shaghoulian:2016gol}
E.~Shaghoulian, {\it {Modular Invariance of Conformal Field Theory on
  $S^1\times S^3$ and Circle Fibrations}},  {\em Phys. Rev. Lett.} {\bf 119}
  (2017), no.~13 131601, [\href{http://arxiv.org/abs/1612.05257}{{\tt
  arXiv:1612.05257}}].

\bibitem{Sen:2005wa}
A.~Sen, {\it {Black hole entropy function and the attractor mechanism in higher
  derivative gravity}},  {\em JHEP} {\bf 09} (2005) 038,
  [\href{http://arxiv.org/abs/hep-th/0506177}{{\tt hep-th/0506177}}].

\bibitem{Morales:2006gm}
J.~F. Morales and H.~Samtleben, {\it {Entropy function and attractors for AdS
  black holes}},  {\em JHEP} {\bf 10} (2006) 074,
  [\href{http://arxiv.org/abs/hep-th/0608044}{{\tt hep-th/0608044}}].

\bibitem{Dias:2007dj}
O.~J.~C. Dias and P.~J. Silva, {\it {Euclidean analysis of the entropy
  functional formalism}},  {\em Phys. Rev.} {\bf D77} (2008) 084011,
  [\href{http://arxiv.org/abs/0704.1405}{{\tt arXiv:0704.1405}}].

\bibitem{Suryanarayana:2007rk}
N.~V. Suryanarayana and M.~C. Wapler, {\it {Charges from Attractors}},  {\em
  Class. Quant. Grav.} {\bf 24} (2007) 5047--5072,
  [\href{http://arxiv.org/abs/0704.0955}{{\tt arXiv:0704.0955}}].

\bibitem{Hosseini:2016tor}
S.~M. Hosseini and A.~Zaffaroni, {\it {Large $N$ matrix models for 3d ${\cal
  N}=2$ theories: twisted index, free energy and black holes}},  {\em JHEP}
  {\bf 08} (2016) 064, [\href{http://arxiv.org/abs/1604.03122}{{\tt
  arXiv:1604.03122}}].

\bibitem{Lorenzen:2014pna}
J.~Lorenzen and D.~Martelli, {\it {Comments on the Casimir energy in
  supersymmetric field theories}},  {\em JHEP} {\bf 07} (2015) 001,
  [\href{http://arxiv.org/abs/1412.7463}{{\tt arXiv:1412.7463}}].

\bibitem{Blazquez-Salcedo:2017ghg}
J.~L. Blazquez-Salcedo, J.~Kunz, F.~Navarro-Lerida, and E.~Radu, {\it
  {Squashed, magnetized black holes in $D=5$ minimal gauged supergravity}},
  2018.

\bibitem{Cassani:2018mlh}
D.~Cassani and L.~Papini, {\it {Squashing the Boundary of Supersymmetric
  AdS$_5$ Black Holes}},  {\em JHEP} {\bf 12} (2018) 037,
  [\href{http://arxiv.org/abs/1809.02149}{{\tt arXiv:1809.02149}}].

\bibitem{Markeviciute:2018yal}
J.~Markeviciute and J.~E. Santos, {\it {Evidence for the existence of a novel
  class of supersymmetric black holes with AdS$_5\times$S$^5$ asymptotics}},
  \href{http://arxiv.org/abs/1806.01849}{{\tt arXiv:1806.01849}}.

\bibitem{Markeviciute:2018cqs}
J.~Markeviciute, {\it {Rotating Hairy Black Holes in AdS$_5\times$S$^5$}},
  \href{http://arxiv.org/abs/1809.04084}{{\tt arXiv:1809.04084}}.

\bibitem{Hosseini:2018dob}
S.~M. Hosseini, K.~Hristov, and A.~Zaffaroni, {\it {A note on the entropy of
  rotating BPS AdS$_7\times S^4$ black holes}},  {\em JHEP} {\bf 05} (2018)
  121, [\href{http://arxiv.org/abs/1803.07568}{{\tt arXiv:1803.07568}}].

\bibitem{Gauntlett:2003fk}
J.~P. Gauntlett and J.~B. Gutowski, {\it {All supersymmetric solutions of
  minimal gauged supergravity in five-dimensions}},  {\em Phys. Rev.} {\bf D68}
  (2003) 105009, [\href{http://arxiv.org/abs/hep-th/0304064}{{\tt
  hep-th/0304064}}]. [Erratum: Phys. Rev.D70,089901(2004)].

\bibitem{Cassani:2015upa}
D.~Cassani, J.~Lorenzen, and D.~Martelli, {\it {Comments on supersymmetric
  solutions of minimal gauged supergravity in five dimensions}},  {\em Class.
  Quant. Grav.} {\bf 33} (2016), no.~11 115013,
  [\href{http://arxiv.org/abs/1510.01380}{{\tt arXiv:1510.01380}}].

\bibitem{Ortin:2002qb}
T.~Ortin, {\it {A Note on Lie-Lorentz derivatives}},  {\em Class. Quant. Grav.}
  {\bf 19} (2002) L143--L150, [\href{http://arxiv.org/abs/hep-th/0206159}{{\tt
  hep-th/0206159}}].

\bibitem{Spiridonov:2010em}
V.~P. Spiridonov, {\it {Elliptic beta integrals and solvable models of
  statistical mechanics}},  {\em Contemp. Math.} {\bf 563} (2012) 181--211,
  [\href{http://arxiv.org/abs/1011.3798}{{\tt arXiv:1011.3798}}].

\bibitem{Friedman_2004}
E.~Friedman and S.~Ruijsenaars, {\it Shintani{\textendash}barnes zeta and gamma
  functions},  {\em Advances in Mathematics} {\bf 187} (oct, 2004) 362--395.

\end{thebibliography}

\providecommand{\href}[2]{#2}\begingroup\raggedright\endgroup

\end{document}